\documentclass[twocolumn]{aastex63}

\newcommand{\angstrom}{\textup{\AA}}
\newcommand{\ML}{M_{\rm dyn}/L}
\newcommand{\MM}{M_{\rm dyn}/M_* }

\usepackage{amsmath}

\shortauthors{De Graaff et al.}
\graphicspath{{./}{figures/}}

\begin{document}

\title{The Fundamental Plane in the LEGA-C Survey: unraveling the $M/L$ variations of massive star-forming and quiescent galaxies at $z\sim0.8$}

\author{Anna de Graaff}
\affiliation{Leiden Observatory, Leiden University, P.O.Box 9513, NL-2300 AA Leiden, The Netherlands; \url{ graaff@strw.leidenuniv.nl}}

\author{Rachel Bezanson}\affiliation{Department of Physics and Astronomy, University of Pittsburgh, Pittsburgh, PA 15260, USA}

\author{Marijn Franx}\affiliation{Leiden Observatory, Leiden University, P.O.Box 9513, NL-2300 AA Leiden, The Netherlands; \url{ graaff@strw.leidenuniv.nl}}

\author{Arjen van der Wel}\affiliation{Sterrenkundig Observatorium, Universiteit Gent, Krijgslaan 281 S9, B-9000 Gent, Belgium}\affiliation{Max-Planck-Institut f\"ur Astronomie, K\"onigstuhl 17, D-69117, Heidelberg, Germany}

\author{Bradford Holden}\affiliation{UCO/Lick Observatory, University of California, Santa Cruz, CA 95064, USA}

\author{Jesse van de Sande}\affiliation{Sydney Institute for Astronomy, School of Physics, A28, The University of Sydney, NSW, 2006, Australia}\affiliation{ARC Centre of Excellence for All Sky Astrophysics in 3 Dimensions (ASTRO 3D), Australia}

\author{Eric F. Bell}\affiliation{Department of Astronomy, University of Michigan, 1085 S. University Avenue, Ann Arbor, MI 48109, USA}

\author{Francesco D'Eugenio}\affiliation{Sterrenkundig Observatorium, Universiteit Gent, Krijgslaan 281 S9, B-9000 Gent, Belgium}

\author{Michael V. Maseda}\affiliation{Leiden Observatory, Leiden University, P.O.Box 9513, NL-2300 AA Leiden, The Netherlands; \url{ graaff@strw.leidenuniv.nl}}

\author{Adam Muzzin}\affiliation{Department of Physics and Astronomy, York University, 4700 Keele St., Toronto, Ontario, M3J 1P3, Canada}


\author{David Sobral}\affiliation{Department of Physics, Lancaster University, Lancaster LA1 4YB, UK}

\author{Caroline M.S. Straatman}\affiliation{Sterrenkundig Observatorium, Universiteit Gent, Krijgslaan 281 S9, B-9000 Gent, Belgium}

\author{Po-Feng Wu}\affiliation{National Astronomical Observatory of Japan, 2-21-1 Osawa, Mitaka, Tokyo 181-8588, Japan}



\begin{abstract}
We explore the connection between the kinematics, structures and stellar populations of massive galaxies at $0.6<z<1.0$ using the Fundamental Plane (FP). Combining stellar kinematic data from the Large Early Galaxy Astrophysics Census (LEGA-C) survey with structural parameters measured from deep \textit{Hubble Space Telescope} imaging, we obtain a sample of 1419 massive ($\log(M_*/M_\odot) >10.5$) galaxies that span a wide range in morphology, star formation activity and environment, and therefore is representative of the massive galaxy population at $z\sim0.8$. We find that quiescent and star-forming galaxies occupy the parameter space of the $g$-band FP differently and thus have different distributions in the dynamical mass-to-light ratio ($\ML_g$), largely owing to differences in the stellar age and recent star formation history, and, to a lesser extent, the effects of dust attenuation. In contrast, we show that both star-forming and quiescent galaxies lie on the same mass FP at $z\sim 0.8$, with a comparable level of intrinsic scatter about the plane. We examine the variation in $M_{\rm dyn}/M_*$ through the thickness of the mass FP, finding no significant residual correlations with stellar population properties, S\'ersic index, or galaxy overdensity. Our results suggest that, at fixed size and velocity dispersion, the variations in $\ML_g$ of massive galaxies reflect an approximately equal contribution of variations in $M_*/L_g$, and variations in the dark matter fraction or initial mass function.

\end{abstract}



\section{Introduction} \label{sec:intro}

{The stellar kinematics, sizes and luminosities of quiescent galaxies are strongly correlated, forming a tight scaling relation known as the Fundamental Plane \citep[FP; e.g.,][]{Djorgovski1987,Dressler1987,Jorgensen1996}. Star-forming galaxies, on the other hand, have been shown to follow a linear scaling relation between the galaxy kinematics and luminosity \citep[the Tully-Fisher relation;][]{TullyFisher1977}. However, with few modifications to the FP, star-forming galaxies may be found to lie on the same planar scaling relation as the quiescent galaxy population, as was first demonstrated at $z\sim0$ by \citet{Zaritsky2008}. These observations raise the question of how galaxies settle onto the FP at higher redshift, and thus how the positions of galaxies within the FP, both at low and high redshifts, are related to different galaxy properties and their assembly histories. }

In the local Universe, galaxies have bimodal distributions in their colors and structures. At high stellar mass, the majority of galaxies have low star formation rates (SFRs) and therefore red colors, in stark contrast with the blue, star-forming population that is dominant at lower stellar masses \citep{Blanton2003}. The color bimodality becomes even more pronounced after correcting for reddening due to dust \citep{Wyder2007,Taylor2015}, and is tightly linked with the morphological type \citep{Roberts1994,Kauffmann2003}, as blue galaxies tend to form flattened disks with exponential surface brightness profiles. Red, quiescent galaxies, on the other hand, are rounder in shape and have more centrally concentrated light profiles. The morphological properties are also correlated with the dynamical structure: on average, quiescent galaxies have a lower (projected) angular momentum, with a subset being pressure-supported entirely, whereas the star-forming disks are dynamically cold and supported primarily by rotation \citep[e.g.,][]{Romanowsky2012,Cappellari2016, vdSande2018}. 

Moreover, at fixed stellar mass quiescent galaxies are systematically smaller than star-forming galaxies, a result which holds up to $z\sim 3$ \citep{Franx2008,vdWel2014,Mowla2019,Suess2019a}. The rate of size growth also differs, pointing toward different growth mechanisms for disks \citep[e.g,][]{Mo1998,Somerville2008} and spheroids \citep[e.g.,][]{Hopkins2009,Naab2009,Bezanson2009}. On the other hand, differences in the colors and structures between the two populations begin to fade toward higher redshifts. The bimodality in color extends at least to $z\sim3$, but with bluer dust-corrected colors overall and with star-forming galaxies forming an increasingly larger fraction of the total population \citep{Brammer2009,Whitaker2011,Muzzin2013b}. Structurally, observations indicate that quiescent galaxies become more similar to the star-forming population at higher redshift, as they are more flattened and have less concentrated light profiles \citep{Chevance2012,vdWel2014b,Hill2019}. 
Consistent with the observed flattened morphologies, \citet{Belli2017}, \citet{Toft2017} and \citet{Newman2018} show that even very massive quiescent galaxies can have significant rotational support at $z\sim 2$, and \citet{Bezanson2018a} find a systematic increase in their rotational support at $z\sim0.8$ with respect to $z\sim0$.

Crucially, this leads to the question of how the evolution in color is coupled to the observed growth in size and change in structure of galaxies.
Scaling relations offer a statistical framework within which we can assess the properties of the bimodal galaxy population as well as possible evolutionary mechanisms. 
{For quiescent galaxies, the most commonly studied relation is the FP, which connects the stellar velocity dispersion, effective radius and surface brightness with a remarkably low scatter \citep[e.g.,][]{Djorgovski1987,Dressler1987,Jorgensen1996}.} The zero point of the FP and the tilt with respect to the virial plane can be interpreted in terms of the dynamical mass-to-light ratio ($M_{\rm dyn}/L$): the zero point is directly proportional to $\log (M_{\rm dyn}/L)$ \citep{Faber1987}, whereas the tilt of the FP reflects a dependence of $M_{\rm dyn}/L$ on mass, which can be due to systematic variations in the galaxy structure or the stellar population properties \citep[e.g.,][]{Bender1992,Trujillo2004,Cappellari2006,HydeBenardi2009,Graves2009II,Graves2010III,Cappellari2013}.

The low-redshift FP has been used extensively to study the properties and formation of the quiescent population. There is a correlation with stellar age and $\alpha$-element abundance through the thickness of the FP of early-type galaxies at $z\sim0$ \citep[e.g.,][]{Forbes1998,Gargiulo2009,Graves2009II}, which \citet{Gargiulo2009} show is consistent with a dissipational merger formation scenario for early-type galaxies. By mapping galaxy properties throughout the FP, \citet{Graves2010III} found that the position perpendicular to the FP depends not only on the star formation history, but also on structural properties, and suggest that the link between these two is most readily explained by differences in the truncation time of star formation, although dissipational mergers may also play a role. 

Studies of the FP at different redshifts provide additional constraints on the evolution of quiescent galaxies. The rapid change in the zero point of the FP, corresponding to a strong decrease in $M_{\rm dyn}/L$ toward higher redshift, has been used to estimate the formation epoch of massive quiescent galaxies \citep[e.g.,][]{vanDokkum1996,vdWel2005,VanDokkum2007,vdSande2014}. On the other hand, the redshift dependence of the tilt of the FP has been subject to debate, with several authors reporting a rotation in the FP at intermediate redshift with respect to the local FP \citep[e.g.][]{diSerego2005,Joergensen2013,Saracco2020}. Others find no significant change in the tilt after taking into account selection effects \citep{Holden2010}, or only very weak evidence \citep{Saglia2010,Saglia2016}, therefore leading to diverging conclusions on the mass dependence of the rate of change in $\ML$ with redshift \citep[e,g.,][]{diSerego2005,Holden2010}, as well as the slope of the stellar initial mass function \citep[IMF;][]{Renzini1993}.

The difficulty of measuring absorption line kinematics for faint sources has thus far restricted studies of the FP at higher redshifts to relatively small numbers of galaxies that are either very bright or reside in high-density environments \citep[e.g.,][]{Holden2010,vdSande2014,Beifiori2017,Prichard2017,Saracco2020}. 
\citet{vdSande2014} demonstrate that, as a result of their selection on luminosity, the colors of their sample are not representative of the main quiescent galaxy population, which steepens the inferred evolution in $\ML$ if left uncorrected. Moreover, the FP differs for galaxies in clusters and in the field at both low and intermediate redshifts \citep[e.g.,][]{LaBarbera2010,Saglia2010,Joachimi2015}, due to a systematic difference in age and possibly structure. These selection criteria, in addition to the effect of progenitor bias \citep{VanDokkum2001_pbias}, can therefore lead to a significant bias in the inferred evolution of quiescent galaxies. The effects of selection biases are often difficult to model, however, particularly when the sample size is small.

Interestingly, \citet{Zaritsky2008}, \citet{Bezanson2015}, and more recently \citet{Aquino2020} have demonstrated that star-forming and quiescent galaxies may lie on the same planar scaling relation at low redshift, provided that both the stellar mass-to-light ratios ($M_*/L$) and rotation velocities are taken into account. The tilt and zero point of the mass FP, which is obtained by substituting the surface brightness in the luminosity FP with the stellar mass surface density, therefore appear to be insensitive to the significant variation in galaxy color and structure. \citet{Bezanson2015} show that this result likely holds out to $z\sim 1$, although with a different zero point from the mass FP at $z\sim 0$. In apparent tension with observations of the Tully-Fisher relation of star-forming galaxies, which is independent of the galaxy size or surface brightness \citep[e.g.,][]{Zwaan1995,Courteau1999}, 
these results suggest that previous FP analyses can be extended to the star-forming population, which would allow for the galaxy population to be studied as a whole and hence minimize the impact of selection effects and progenitor bias.

In this paper, we present the luminosity and mass FP of both star-forming and quiescent galaxies at $z\sim0.8$ from
the Large Early Galaxy Astrophysics Census (LEGA-C) survey \citep{vdWel2016,Straatman2018}, which provides deep continuum spectroscopy for a large, $K_{\rm s}$-band selected sample of galaxies at $0.6<z<1.0$. 
We explore systematic variations in the structural, environmental and stellar population properties within the scatter of the FP, to study the connection between the stellar populations and structures of massive galaxies at $z\sim0.8$.

The paper is structured as follows. We describe the data sets used, the sample selection criteria and our spectral energy distribution (SED) modeling in Section~\ref{sec:data}. We examine the dependence of the scatter in the luminosity FP on variations in $M_*/L$ in Section~\ref{sec:lfp_results}. We present the mass FP in Section~\ref{sec:mfp_results} and discuss correlations with galaxy structure and environment. The implications of our findings are discussed in Section~\ref{sec:discussion} and summarized in Section~\ref{sec:conclusion}.

We assume a flat $\Lambda$CDM cosmology throughout, with $\Omega_{\rm m}=0.3$ and $H_0=70\,\rm km\,s^{-1}\,Mpc^{-1}$. All magnitudes are in the AB photometric system.

\section{Data}\label{sec:data}

\subsection{The LEGA-C Survey}\label{sec:legac}

The LEGA-C survey \citep{vdWel2016,Straatman2018} is a deep spectroscopic survey conducted with the VIMOS spectrograph on the Very Large Telescope, targeting massive galaxies at redshifts $0.6<z<1.0$ in the COSMOS field. The primary sample of the survey consists of $\sim 3000$ $K_{\rm s}$-band magnitude selected objects, with a redshift-dependent limit $K_{\rm s} = 20.7 - 7.5\log[(1+z)/1.8]$, corresponding to stellar masses of $\log(M_*/M_\odot) \gtrsim 10$. Each target was observed for a total of $\sim 20\,$h at a resolution of $R\sim 2500$ in the wavelength range $\sim 6300-8800\,\angstrom$, resulting in spectra which reach a typical continuum signal-to-noise level of $S/N\approx 20\,\angstrom^{-1}$. Here, we use the third data release of the LEGA-C survey, comprising 4209 spectra (including duplicate observations) which were reduced in a similar fashion to \citet{Straatman2018}.

Integrated stellar velocity dispersions are measured from the absorption linewidths in the 1D optimally extracted spectra using the Penalized Pixel-Fitting code \citep[pPXF;][]{Cappellari2004,Cappellari2017}. As described in full detail in \citet{Straatman2018} and \citet{Bezanson2018b}, the continuum emission of each spectrum is modeled using a set of high-resolution synthetic stellar population templates, and the observed stellar velocity dispersion is measured as the Gaussian broadening of the best-fitting combination of templates. We note that this measurement differs from the \textit{intrinsic} stellar velocity dispersion: absorption lines in the 1D, spatially-integrated spectrum can also be broadened by the (projected) rotational motions of a galaxy, and hence both the intrinsic velocity dispersion and rotational velocity contribute to the integrated velocity dispersion. The inclusion of rotational motion is important, as the resulting integrated velocity dispersion approximates the second velocity moment in the virial theorem \citep[see][]{Cappellari2006}. These integrated velocity dispersions are, however, dependent on the inclination of galaxies with respect to the line of sight, especially for rotationally-supported systems. We explore the effect of inclination on our results in Section~\ref{sec:axisratio}. We correct all measured dispersions to an aperture of one effective radius using the typical correction derived by \citet{vdSande2013}, $\sigma = 1.05\times\sigma_{\rm obs}$. The same, constant correction is applied to all galaxies, which may be incorrect if there is a strong radial gradient in the profile of the velocity dispersion. However, since the aperture of the slit is on average only slightly larger than the typical effective radius of the LEGA-C galaxies, the choice of aperture correction does not have a large effect: our results and conclusions do not change significantly if we instead use the commonly adopted aperture correction by \citet{Cappellari2006}, which takes into account the ratio of the slit aperture and the effective radius.

\subsection{Ancillary data to LEGA-C}\label{sec:photometry}
Morphological information in the rest-frame optical is available for nearly all LEGA-C galaxies from \textit{HST} ACS F814W imaging in the COSMOS field \citep{Scoville2007}. Structural parameters are derived by fitting S\'ersic profiles to the ACS imaging using \textsc{Galfit} \citep{Peng2010}, following the procedures described in \citet{vdWel2012,vdWel2016}. 
The S\'ersic profile is parameterized by the S\'ersic index $n$, the effective radius along the major axis $a$, and the ratio of the minor to major axis $b/a$. In the following, we consider only the circularized effective radius $r_{\rm e}=\sqrt{ba}$, and correct all sizes to a rest-frame wavelength of $5000\,\angstrom$, following \citet{vdWel2014}. {We note that the circularized radius may not provide a good estimate of the galaxy size for disk-like morphologies, as it is dependent on the inclination angle of the system. However, as will be further discussed in Section~\ref{sec:axisratio}, by using the circularized radius (as opposed to the major axis radius) we are able to approximately counterbalance the dependence of the integrated velocity dispersion on the galaxy inclination, and thus mitigate the effects of galaxy inclination on the FP.} Lastly, we assume a nominal uncertainty of $10\%$ on the measured sizes, and $5\%$ on the integrated luminosity of the S\'ersic profile \citep[motivated by][Fig. 7]{vdWel2012}.

The LEGA-C targets were selected from the $K_{\rm s}$-selected UltraVISTA catalog constructed by \citet{Muzzin2013a}, which consists of PSF-matched photometry in 30 bands ranging from $0.15-24\,\micron$. We measure rest-frame $ U-V$ and $ V-J$ colors from the multi-wavelength photometry using the EAZY template fitting code \citep{Brammer2008} with redshifts fixed to the spectroscopic redshifts, as described in detail in \citet{Straatman2018}.

We use the MAGPHYS code \citep{daCunha2008} to fit the photometric SEDs and derive stellar population properties. MAGPHYS uses an energy balance recipe, which accounts for light absorbed by dust in the stellar birth clouds being re-radiated in the infrared. To fit the SEDs, we use the infrared libraries from \citet{daCunha2008} and the \citet{Bruzual2003} stellar population templates, and assume a Chabrier IMF \citep{Chabrier2003}, an exponentially declining star formation history (SFH) with random bursts of star formation superimposed, and a two-component dust model \citep{Charlot2000}. We fix the redshift to the spectroscopic redshift and use only a subset of the UltraVISTA photometry, consisting of all available broad bands ($uBgVrizYJHK_s$ as well as the \textit{Spitzer}/IRAC and \textit{Spitzer}/MIPS photometry).
For all SED-derived properties, we use the median of their posterior likelihood distribution and treat the 16th and 84th percentiles as $1\sigma$ uncertainties. We provide our catalog of SED properties used in this work in Appendix~\ref{sec:apdx_mstar} (Table~\ref{tab:sed}), and also show a comparison between our stellar mass estimates and those presented in \citet{vdWel2016}. Finally, we scale the stellar mass to a total stellar mass using the total luminosity of the best-fit S\'ersic profile \citep[e.g.,][]{Taylor2010}, a small correction that typically increases the stellar mass by $\sim 2\%$.

\subsection{Sample selection at $z\sim0.8$}\label{sec:sample_selection}

We select galaxies from the primary LEGA-C sample, using the flag $f_{\rm primary} = 1$ and redshift restriction $0.6\leq z\leq1.0$ (2915 spectra, of which 294 are duplicate observations). Of this sample, we select all (2477) galaxies of stellar mass $\log(M_*/M_\odot) \geq 10.5$. We exclude 51 spectra which do not meet the quality criteria described in \citet{Straatman2018} (e.g., flaws in the data reduction), as well as those (365) with a $>15\%$ uncertainty on the integrated stellar velocity dispersion. Moreover, we require that the \textsc{Galfit} fit has converged within the parameter constraints, leaving 1656 objects, of which 167 are duplicate observations. We visually inspect the model and residual images and flag galaxies with significant residual flux ($f_{\rm morph}$), which show merger activity or for which a two-component fit would be more appropriate (e.g., due to the presence of a point-source AGN, or star-forming clumps). Lastly, we flag objects that appear to be one system in the ground-based imaging, but are found to be close pairs of galaxies in the \textit{HST} image. The resulting sample consists of 1489 unique objects, of which 66 are flagged as $f_{\rm morph}=1$ and 28 are close pairs. We inspect the close galaxy pairs in this sample, and remove pairs (4) where the line broadening in the spectrum reflects their velocity offset, rather than the internal galaxy kinematics. Our final sample, for which $f_{\rm morph}=0$, comprises 1419 galaxies. We note that including objects for which $f_{\rm morph}=1$ introduces additional scatter, but does not change the results and conclusions in this paper.

\begin{figure}
    \centering
    \includegraphics[width=\linewidth]{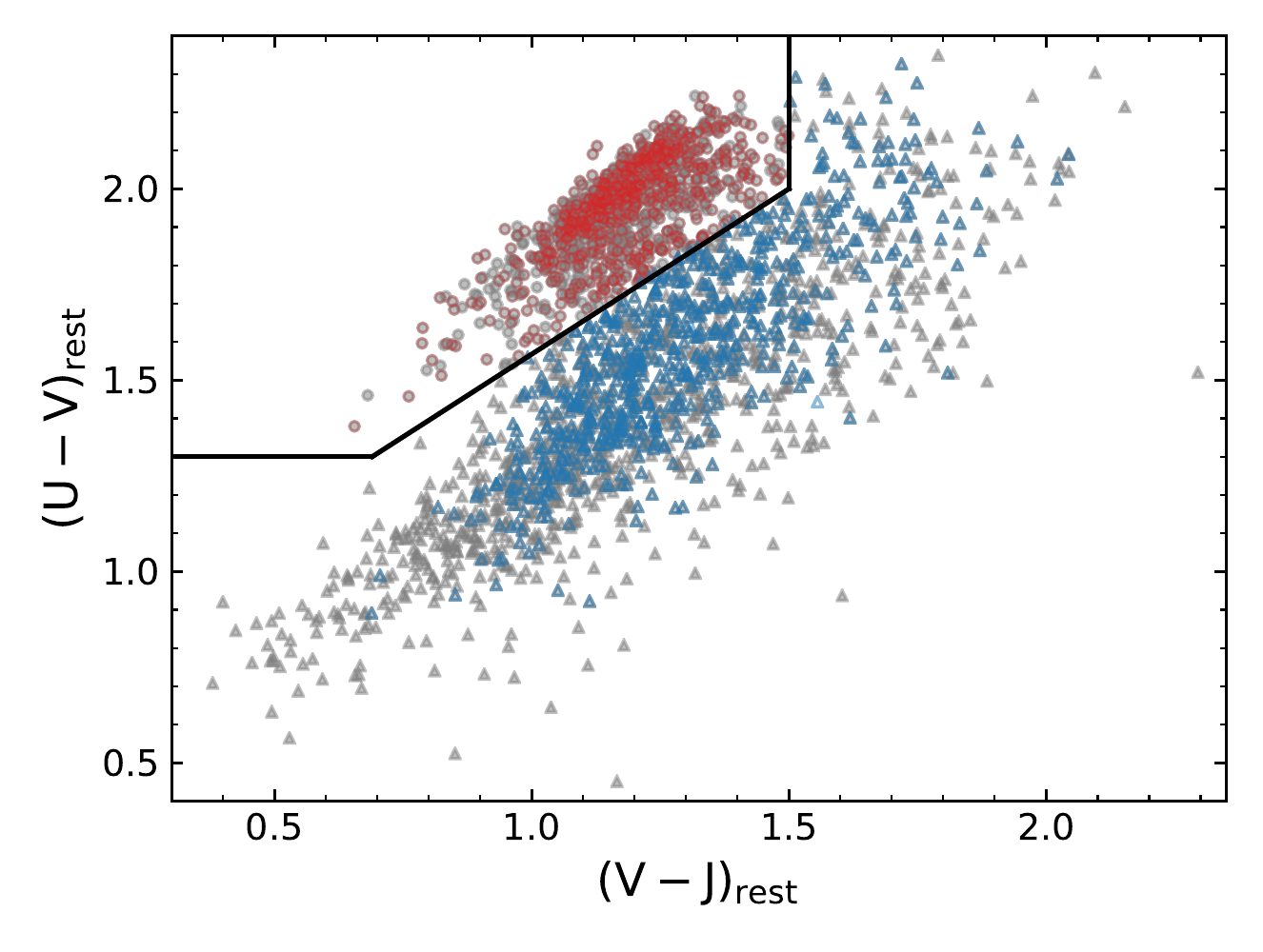}
    \caption{Rest-frame UVJ colors of galaxies in the primary sample of the LEGA-C survey at $0.6<z<1.0$. The selected sample of 1419 galaxies are highlighted in red (quiescent galaxies) and blue (star-forming galaxies), with solid lines showing the quiescent criteria from \citet{Muzzin2013b}.}
    \label{fig:uvj}
\end{figure}

We show the UVJ diagram of all (2621) primary LEGA-C galaxies at $0.6<z<1.0$ in Fig.~\ref{fig:uvj}, with the selected sample marked in red (quiescent) and blue (star-forming); we classify galaxies as quiescent and star-forming using the rest-frame $U-V$ and $V-J$ colors, following the \citet{Muzzin2013b} criteria:
\begin{align}
    U-V &> 1.3 \\
    V-J &< 1.5 \\
    U-V &> 0.69 + 0.88\,(V-J)\, .
\end{align}
Our selected sample populates a large region in the color-color space, and is therefore representative of the massive galaxy population. It does not sample the bluest colors, which can be attributed to our selection on stellar mass: LEGA-C galaxies in the lower left corner of the UVJ diagram have a typical stellar mass of $\log(M_*/M_\odot) \approx 10.0$, and are therefore excluded. The S/N criterion imposed on the velocity dispersion does introduce some bias against (massive) galaxies with very red rest-frame $V-J$ colors, typically corresponding to galaxies that are more strongly attenuated by dust and thus have a lower continuum S/N level in the spectra.

\begin{figure}
    \centering
    \includegraphics[width=\linewidth]{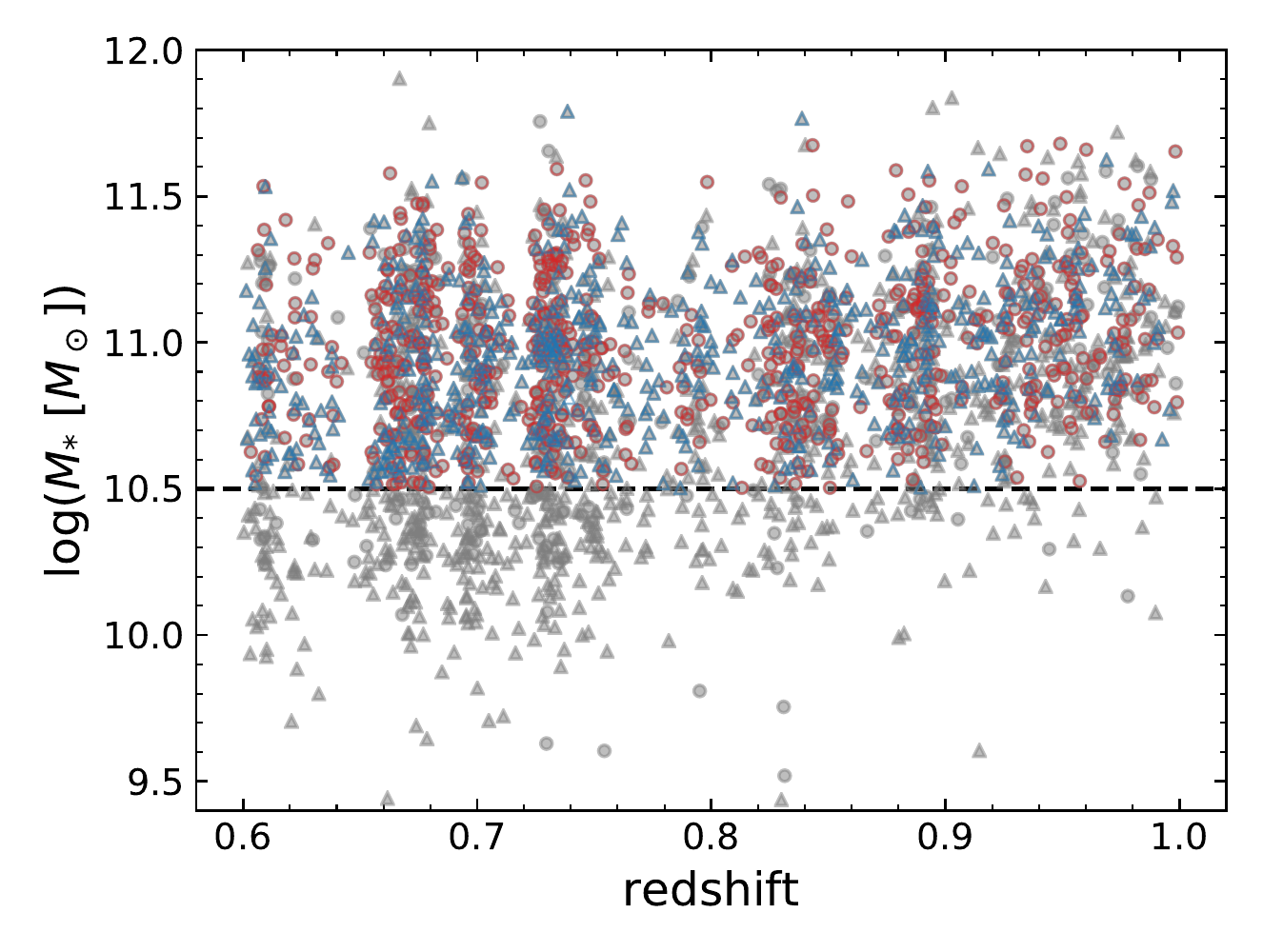}
    \caption{Stellar mass versus spectroscopic redshift of galaxies in the primary sample of the LEGA-C survey. The selected galaxies are marked in red and blue, indicating the UVJ quiescent and star-forming sample respectively. The dashed line shows the stellar mass criterion used to construct a representative sample of massive galaxies at $z\sim0.8$. There are two discernible overdensities at $z\approx0.67$ and $z\approx0.73$, comprising $\sim 40\%$ of the sample.}
    \label{fig:mstar_z}
\end{figure}

Fig.~\ref{fig:mstar_z} shows the distribution of the selected sample as a function of the stellar mass and redshift. The dashed line marks a stellar mass of $\log(M_*/M_\odot) =10.5$, above which we define our sample of LEGA-C galaxies (1419 objects) that is representative of galaxies of stellar mass $\log(M_*/M_\odot) \geq 10.5$ at $z\sim0.8$.

\subsection{Comparison sample at $z\sim0$}\label{sec:sdss}

We compile a reference sample of local galaxies by selecting galaxies in the redshift range $0.05<z<0.07$ from the 7$^{\rm th}$ data release of the SDSS \citep[DR7;][]{sdss:dr7}, for which $sciencePrimary =1$, $reliable=1$, $z\_warning=0$, $sn\_median>15$ and the uncertainty on the stellar velocity dispersion is $<15\%$. 
To obtain stellar mass estimates that are comparable with the LEGA-C SED fits, we match the selected SDSS sample with the MAGPHYS-derived stellar mass catalog by \citet{Chang2015}. This has the advantage that (i) the same models and fitting method are used as in Section~\ref{sec:legac}, and (ii) the photometry used spans a range in wavelength ($0.4-22\,\micron$) that is similar to the UltraVISTA photometry, since \citet{Chang2015} cross-match the SDSS photometry with WISE. 
We use the structural parameters derived by \citet{Simard2011} from the 2D single S\'ersic profile fits in the $r$-band. {As there are multiple structural parameter catalogs available for the SDSS, we examine the effect of our choice of the catalog used in Appendix~\ref{sec:apdx_sersic}, finding no significant differences in the resulting FP. } Following Section~\ref{sec:legac}, we consider only the circularized effective radius, and correct the stellar mass estimates for missing flux using the total luminosity of the S\'ersic profile. Selecting only galaxies of total stellar mass $\log(M_*/M_\odot) \geq 10.5$, our final sample contains 23,036 galaxies.

Moreover, we calculate rest-frame colors and luminosities using \textsc{kcorrect} \citep{Blanton2007}, and distinguish between quiescent and star-forming galaxies using the rest-frame $u-r$ and $r-z$ colors and the color cuts from \citet{Holden2012}:
\begin{align}
    u-r &> 2.26\,, \\
    r-z &< 0.75\,, \\
    u-r &> 0.76 + 2.5\,(r-z)\,.
\end{align}

Lastly, we consider the fact that the SDSS fiber spectra have an aperture diameter of $3\arcsec$, which covers only the central region of a galaxy at $z\approx0.06$. We use publicly available data from the Mapping Nearby Galaxies at Apache Point Observatory survey \citep[MaNGA;][]{Bundy2015} of the SDSS DR15 \citep{sdss:dr15} to assess the effect of aperture size on the integrated stellar velocity dispersion, taking into account the dependence on the effective radius, S\'ersic index, and axis ratio. As further detailed in Appendix~\ref{sec:apdx_apcor}, we hence derive a statistical aperture correction (typically $\sim 3\%$) to calculate the integrated stellar velocity dispersion within the effective radius from the fiber-derived SDSS DR7 velocity dispersions.

%

\section{Luminosity fundamental plane }\label{sec:lfp_results}

We begin by focusing on the fundamental plane in luminosity, specifically the luminosity measured in the rest-frame $g$-band. We measure the correlation between the residuals of the FP and various SED properties to explore the origin of the scatter in the FP and the differences between the star-forming and quiescent galaxy populations.

The FP describes the relation between the (integrated) stellar velocity dispersion ($\sigma$), surface brightness ($I_{\rm e}$), and effective radius ($R_{\rm e}$):
\begin{equation}
    \log R_{\rm e} = a \, \log \sigma  + b \,\log I_{\rm e} + c\,,
     \label{eq:lfp}
\end{equation}
where the coefficients $a$ and $b$ describe the tilt of the plane, and $c$ is the zero point. The parameters $R_{\rm e}$ and $\sigma$ have units of kpc and $\rm km\,s^{-1}$ respectively, and $\log I_{\rm e} \equiv -0.4\,\mu_{\rm e}$, where $\mu_{\rm e}$ is the mean surface brightness within the effective radius \citep[see, e.g.,][]{HydeBenardi2009}:
\begin{equation}
    \mu_{\rm e} = m + 2.5\,\log\left( 2\pi r_{\rm e}^2 \right) - 10\,\log(1+z)\,,
\end{equation}
where $m$ is the (rest-frame) apparent magnitude, and $r_{\rm e}$ is the effective radius in arcseconds.

\subsection{Tilt of the FP} \label{sec:lfp_tilt}

An accurate measurement of the tilt, such as in \citet{HydeBenardi2009}, requires a detailed analysis of the sample completeness in both $M_*$ and $\sigma$, as well as the uncertainties on all observed parameters. A full analysis of the tilt of the FP is beyond the scope of the current paper,
and we will therefore assume minimal evolution in the tilt of the FP throughout, adopting the measurement of the rest-frame $g$-band plane ($I_{\rm e}=I_{\rm e,\,g}$) by \citet{HydeBenardi2009} for galaxies at $z\sim 0$, of $a=1.404$ and $b=-0.761$. 

However, as discussed in Section~\ref{sec:intro}, there are several previous studies at variance with this assumption, 
{as less massive galaxies of low $\ML$ are likely to cause the FP to deviate more strongly from the virial plane toward higher redshift \citep[see, e.g.,][]{Joergensen2013}. }
Therefore, we consider here the possibility of an evolution in the tilt and its effect on the results presented in the following sections.

Following an approach similar to \citet{Jorgensen1996} and \citet{Holden2010}, we determine the best-fit values of $a$ and $b$ of the FP by minimizing the sum of the absolute orthogonal deviations,
\begin{equation}
    \Delta_{\rm LFP} = \frac{| \log\,R_{\rm e} -  a\, \log\,\sigma -  b\,\log\,I_{\rm e,\,g} - c \, |}{\sqrt{1+a^2 + b^2}}\,.
    \label{eq:deltalFP}
\end{equation}
We use the total completeness correction \citep[`Tcor', see][]{Straatman2018} as weights in the minimization, such that less luminous galaxies receive a greater weight in the fitting procedure. This completeness correction accounts for the selection function of LEGA-C galaxies with respect to the full parent sample of $K_{\rm s}$-band selected objects from the UltraVISTA catalog, and includes a $V_{\rm max}$ correction. We note, however, that this completeness correction does not correct for the additional selection criteria imposed in Section~\ref{sec:sample_selection}, such as the maximum allowed uncertainty on the integrated velocity dispersion. To mitigate a bias against low-mass galaxies of high $\ML$, we impose a minimum velocity dispersion of $\log (\sigma/ {\rm km\,s^{-1}}) >2.1$\,: this limit corresponds to a completeness in $\log\,\sigma$ of $>50\%$ up to $K_{\rm s}=20.1$ (the magnitude limit comprising $90\%$ of our sample).

As our data span a wide range in redshift and the zero point $c$ changes significantly within $0.6<z<1.0$ \citep{deGraaff2020}, we restrict our fitting to a redshift range of $\Delta z=0.10$. We measure the tilt in the range $0.65<z<0.75$, which encompasses the largest fraction of galaxies in our selected sample within the narrow window of $\Delta z=0.10$ (602 objects; see Fig.~\ref{fig:mstar_z}). For comparison with previous studies, we use only the 325 quiescent galaxies within this redshift range. 
The best-fit parameters are $a=1.29\pm0.18$ and $b=-0.62\pm0.04$ (where errors are estimated by bootstrapping the data).

This value of $a$ is in good agreement with the value of $a=1.40\pm0.05$ found by \citet{HydeBenardi2009}, the measurement by \citet{Jorgensen1996} ($a=1.24\pm0.07$), as well as the results by \citet{Holden2010}, who found $a=1.18\pm0.08$ and $a=1.19\pm0.13$ at $z\sim0$ and $z\sim 0.8$ respectively. The other parameter, $b$, appears to be in tension with these studies, including the assumed value of $b=-0.76\pm0.02$ by \citet{HydeBenardi2009} (a discrepancy of $\approx 4\sigma$).

To evaluate the dependence of the measured tilt of the FP on the fitting method used, we apply our method to the selected reference sample of $z\sim0$ galaxies (Section~\ref{sec:sdss}). Imposing the same criterion of $\log (\sigma/ {\rm km\,s^{-1}}) >2.1$, we find $a=1.296\pm0.015$ and $b=-0.732\pm0.004$. {This is indeed slightly lower than the measurement by \citet{HydeBenardi2009},} who used a more comprehensive fitting technique, and leaves a difference of $\approx 3\sigma$ in $b$ with respect to the LEGA-C measurement.

{In agreement with previous measurements of the FP of quiescent galaxies \citep[e.g.,][]{Joergensen2013,Saracco2020} we thus find a slight change in the tilt toward higher redshift. We note that there may be small systematic effects contributing to this observed evolution, as the SDSS data and LEGA-C data differ systematically in their measurements of $R_{\rm e}$, $\sigma$ and $L_g$, as well as the galaxy selection function. We further investigate the redshift dependence of the tilt in Appendix~\ref{sec:apdx_tilt},} where we consider the full redshift range of LEGA-C as well as the effects of measurement uncertainties and selection bias.

\begin{figure*}
    \centering
    \includegraphics[width=\linewidth]{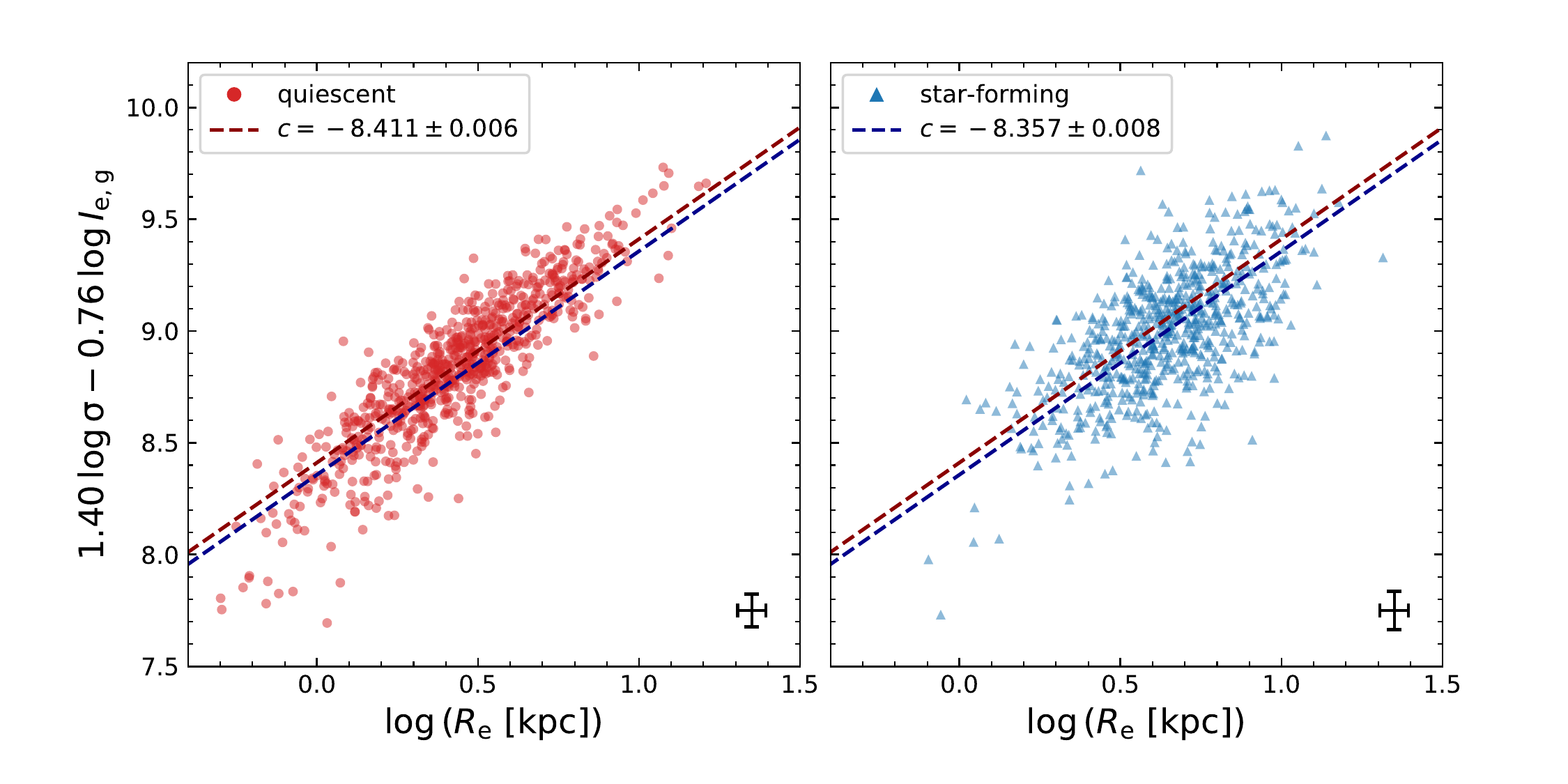}
    \caption{Edge-on view of the rest-frame $g$-band fundamental plane of quiescent (left) and star-forming (right) LEGA-C galaxies, assuming a fixed tilt from \citet{HydeBenardi2009}. Star-forming and quiescent galaxies occupy different parts of the parameter space, as they differ in their best-fit zero points (dashed lines), effective radii, and scatter about the plane {($0.139\pm0.006$\,dex and $0.085 \pm 0.004$\,dex respectively)}. Observational uncertainties are similar for both populations, therefore indicating a significantly higher intrinsic scatter for star-forming galaxies.}
    \label{fig:fp_gband}
\end{figure*}

Importantly, however, we have used the tilt measured in this section to verify that our assumption of no evolution does not affect our conclusions. If we adopt our measurement of the tilt, only the measurements of the zero points change significantly ($>3\sigma$), although the relative difference between the zero points of the quiescent and star-forming populations remains. The observed correlations within the residuals from the FP in the following sections are also largely unchanged, as the correlation coefficients change only minimally in value.

\subsection{Correlations between residuals from the FP and stellar population properties} \label{sec:ML_corr}

We fit the zero point ($c$) of the plane for the quiescent and star-forming samples separately by minimizing the mean absolute orthogonal deviation (Eq.~\ref{eq:deltalFP}) at fixed $a$ and $b$.
{We calculate the scatter about the best-fit zero point as the normalized median absolute deviation (NMAD) in $\Delta_{\rm LFP}$ (Eq.~\ref{eq:deltalFP})}, and estimate uncertainties on both quantities using bootstrap resampling. 

Fig.~\ref{fig:fp_gband} shows an edge-on projection of the $g$-band FP, for both quiescent (red) and star-forming (blue) galaxies, with dashed lines indicating the respective best-fit zero point. Traditionally, studies of the FP have focused on quiescent galaxies only \citep[e.g.,][]{Dressler1987,Jorgensen1996,vdWel2004}, as they form a tight sequence and can therefore be used as a distance indicator, or to study the evolution of the mass-to-light ratio ($M_{\rm dyn}/L$). {We confirm this result for the LEGA-C sample of quiescent galaxies, which has a scatter of $0.085 \pm 0.004$\,dex. However, we also show that star-forming galaxies seem to follow the same tilt, albeit with a larger scatter, of $0.139\pm0.006$\,dex.}  The star-forming galaxies occupy a different area of the parameter space: they are typically larger in size, consistent with the findings by \citet{vdWel2014}, { and their best-fit zero point ($c=-8.411\pm0.006$) is slightly lower than that of the quiescent population ($c=-8.357\pm0.008$, a difference of $5.4\sigma$), which corresponds to a systematic offset of $\Delta\log I_{\rm e,g} = 0.071 \pm 0.013\,$dex between the two populations.}

We estimate the intrinsic scatter in the FP using Monte Carlo simulations: assuming a FP of zero intrinsic scatter, we self-consistently vary $R_{\rm e}$, $I_{\rm e,g}$ and $\sigma$ within the observational uncertainties {(i.e., taking into account covariances between the different quantities), and calculate the resulting scatter in $\Delta_{\rm LFP}$}. By doing so for 1000 simulations, we obtain a robust estimate of the scatter in {$\Delta_{\rm LFP}$} due to observational uncertainties alone. The remaining contribution to the observed scatter then is due to intrinsic variation about the plane. {We find that the intrinsic scatter is slightly lower than the observed scatter, at $0.082\pm 0.005$\,dex and $0.134\pm0.006$\,dex for the quiescent and star-forming samples respectively}, indicating that the observed scatter is dominated by physical differences between galaxies. The value of $M_{\rm dyn}/L_g$ for the star-forming and quiescent populations therefore differs not only in the mean value, but also in the variance. This can reflect both (i) a difference in the structural properties, i.e. a systematically lower value of $\MM$ for star-forming galaxies as well as an increased intrinsic scatter in $\log \MM$, and (ii) a systematic difference in the \textit{stellar} mass-to-light ratio ($M_*/L_g$) between the two populations. Considering the UVJ color selection, a difference in $M_*/L_g$ may be expected to contribute the systematic offset between the two zero points. Moreover, the wide range in color spanned by the star-forming galaxies (Fig.~\ref{fig:uvj}) suggests that they are more strongly affected by dust attenuation, thus leading to a larger intrinsic scatter in the FP.

\begin{figure*}
    \centering
    \includegraphics[width=0.95\linewidth]{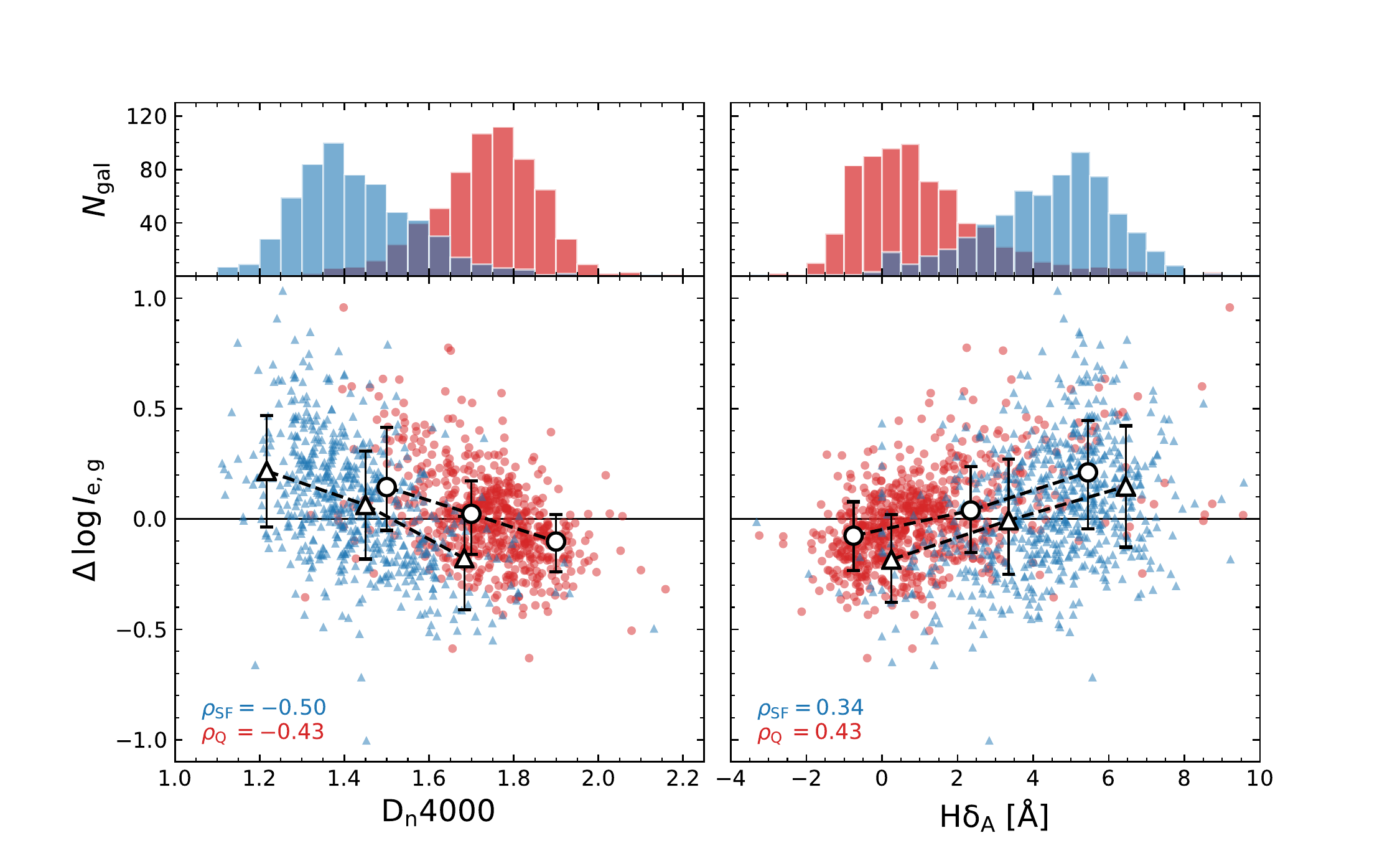}
    \caption{FP residual in $\log\,I_{\rm e, g}$ versus the spectral age indices $\rm D_n 4000$ (left) and $\rm H\delta_A$ (right). Red and blue markers indicate the quiescent and star-forming population respectively, with black open markers showing the running median and 16$^{\rm th}$ and 84$^{\rm th}$ percentiles. There is a strong correlation with $\Delta\log\,I_{\rm e, g}$ in both panels (Spearman rank correlation coefficients, $\rho$, are denoted in each panel), albeit with large scatter, such that at fixed $\sigma$ and $R_{\rm e}$ galaxies with higher surface brightness are younger. Since the distributions in $\rm D_n 4000$ and $\rm H\delta_A$ differ for the star-forming and quiescent galaxies, with the latter being older, this shows that stellar age is an important driver of the differences between the two populations in the $g$-band FP (Fig.~\ref{fig:fp_gband}).}
    \label{fig:lfp_age}
\end{figure*}

\begin{figure*}
    \centering
    \includegraphics[width=0.85\linewidth]{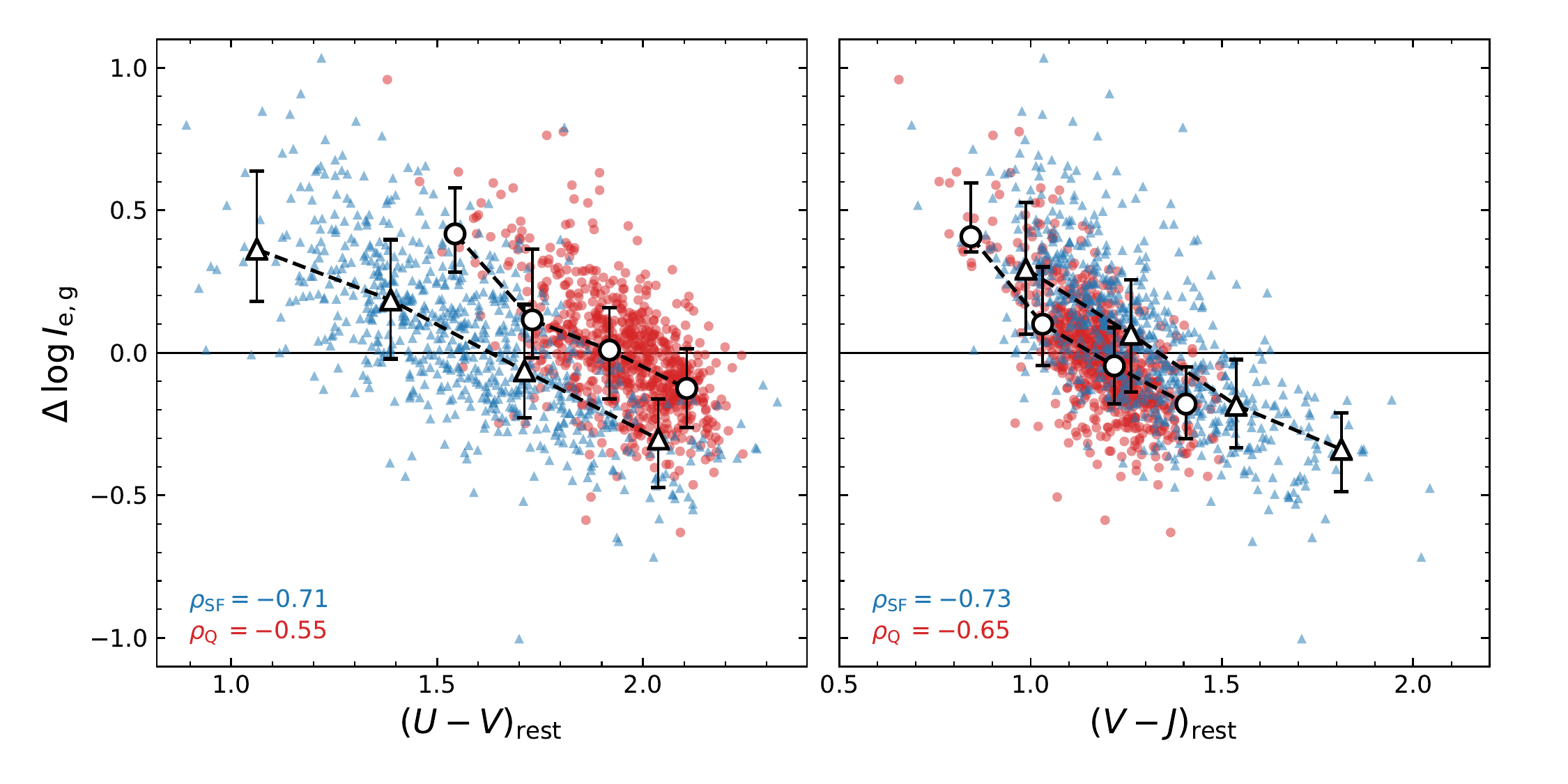}
    \caption{Correlation in the residuals from the FP with the rest-frame $U-V$ and $V-J$ colors. Symbols indicate the same as in Fig.~\ref{fig:lfp_age}. Since $U-V$ and $V-J$ in turn correlate with properties of the stellar mass-to-light ratio ($M_*/L_g$), the strong correlations through the thickness of the FP suggest that variations in stellar age, dust attenuation, and star formation activity contribute significantly to the scatter in the FP, which we explore in Fig.~\ref{fig:lfp_sed}. }
    \label{fig:lfp_colour}
\end{figure*}

We demonstrate the dependence of the scatter on $M_*/L_g$ in Figs.~\ref{fig:lfp_age} \& \ref{fig:lfp_colour}, where we show the residual from the FP in $\log\,I_{\rm e,g}$ as a function of the $4000\,\angstrom$ break \citep[D$\rm _n$4000;][]{Wu2018} and the Lick index $\rm H\delta_A$, which are age indicators measured directly from the spectra, as well as the rest-frame $U-V$ and $V-J$ colors.  Similar to the results from \citet{Graves2009II} for quiescent galaxies at $z\sim0$, we find a correlation with age (D$\rm _n$4000, $\rm H\delta_A$) through the thickness of the FP, which continues down toward younger, star-forming galaxies. Since $\Delta\log I_{\rm e,g} \approx -\Delta\log M_{\rm dyn}/L_g$, this correlation translates to a lower (higher) value of $M_{\rm dyn}/L_g$ for younger (older) galaxies. Our findings are also consistent with results by \citet{Jorgensen2019}, who find increased Balmer line absorption ($\rm H\zeta_{A}$) and lower $\ML$ for quiescent galaxies in clusters at $z\sim1$ with respect to early-type galaxies at $z\sim0$, which they interpret as being due to a difference in age.

The residuals of the FP correlate even more strongly with the rest-frame $U-V$ and $V-J$ colors (Fig.~\ref{fig:lfp_colour}), which in turn depend on a combination of dust attenuation, specific star formation rate (sSFR) and age \citep[see, e.g.,][]{Leja2019b}. Galaxies with positive values of $\Delta\,\log\,I_{\rm e,g}$ are therefore not only younger on average, they may also have a higher sSFR or be less dust-obscured, or, a combination of both. 

\begin{figure*}
    \centering
    \includegraphics[width=0.9\linewidth]{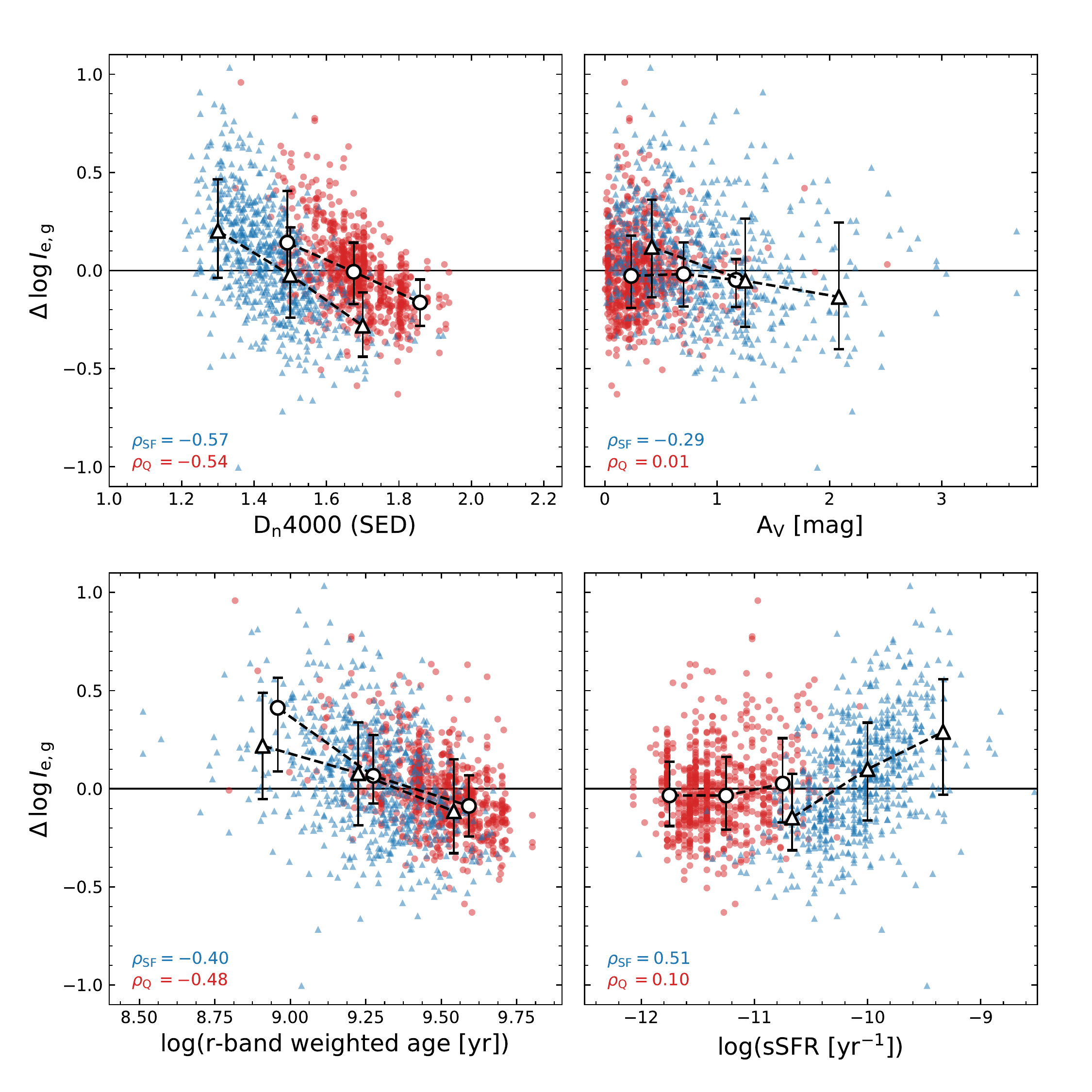}
    \caption{Correlation in the residuals from the FP with properties from the SED modeling (D$\rm _n$4000 break, dust attenuation, luminosity-weighted stellar age, and specific star formation rate), which drive the stellar mass-to-light ratio. Symbols indicate the same as in Fig.~\ref{fig:lfp_age}. For the quiescent galaxies, only the variation in stellar age and D$\rm _n$4000 (which, apart from the stellar age, is also dependent on the metallicity) contribute significantly to the intrinsic scatter of the $g$-band FP. Therefore, the increased intrinsic scatter for the star-forming population can, at least partially, be attributed to additional effects from variations in the sSFR and dust attenuation.}
    \label{fig:lfp_sed}
\end{figure*}

We explore these different contributions to the scatter using the results from our SED modeling (Section~\ref{sec:photometry}). Firstly, the upper left panel of Fig.~\ref{fig:lfp_sed} shows that the residual correlation with D$\rm _n$4000 obtained from the best-fit SED models agrees well with that from the spectra (Fig.~\ref{fig:lfp_age}): in both cases there is a strong anti-correlation between $\Delta\log I_{\rm e,g,}$ and D$\rm _n$4000, and the models are able to reproduce the observed bimodality, such that at fixed value of $\Delta\log I_{\rm e,g}$ star-forming galaxies have a lower value of D$\rm _n$4000. On the other hand, the models do not reproduce the observed, broad distribution in D$\rm _n$4000, which may be due to incompleteness in the modeling itself or the result of degeneracy between the effects of age and dust on the observed SED.

The other panels of Fig.~\ref{fig:lfp_sed} show the residual correlations with the dust attenuation ($A_{\rm V}$; measured from the best-fit SED model), the luminosity-weighted age (in the $r$-band) and the sSFR averaged over the last $100\,\rm Myr$. For quiescent galaxies, the only significant correlation is with the stellar age. On the other hand, the scatter within the star-forming population correlates not only with age, but also weakly with the dust attenuation and, more strongly, with the sSFR.

The different intrinsic scatter for the star-forming and quiescent populations as well as the offset between the FP zero points (Fig.~\ref{fig:fp_gband}) are therefore, at least in part, due to significant differences in $M_*/L_g$ between the two populations. We note that we also find the spread in all four observed properties (D$\rm _n$4000, $\rm H\delta_A$, $U-V$, $V-J$) to be slightly larger for the star-forming population than the quiescent population, which is consistent with their observed increased scatter in the FP.  {Interestingly, whereas the deviation between the best-fit zero points of the quiescent and star-forming samples is relatively small ($\Delta\log I_{\rm e,g} \approx 0.07\,$dex), we find that at a fixed value of D$\rm _n$4000 or $(U-V)_{\rm rest}$ the differences between the two populations can be up to three times greater ($\Delta\log I_{\rm e,g}\sim 0.2\,$dex), which may be due to variation in $M_*/L_g$, or differences in the structural properties. Thus far, we have neglected the effects of potential structural differences between the two populations, which we explore in full detail in the following section.}

\section{Mass fundamental plane }\label{sec:mfp_results}

In this section we use the mass FP to explore the structural properties of galaxies within the parameter space of the FP, as well as the effect of environment.
If we multiply the surface brightness of Eq.~\ref{eq:lfp} by the $M_*/L_g$ estimated from the SED modeling (Section~\ref{sec:photometry}), we obtain the stellar mass surface density ($\Sigma_*$), and hence the mass FP:
\begin{equation}
     \log R_{\rm e} = \alpha \, \log \sigma  + \beta \,\log \Sigma_* + \gamma,
     \label{eq:mfp}
\end{equation}
where $\alpha$ and $\beta$ describe the tilt of the mass FP, and $\gamma$ is the zero point.

\subsection{Tilt of the mass FP}\label{sec:mfp_tilt}

As in Section~\ref{sec:lfp_results}, we assume that the tilt of the FP does not vary significantly with redshift and adopt the results for the mass FP from \citet{HydeBenardi2009} of $\alpha = 1.629$ and $\beta=-0.84$, which was derived with an orthogonal fit to a large ($N\sim 50,000$) sample of early-type galaxies that takes into account both the measurement uncertainties and sample completeness. We again test the effect of this assumption using a more simple, orthogonal fit of the FP, and examine the possible redshift evolution of the tilt in more detail in Appendix~\ref{sec:apdx_tilt}.

We follow the same methodology as in Section~\ref{sec:lfp_tilt}, minimizing the sum of the orthogonal deviations,
\begin{equation}
    \Delta_{\rm MFP} = \frac{| \log\,R_{\rm e} -  \alpha\, \log\,\sigma -  \beta\,\log\,\Sigma_* - \gamma \, |}{\sqrt{1+\alpha^2 + \beta^2}}\,,
    \label{eq:deltaMFP}
\end{equation}
and using the total completeness corrections (Tcor) as weights. We include only quiescent galaxies in our fits for comparison with other FP studies, and exclude galaxies for which $\log( \sigma/ {\rm km\,s^{-1}})<2.1$. 

In the redshift range $0.65< z < 0.75$ we measure a best-fit tilt of $\alpha= 1.56\pm 0.12$ and $\beta= -0.68\pm 0.03 $ {(where error bars are estimated through bootstrap resampling), which is significantly different from the assumed values by \citet{HydeBenardi2009}.} However, as in Section~\ref{sec:lfp_tilt}, we find that our measurement for the SDSS differs from the tilt found by \citet{HydeBenardi2009} due to differences in the methodology used. Both $\alpha$ and $\beta$ measured from the LEGA-C data are consistent within $<2\sigma$ with our best-fit parameters for the SDSS, of $\alpha=1.432\pm 0.012$ and $\beta=-0.736 \pm 0.003$. This remains the case even when we fit the entire LEGA-C sample combined (i.e., $0.6<z<1.0$), for which we find $\alpha=1.49\pm 0.10$ and $\beta= -0.70\pm 0.02$, suggesting no significant rotation of the mass FP at $z\sim0.8$ with respect to $z\sim0$. Our results are in agreement with measurements by \citet{Zahid2016}, who found no change in the tilt of the FP with respect to the SDSS for a sample of massive quiescent galaxies at $0.1<z<0.6$. {Interestingly, these results seem to suggest that the measurement of the tilt of the mass FP, unlike the $g$-band FP, is not strongly dependent on the selection function, as was also recently shown by \citet{Bernardi2020} at low redshift. }

{However, as we have omitted the effect of measurement uncertainties in addition to a careful analysis of the selection function in our measurement of the tilt, we choose to use the values by \citet{HydeBenardi2009} rather than our own measurement. We note that we do not use the more recent measurements by \citet{Bernardi2020}, to adhere to the common convention of using circularized sizes in the FP, and to refrain from making assumptions on the effects of non-homology on the mass FP at higher redshifts. Although we do not use the tilt measured from the LEGA-C data in the rest of this paper, we have used this measurement to test the robustness of our results in the following sections against a different tilt, finding no qualitative differences.}

\subsection{Edge-on view of the mass FP}\label{sec:mfp_edgeon}

\begin{figure}
    \centering
    \includegraphics[width=\linewidth]{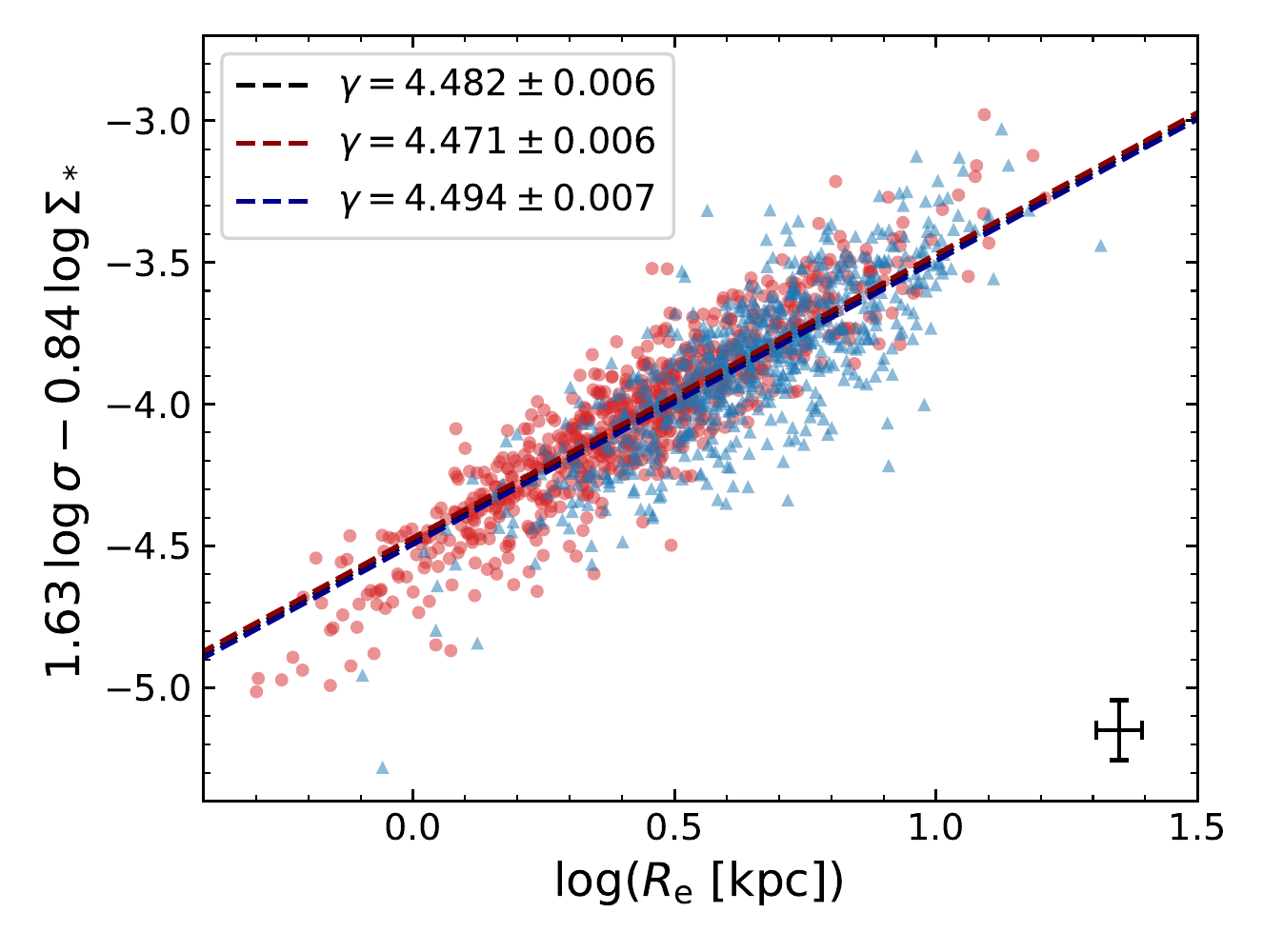}
    \caption{Edge-on view of the mass FP of quiescent (red) and star-forming (blue) LEGA-C galaxies. Dashed lines show the best-fit zero points for the star-forming, quiescent, and combined (black) samples, assuming a fixed tilt from \citet{HydeBenardi2009}. The two populations lie on the same plane: the zero points differ by only $\approx 0.02\,$dex, and the intrinsic scatter is comparable for the quiescent and star-forming samples ($0.107\pm 0.005$\,dex and $0.130\pm0.009$\,dex in $\Delta\log R_{\rm e}$ respectively). }
    \label{fig:fp_mass}
\end{figure}

Analogous to Section~\ref{sec:ML_corr}, we fit the zero point ($\gamma$) by minimizing the mean absolute orthogonal residuals at fixed $\alpha$ and $\beta$. Fig.~\ref{fig:fp_mass} shows an edge-on projection of the mass FP, for both the star-forming (blue) and quiescent (red) galaxies. The best-fit zero points are indicated by dashed lines for the two populations separately (red, blue), as well as for the joint sample (black). Not only do both populations follow the same tilt, the star-forming and quiescent galaxies also have nearly equal zero points, with the two zero points deviating by $0.023\pm0.009\,$dex {(a systematic offset of $\Delta \log\Sigma_* = 0.027 \pm 0.011\,$dex)}. This is consistent with results at low redshift by \citet{Zaritsky2008} and \citet{Bezanson2015}, although \citet{Bezanson2015} find a slightly larger offset ($\approx 0.05\,$dex) between the zero points of the two populations at both $z\sim0$ and $z\sim0.7$. At $z\sim0.7$, however, their offset is not statistically significant due to the sample size.

\begin{figure*}
    \centering
    \includegraphics[width=\linewidth]{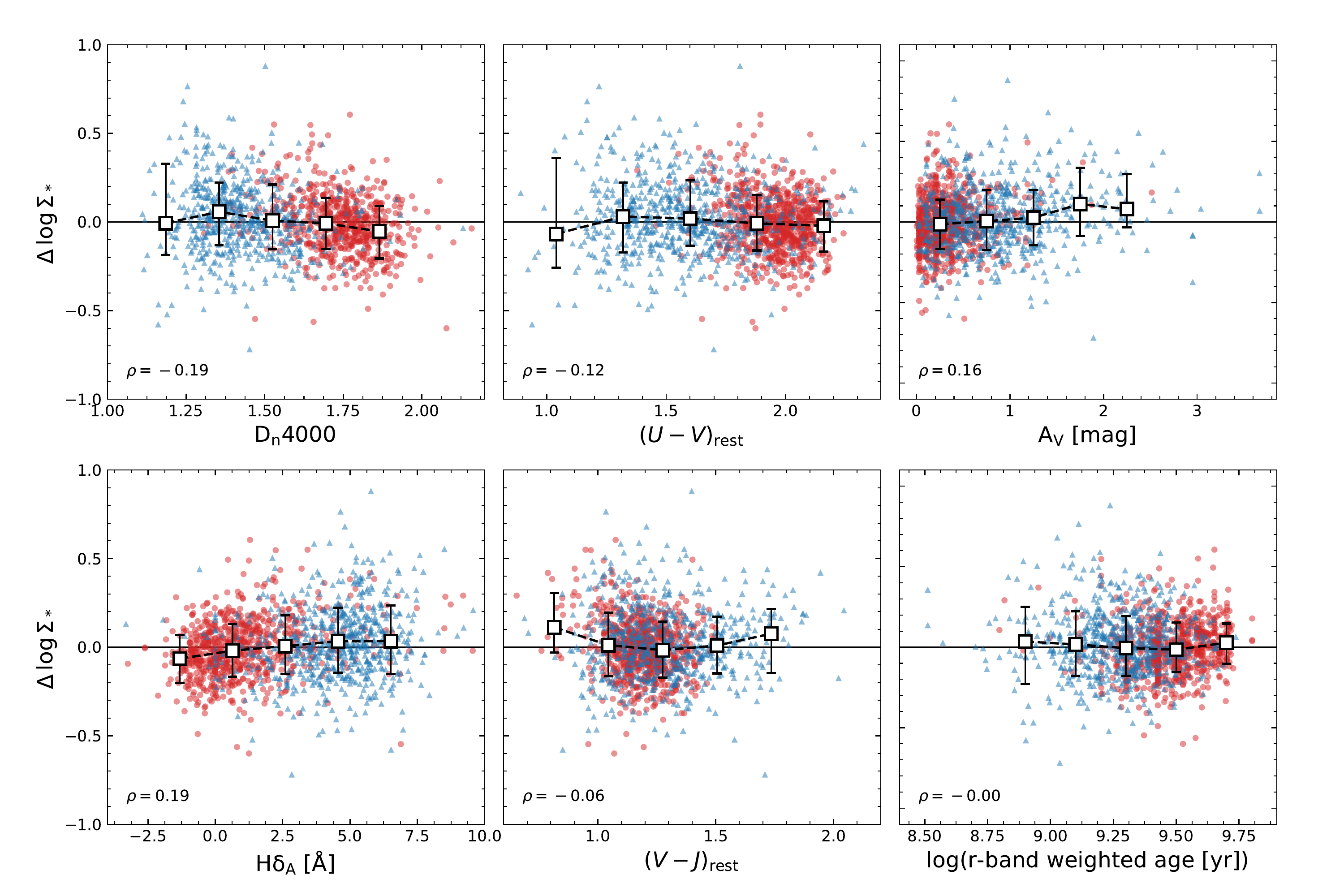}
    \caption{Residual in the mass fundamental plane in $\log\Sigma_*$ as function of the spectral age indices $\rm D_n 4000$ and $\rm H\delta_A$ (top panels), and the rest-frame $U-V$ and $V-J$ colors. Red and blue markers indicate the quiescent and star-forming population respectively, with white squares showing the median and 16$^{\rm th}$ and 84$^{\rm th}$ percentiles of the total sample (with Spearman rank correlation coefficients, $\rho$, denoted in each panel). Contrary to the results in Fig.~\ref{fig:lfp_age} for the $g$-band FP, we find no significant correlation with stellar population properties through the thickness of the mass FP.}
    \label{fig:fp_mass_age}
\end{figure*}

We find that the scatter in the mass FP is lower in comparison with the $g$-band FP, particularly so for the star-forming galaxies: {the NMAD in $\Delta_{\rm MFP}$ (Eq.~\ref{eq:deltaMFP}) is $0.069\pm0.003\,$dex and $0.085\pm0.005\,$dex for the quiescent and star-forming samples respectively, and is consistent with the findings by \citet{Bezanson2015}. Using Monte Carlo simulations, we estimate the intrinsic scatter for the quiescent and star-forming samples to be $0.058\pm 0.003$\,dex and $0.069\pm0.005$\,dex respectively.} Clearly, accounting for the $M_*/L$ dramatically lowers both the total and intrinsic scatter of the star-forming population, although it is still slightly higher than the scatter within the quiescent population. Thus, unlike the $g$-band FP, all massive galaxies occupy the same region within the 3D parameter space of the effective radius, stellar mass surface density, and stellar velocity dispersion, regardless of their color.

The remaining intrinsic scatter is low, but non-zero. In principle, a large number of galaxy properties may drive the intrinsic scatter in the mass FP: we discuss the effect of stellar populations on the FP in Section~\ref{sec:MM_corr}, the structural properties in Sections~\ref{sec:homology}~\&~\ref{sec:axisratio}, and the effect of environment in Section~\ref{sec:environment}. 

\subsection{Are the residuals from the mass FP correlated with stellar population properties?} \label{sec:MM_corr}

In Fig.~\ref{fig:fp_mass_age} we show the residual from the FP in $\log\Sigma_*$ as a function of the spectral properties D$\rm _n$4000 and $\rm H\delta_A$ (left-hand panels), the rest-frame colors $U-V$ and $V-J$ (middle panels), and the SED-derived dust attenuation and stellar age (right-hand panels). Unlike the results of Figs.~\ref{fig:lfp_age}--\ref{fig:lfp_sed}, we find no significant correlations with the different SED properties through the mass FP.
There is only a very weak correlation with the spectral age indicators (D$_n$4000 and $\rm H\delta_A$), which may correspond to the very weak residual correlation between $\Delta\log\Sigma_*$ and $A_{\rm V}$ (upper right panel) or the sSFR \citep[Spearman $\rho = 0.12$; shown in][Fig.~3]{deGraaff2020}.

To first order, the lack of residual correlations within the scatter of the mass FP demonstrates the success of our SED modeling: if we neglect a potential correlation between structural and stellar population properties, and interpret the mass FP as arising from the virial theorem, then we would expect to find no correlation between the zero point $\gamma$ and the stellar population properties of galaxies that are in virial equilibrium.

Thus far, we have simply used our SED models without questioning the underlying model assumptions, although we did show in Fig.~\ref{fig:lfp_sed} that the D$\rm _n$4000 index measured from the best-fit SEDs agree reasonably well with the measurements from the LEGA-C spectra. However, there are a large number of available SED fitting codes, with an even a larger parameter space of, e.g., possible star formation histories, dust laws and IMFs. For instance, in Appendix~\ref{sec:apdx_mstar} we compare our MAGPHYS masses to those derived with FAST \citep{Kriek2009} and find significant, systematic differences between the two, casting doubt on the accuracy of the various stellar mass estimates.

\begin{figure}
    \centering
    \includegraphics[width=\linewidth]{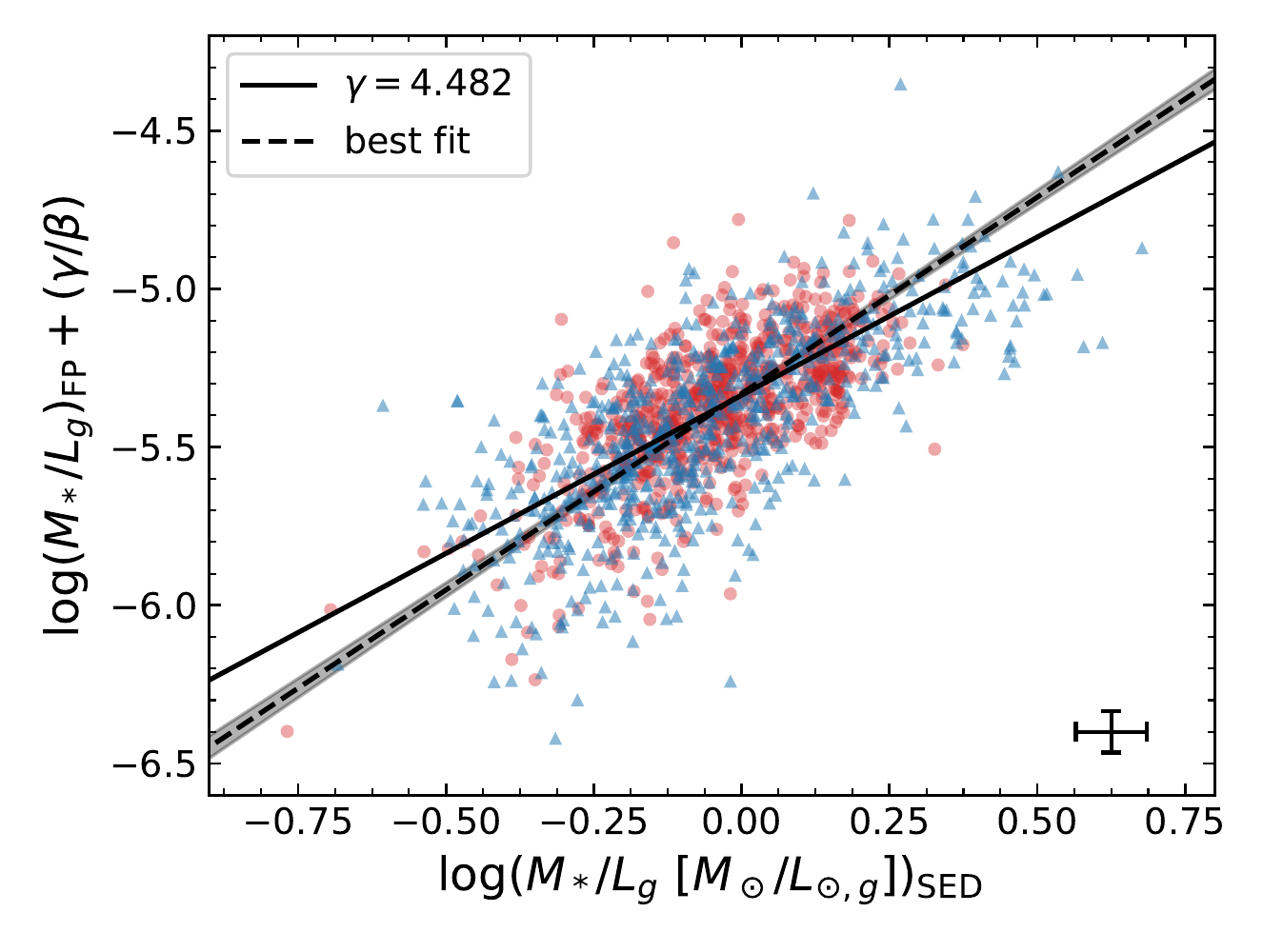}
    \caption{Comparison of the stellar mass-to-light ratio ($M_*/L_g$) predicted from the mass FP, and $M_*/L_g$ estimated from multi-wavelength SED fitting with MAGPHYS, demonstrating that the SED modeling provides a reasonable approximation of $M_*/L_g$.}
    \label{fig:ML_pred}
\end{figure}

Instead of using our modeled stellar masses to shed light on the FP, we can also ask whether the FP itself can provide information on the accuracy of the modeled $M_*/L$ \citep[see also][who discuss the constraining power of $\ML$ on stellar population properties]{vdSande2015}. To do so, we calculate the $M_*/L$ predicted by the mass FP:
\begin{equation}
    \log \left(\frac{M_*}{L_g}\right)_{\rm FP} = \log \Sigma_{\rm *,FP} - \log I_{\rm e,g}\,, 
    \label{eq:MstarL}
\end{equation}
where $I_{\rm e,g,}$ is the observed surface brightness and
\begin{equation}
     \log \Sigma_{\rm *,FP} = \left( \frac{1}{\beta} \right) \log R_{\rm e} -  \left( \frac{\alpha}{\beta} \right) \log\sigma - \left( \frac{\gamma}{\beta} \right)\,.
\end{equation}
The FP does not provide an absolute scaling of $M_*/L_g$, unless the value of $\gamma$ is constrained otherwise \citep[as done by][]{Schechter2014}. In Fig.~\ref{fig:ML_pred} we therefore show $\log(M_*/L_g)_{\rm FP} + (\gamma/\beta)$ versus the $M_*/L_g$ estimated with MAGPHYS. The solid line has a unit slope, with the intercept set equal to the best-fit zero point of the mass FP (Fig.~\ref{fig:fp_mass}). The dashed line on the other hand shows the best fit from an orthogonal distance regression that takes into account uncertainties in both variables, which gives a slope of $m=1.24\pm0.03$. 
We note that the measured slope is only weakly dependent on the adopted tilt: if we instead use the measured tilt from Section~\ref{sec:mfp_tilt}, we find variations of order $\sim 1-2\sigma$ (e.g., $m=1.27\pm0.04$ for the best-fit tilt at $0.65<z<0.75$).

Although the best-fit relation is statistically significantly different from a unit slope, the two different estimates of $M_*/L_g$ agree remarkably well, considering that the only assumption made in calculating $(M_*/L_g)_{\rm FP}$ is that the mass FP has zero intrinsic scatter. Both estimates show a similar, large spread in $M_*/L_g$, and the scatter about the solid line is $\sigma_{\rm NMAD} = 0.117\pm0.004\,$dex, partially driven by the uncertainties (of $\sim 0.06\,$dex in either axis). Moreover, the systematic offset between the solid and dashed lines is $<0.05\,$dex for $\approx 75\%$ of the sample. Only toward extreme values of $M_*/L_g$ do the systematic discrepancies become larger ($\sim 0.1\,$dex), where the SED modeling also becomes more difficult (e.g., accurately predicting the effects of dust, or the recent star formation history) and the intrinsic scatter in the mass FP may become important.

\subsection{Structural non-homology}\label{sec:homology}

The zero point of the mass FP is inversely proportional to ratio of the dynamical and stellar mass (i.e., $\gamma \propto \log(M_*/M_{\rm dyn})$), and therefore depends on the dark matter fraction within the effective radius, as well as the assumed IMF in the SED modeling. Considering structural properties only, one may expect a dependence of the zero point on the S\'ersic index ($n$): $n$ reflects the distribution of the stellar light, and hence the density profile of the stellar mass. Systematic differences in these density profiles may therefore lead to S\'ersic-dependent variations in the velocity dispersion or the dark matter fraction within one $R_{\rm e}$. 
\citet{Bezanson2015} find a weak correlation between $\gamma$ and $n$ at $z\approx 0.06$; however, their sample at $z\sim0.7$ contains too few objects to draw a conclusion on the non-homology of galaxies at higher redshift.

\begin{figure}
    \centering
    \includegraphics[width=\linewidth]{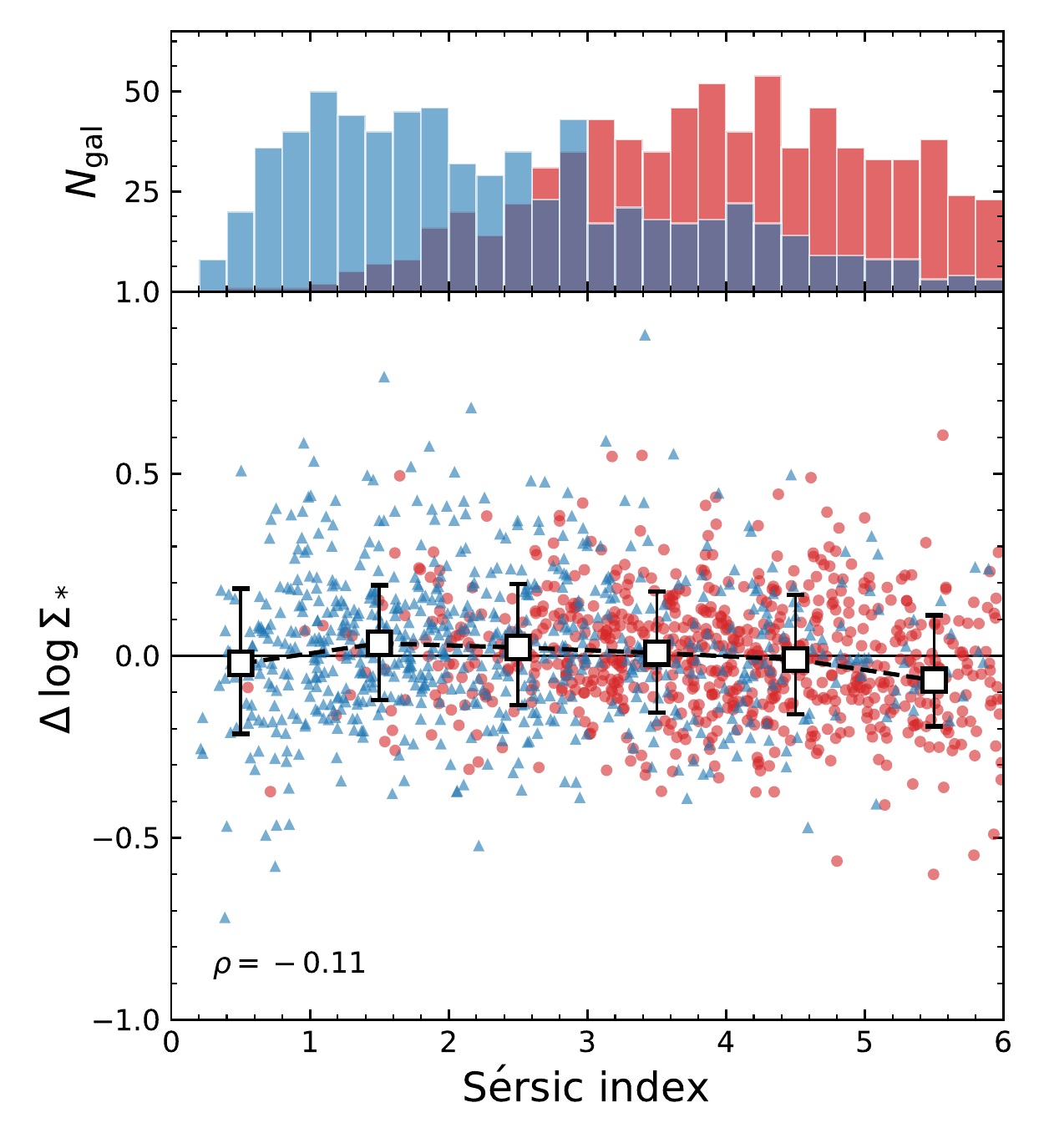}
    \caption{Residual in the mass FP in $\log \Sigma_*$ as a function of the S\'ersic index. Symbols indicate the same as in Fig.~\ref{fig:fp_mass_age}. The star-forming and quiescent galaxies follow very different distributions in S\'ersic index, yet, this has no significant effect on the scatter of the mass FP.}
    \label{fig:legac_nsersic}
\end{figure}

In Fig.~\ref{fig:legac_nsersic}, we show the residual from the FP in $\log\Sigma_*$ (for which $\Delta\log\Sigma_* \approx\Delta\log M_{\rm dyn}/M_*$) as a function of the best-fit S\'ersic index for the significantly larger sample of LEGA-C galaxies. The median of the combined star-forming (blue) and quiescent (red) population, plotted as open squares, shows no dependence on the S\'ersic index, except for the highest bin in S\'ersic index. We confirm this result by performing a linear fit to the data, which indicates a very weak correlation of $\Delta\log\Sigma_* \propto (-0.020\pm 0.004)\,n$ (Spearman rank correlation coefficient $\rho=-0.11$). The lack of an effect due to structural non-homology on the mass FP appears to be contradictory with previous measurements at $z\sim0$, of both the FP \citep{Bezanson2015} and direct measurements of $M_{\rm dyn}/M_*$ \citep[e.g.,][]{Taylor2010}. We discuss the implications of this result in Section~\ref{sec:discussion}.

\subsection{Inclination effects}\label{sec:axisratio}

The third structural parameter of our S\'ersic model is the observed ratio of the major and minor axes ($b/a$), which depends strongly on both the intrinsic morphology and the inclination angle of the system. For example, it provides an estimate of the inclination for systems that are intrinsically flat and axisymmetric. 

Correlations between the projected axis ratio and $M_{\rm dyn}/L$ of quiescent galaxies have been predicted using the luminosity FP and Jeans modeling \citep[e.g.,][]{Jorgensen1996,Cappellari2006}, however, the effect on the observed FP is unclear. \citet{Bezanson2015} find a weak dependence of the integrated velocity dispersion on the projected axis ratio at $z\sim0$, particularly for star-forming and low S\'ersic index systems: flattened (low $b/a$) objects have an elevated integrated velocity dispersion, whereas the opposite is the case for round (high $b/a$) objects. This reflects the fact that for flattened, rotationally supported systems, the integrated velocity dispersion is a combination of both the intrinsic velocity dispersion and the rotation along the line of sight, which is inclination-dependent.

\begin{figure}
    \centering
    \includegraphics[width=\linewidth]{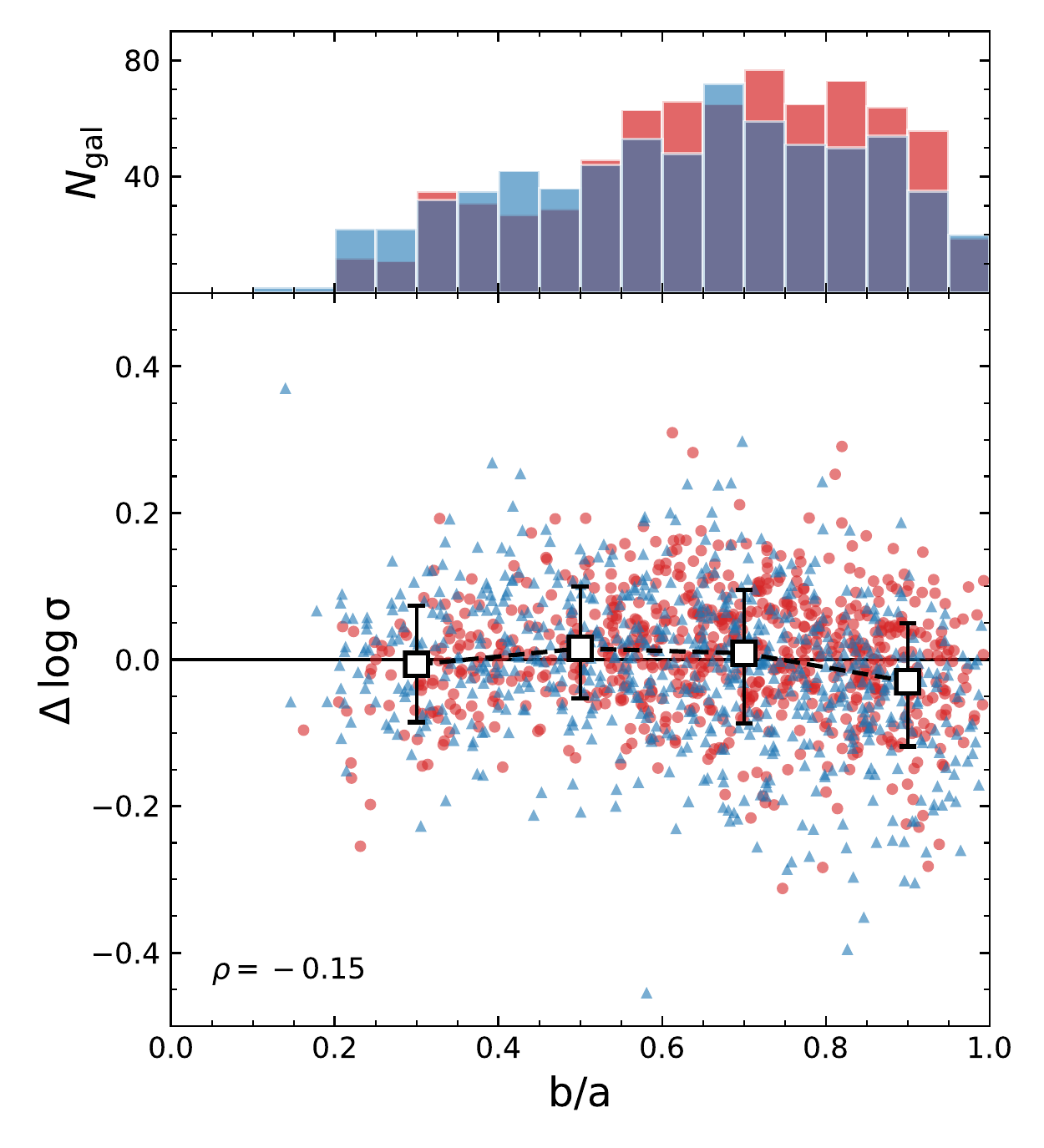}
    \caption{Residual from the mass FP in $\log\,\sigma$ as a function of the projected axis ratio ($b/a$). Symbols indicate the same as in Fig.~\ref{fig:fp_mass_age}. The integrated velocity dispersion is slightly lower than that predicted by the mass FP for rounder (higher $b/a$) systems, reflecting a minimal contribution of rotational motion to the integrated velocity dispersion for objects at low inclination angles. The effect of inclination is therefore a marginal increase in the intrinsic scatter in the FP. }
    \label{fig:legac_axisratio}
\end{figure}

Since flattened galaxies become more common at higher redshift \citep{vdWel2014b,Hill2019}, the effect of the inclination angle on the FP may become important. We evaluate this effect in Fig.~\ref{fig:legac_axisratio}, where we show the residual from the mass FP in $\log\sigma$ (rather than $\log\Sigma_*$) as a function of the projected axis ratio. Quiescent and star-forming galaxies are again indicated by red and blue symbols respectively, with the median of the full sample shown in black. For high values of $b/a$, the residual $\Delta \log\sigma$ is slightly negative: for round or face-on objects, the integrated velocity dispersion is lower than the velocity dispersion predicted from the mass FP in Eq.~\ref{eq:mfp}, since the contribution from rotational motion to the integrated velocity dispersion is minimized for systems at low inclination. Notably, this applies to both the quiescent and star-forming sample, suggesting that rotation is important for quiescent galaxies as well, and is further supported by the large number of highly flattened quiescent galaxies. {The similarity between the projected axis ratio distributions of the star-forming and quiescent galaxies likely reflects a mixture of different intrinsic shapes within these galaxy populations, with both the star-forming and quiescent samples containing a significant fraction of disk-like morphologies as well as more spheroidal structures \citep[see also][]{Chang2013,vdWel2014b}. Additionally, the number of star-forming galaxies with low values of $b/a$ may be slightly reduced by our selection on the SNR of the velocity dispersion (Section~\ref{sec:sample_selection}), as this results in a slight bias against highly reddened star-forming galaxies, which are more likely to be edge-on projections.} 

{The anti-correlation between $b/a$ and $\Delta\log\sigma$}, however, does not continue toward low axis ratios, where we would expect the integrated velocity dispersion to be higher than the FP prediction due to an increased contribution from the rotational velocity. This can be attributed to our use of the circularized effective radius (Section~\ref{sec:photometry}), which is proportional to the square root of the axis ratio. For flattened objects, the smaller effective radius counteracts the increased velocity dispersion, resulting in a predicted velocity dispersion that is approximately equal to the observed value. The net effect of the random inclination angle on the FP therefore is to slightly enhance the scatter about the FP, contributing to the intrinsic scatter derived in Section~\ref{sec:mfp_edgeon}.

Indeed, \citet{Bernardi2020} show that the residuals of the FP correlate strongly with the axis ratio, if the major axis size is used rather than the circularized size. They hence demonstrate {the importance of inclination effects on the FP, and show} that the scatter in the FP can be further reduced by treating $b/a$ as an additional variable in Eq.~\ref{eq:lfp} or Eq.~\ref{eq:mfp}: by fitting a hyperplane to a sample of low-redshift elliptical and lenticular galaxies, {they find that the tilt of the FP, i.e. the values of $a$ and $b$, can differ by $\sim2-3\sigma$ from the traditional (three parameter) FP, and that the scatter about the best-fit FP is decreased by up to $0.009\,$dex. Still, even after accounting for $b/a$ as a separate variable, the effect of galaxy inclination remains apparent in the FP, as more highly inclined galaxies have a lower scatter about the plane than galaxies that are near face-on.} {These different effects are largest for S0 galaxies, and thus potentially even larger for star-forming disks.}

\subsection{Environment}\label{sec:environment}

Many previous studies of the luminosity FP have focused on clusters of galaxies \citep[e.g.,][]{Jorgensen1996,VanDokkum2007,Holden2010, Beifiori2017,Saracco2020}, and explored differences in the properties of the FP between low and high density environments \citep[e.g.,][]{VanDokkum2001,Cappellari2006, LaBarbera2010, Saglia2010,Joachimi2015}. \citet{Burstein1990} first demonstrated that the effect of environment on the FP is expected to be small, as they found no dependence of the zero point on cluster richness. Using a large sample of early-type galaxies in the SDSS, \citet{LaBarbera2010} showed that the zero point of the luminosity FP indeed correlates weakly with the local galaxy density, regardless of the chosen passband. \citet{Joachimi2015} obtained similar results by considering the spatial correlation function of residuals in the $r$ and $i$-band FP with the galaxy density field, and additionally find small systematic differences between central galaxies and satellites. 

Interpreting the zero point of the plane as $M_{\rm dyn}/L$, these results imply that galaxies in lower density environments have lower values of $M_{\rm dyn}/L$ than those in high density environments, and that central galaxies have higher $M_{\rm dyn}/L$ than satellites. A systematically lower luminosity-weighted age for field galaxies can explain their lower values in $M_{\rm dyn}/L$ as compared to cluster galaxies \citep{VanDokkum2007,LaBarbera2010}, and is broadly consistent with the picture of hierarchical structure formation, from which we would expect galaxies to form earlier in highly dense environments. \citet{Joachimi2015} suggest that the lower value of $M_{\rm dyn}/L$ for satellite galaxies, which is not only lower than that of central galaxies, but also of field galaxies, can be attributed to the tidal stripping of dark matter and hot gas in the subhaloes as they fall into more massive haloes.

\begin{figure}
    \centering
    \includegraphics[width=\linewidth]{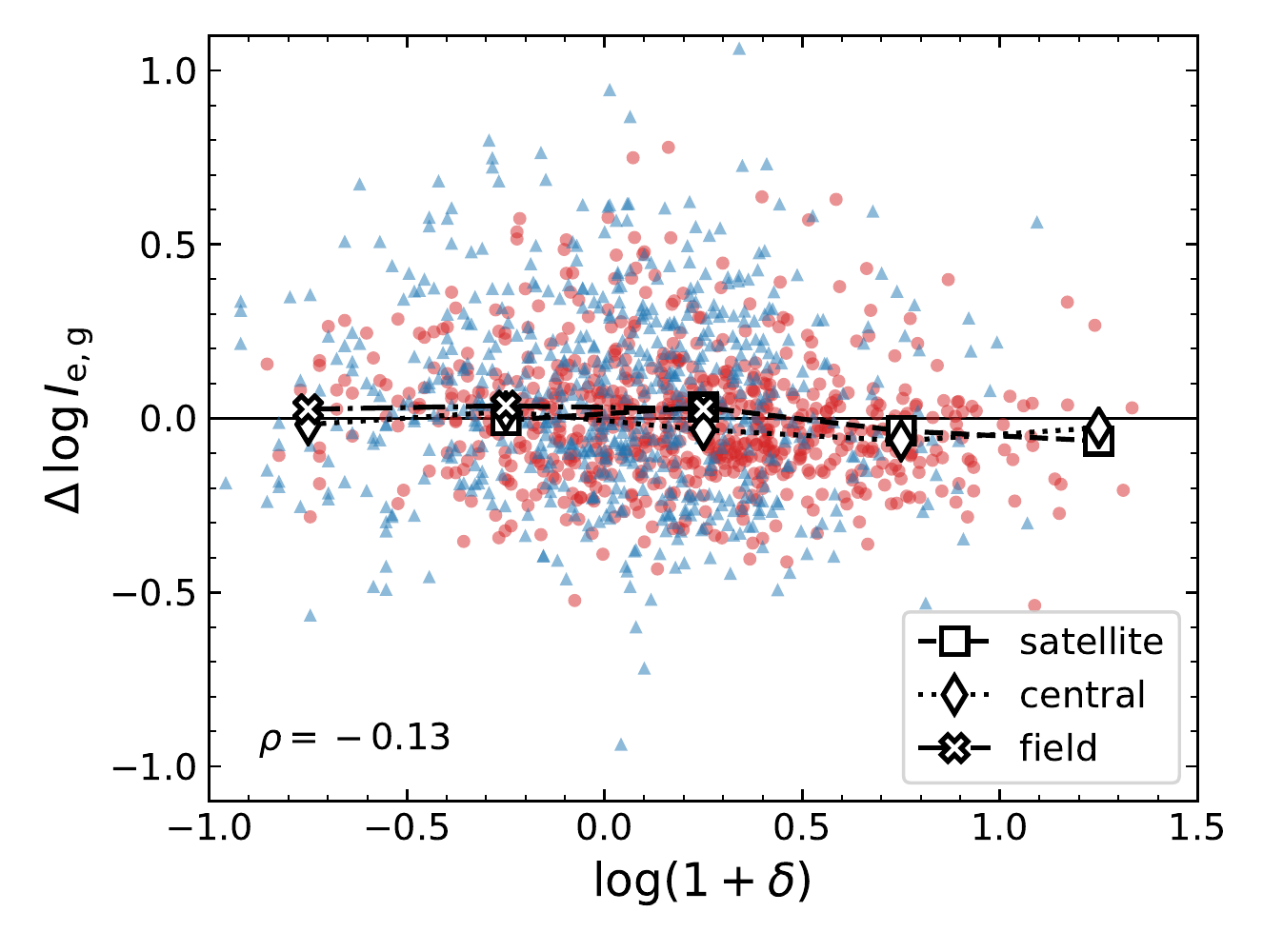}
    \includegraphics[width=\linewidth]{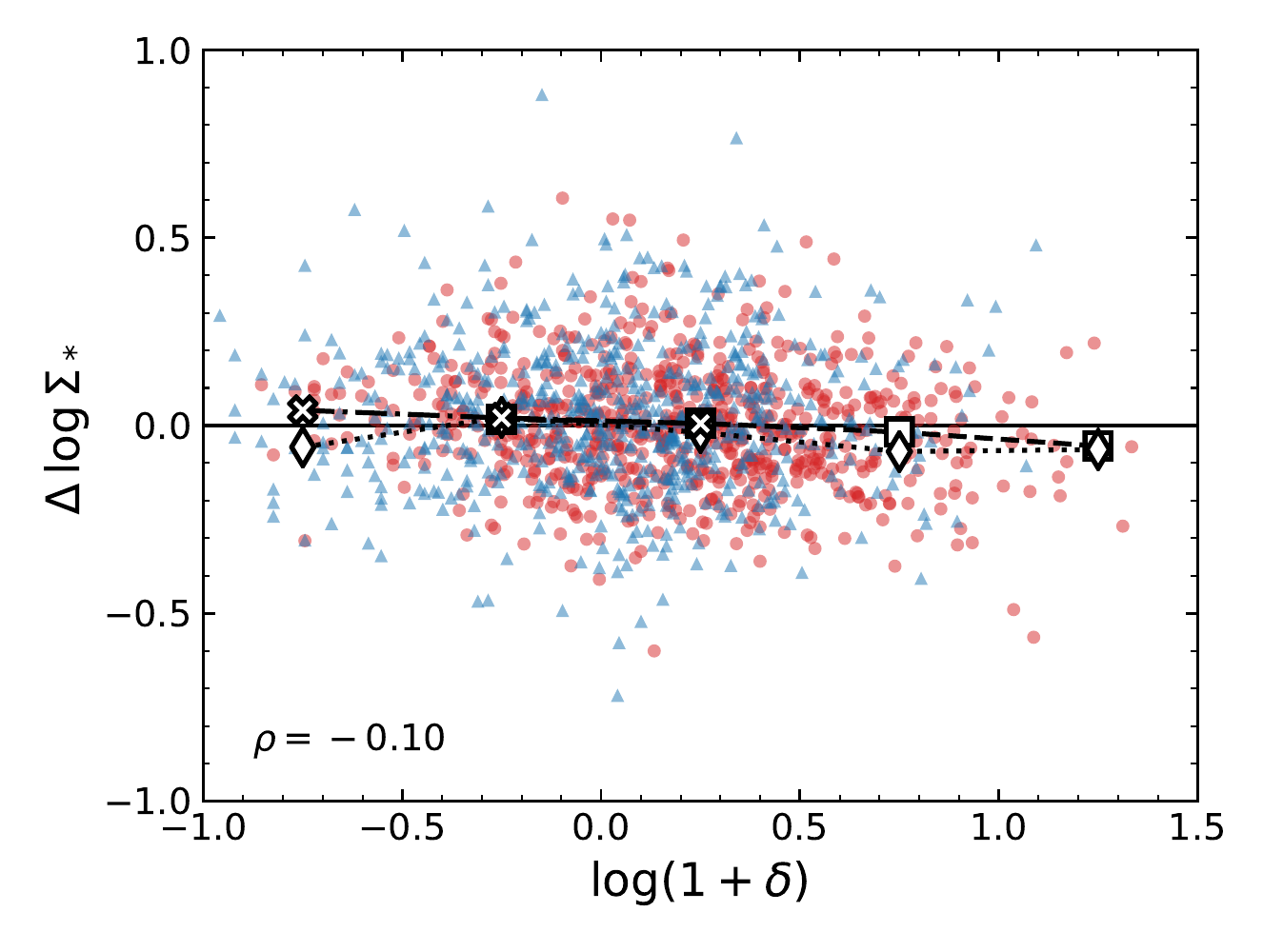}
    \caption{Residual in the $g$-band (top) and mass (bottom) FP as a function of the local overdensity \citep{Darvish2017}. Red and blue symbols indicate the quiescent and star-forming population respectively. White markers show the median of galaxies that are classified as central (diamonds), satellite (squares) or field (crosses) galaxies. We find no significant environmental dependence within the LEGA-C data for both the $g$-band and mass FP.}
    \label{fig:fp_env}
\end{figure}

We explore the effect of environment on the FP by matching the LEGA-C sample with the \citet{Darvish2017} cosmic web catalog (with a maximum matching radius of $1\arcsec$), which contains measurements of the projected density field of the COSMOS field out to $z=1.2$, and categorizes galaxies as `central', `satellite' or `isolated'. This catalog was constructed using the COSMOS2015 photometric redshift catalog \citep{Laigle2016} in the UltraVISTA-DR2 region \citep{McCracken2012,Ilbert2013} following the adaptive weighted kernel smoothing method described in \citet{Darvish2015}. In Fig.~\ref{fig:fp_env} we show in the top panel the residual from the $g$-band FP in $\log I_{\rm e,g}$ as a function of the projected overdensity, for both the quiescent (red) and star-forming (blue) sample. Since the redshift distribution of the few galaxies at high overdensity is not representative of the full sample, we have corrected the values of $\Delta\log I_{\rm e,g}$ for the redshift evolution derived in \citet{deGraaff2020}. The medians for galaxies classified as central, satellite or field \citep[`isolated' in the catalog by][]{Darvish2017} are indicated by white symbols. There is a very weak anti-correlation between the residual in $\log I_{\rm e,g}$ and the overdensity, such that $\Delta\log I_{\rm e,g}\propto (-0.085\pm0.015)\log(1+\delta)$. Since this residual is inversely proportional to $M_{\rm dyn}/L$, it is consistent with previous findings that galaxies in higher density environments have a higher value of $M_{\rm dyn}/L$. When dividing our sample into satellites, centrals and field galaxies, we do not find any significant systematic differences between the subsamples, in contrast with the weak, but significant, effect found by \citet{Joachimi2015}. However, our sample contains far fewer objects than these studies at low redshift, particularly so at high overdensity. Moreover, our measurements do not account for uncertainties in the density field estimation, which is particularly difficult to constrain precisely at low overdensities, and we therefore cannot draw any strong conclusions on the effect of environment on the FP.

Analogous to the top panel of Fig.~\ref{fig:fp_env}, in the bottom panel we show the residual in $\log\Sigma_*$ of the mass FP as a function of the overdensity. We find an even weaker dependence of the zero point of the mass FP on environment, both in terms of overdensity, with $\Delta\log\Sigma_* \propto (-0.052\pm0.014)\log(1+\delta)$, and galaxy type (satellite, central, field). Within the current galaxy sample and level of uncertainty, this suggests that at fixed $R_{\rm e}$ and $\sigma$ the structural properties of galaxies in high density environments do not differ significantly from those in the field.

\section{Discussion}\label{sec:discussion}

\subsection{Stellar populations}

In agreement with many other studies \citep[e.g.,][]{Jorgensen1996,Forbes1998,Wuyts2004,Gargiulo2009}, we have shown that there is significant scatter in the luminosity FP, which cannot be attributed to measurement uncertainties alone. We find that the residuals from the FP correlate strongly with spectral features (D$\rm _n$4000, H$\rm\delta_A$) as well as rest-frame colors ($U-V$, $V-J$). These correlations can be interpreted as systematic variations in $M_*/L$ due to varying stellar ages, and in the case of the star-forming population, also different sSFRs and dust attenuation.

Previous results at low redshift, where significant residual correlations with stellar age are present in the FP \citep{Forbes1998,Gargiulo2009,Graves2009II}, thus also hold at $z\sim 1$. Moreover, this correlation appears to be stronger in our sample as compared with both \citet{Gargiulo2009} and \citet{Graves2009II}. \citet{Graves2010III} show that variations in $M_*/L$ contribute approximately $22\%$ to the intrinsic thickness of the FP (i.e., ${\rm d}\log(M_*/L)/{\rm d}(\Delta\log I_{\rm e}) \approx -0.22$), although depending on the stellar population modeling method used this value may be anywhere between $2\% - 53\% $. 

However, these studies at low redshift focus on early-type galaxies alone, which are selected by morphology as well as insignificant H$\alpha$ or [\ion{O}{2}]3727 line emissivity, whereas we here have extended the analysis to the full population of massive galaxies. The selected samples of early-types at $z\sim 0$ therefore likely consist of galaxies that span a narrower range in age and $M_*/L$. Moreover, at $z\sim 0$ the FP is often studied in the $r$-band, which may differ significantly from the rest-frame $g$-band considered here.

\begin{figure}
    \centering
    \includegraphics[width=\linewidth]{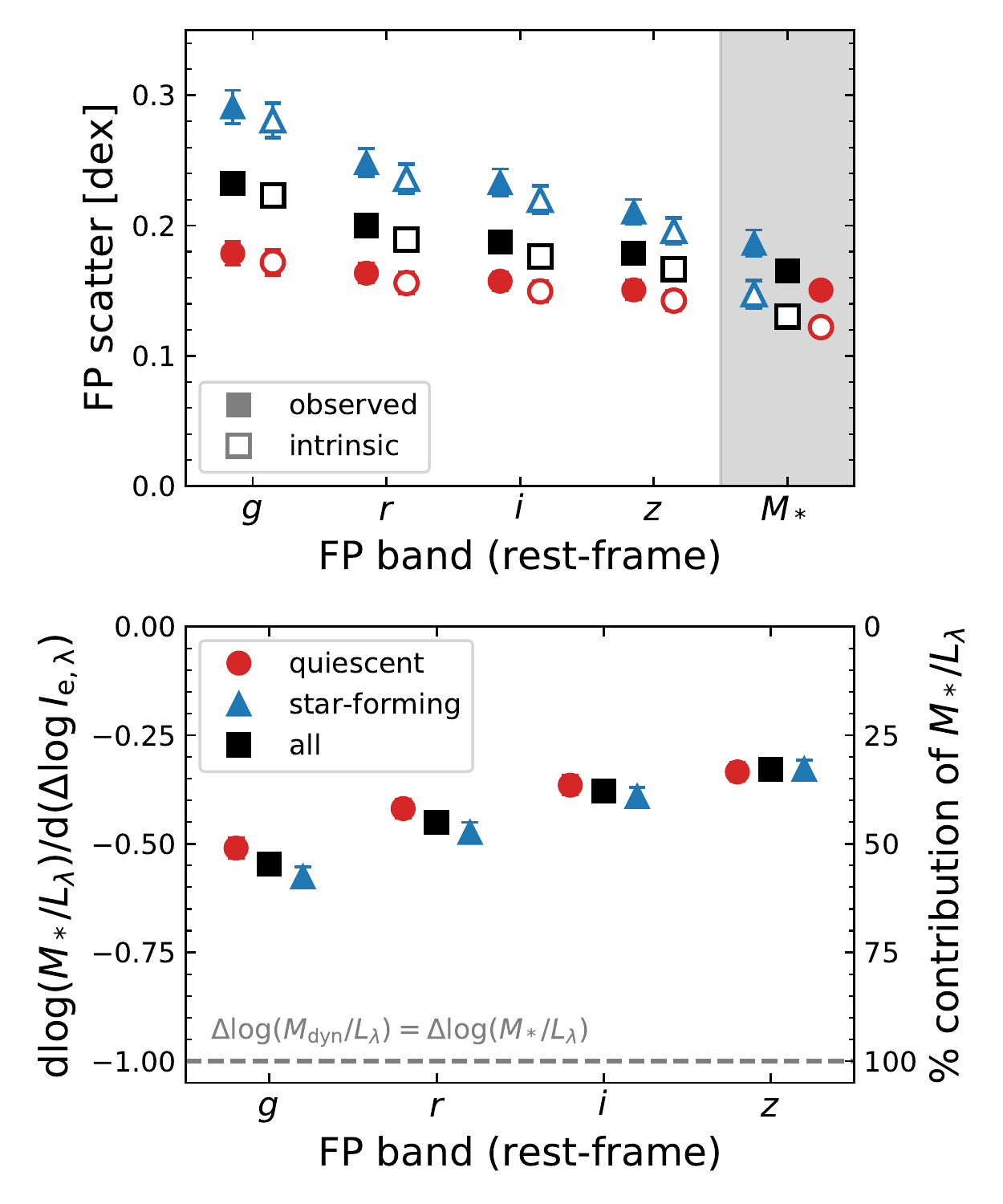}
    \caption{Effect of variation in the stellar mass-to-light ratio ($M_*/L_\lambda$) on the thickness of the FP. \textbf{Top}: Scatter in the FP in $\Delta\log I_{\rm e,\lambda}$ for different rest-frame wavelengths, with solid and open symbols showing the observed and intrinsic scatter respectively. The scatter in the mass FP (in $\Delta\log\Sigma_*$) is shown for reference. \textbf{Bottom:} Contribution of $M_*/L_\lambda$ to the residual from the FP in $\log I_{\rm e,\lambda}$. The dashed line shows the maximum value, since $\Delta\log I_{\rm e,\lambda} \approx \Delta\log (M_{\rm dyn}/L_\lambda)$. Both the observed and intrinsic scatter in the FP decrease toward longer wavelength, due to a decrease in the contribution from variations in $M_*/L_\lambda$.}
    \label{fig:ml_comp}
\end{figure}

We evaluate the contribution of variations in $M_*/L$ to the thickness of the FP in Fig.~\ref{fig:ml_comp}, using the different measurements of the tilt by \citet[][Table 2]{HydeBenardi2009} to obtain the FP in different rest-frame passbands. Firstly, we consider the observed (filled symbols) and intrinsic (open symbols) scatter in $\Delta \log I_{\rm e,\lambda}$ at different wavelengths, for the quiescent (red), star-forming (blue), and combined (black) subsamples. The scatter about the mass FP (in $\Delta \log \Sigma_*$) is shown for reference. For the quiescent galaxies the observed scatter in the mass FP is approximately equal to that in the $r$, $i$ and $z$-band FPs, whereas the intrinsic scatter in the mass FP is significantly lower than the luminosity FP, reflecting the relatively large uncertainty on the SED modeling in comparison with the observational error on the luminosity. 
More importantly, there is a significant decrease in both the observed and intrinsic scatter toward longer wavelength, particularly so for the star-forming subsample. This reflects a lower contribution of $M_*/L$ to the intrinsic scatter and suggests, unsurprisingly, that variations in the dust attenuation and recent star formation are most apparent at short wavelengths.

In the bottom panel we quantify the contribution of $M_*/L$ variations using the SED-derived $M_*/L$ estimates and a simple least-squares fit \citep[to match the methods by][]{Graves2010III}. We note that we do not subtract the mean value of $M_*/L$ along the (face-on) midplane, because the face-on FP is sparsely populated in comparison to the low-redshift studies, which together with the large uncertainties on $M_*/L$ makes a robust estimate of the mean $M_*/L$ difficult. However, this mainly affects the uncertainty on the fit, and is unlikely to lead to a significant bias on the measured contribution of $M_*/L$. 

We find that in the rest-frame $g$-band approximately $55\%$ of the thickness of the FP is due to variations in $M_*/L_g$, with the contribution being slightly higher for star-forming galaxies ($\sim58\%$, versus $\sim51\%$ for quiescent galaxies). {Stellar populations thus are the main driver of the intrinsic scatter in the $g$-band FP, exceeding the contributions of all other quantities examined in Section~\ref{sec:mfp_results}. On the other hand, \citet{Bernardi2020} recently showed that, for rotating systems, the use of the integrated velocity dispersion rather than the luminosity-weighted average of the second moment of the velocity (which is attainable from IFU data only; see Eq.~\ref{eq:sigma_aper}) may also be a cause of substantial scatter in the FP. However, this additional scatter of approximately $\Delta \log\sigma\sim 0.03\,$dex (based on their Fig. A1) is still at least a factor $\sim 3$ lower than the contribution from stellar populations found here, and is further mitigated by the fact that this effect only becomes apparent in the case of very high S/N spectra. }

{Fig.~\ref{fig:ml_comp} also shows that the dependence on $M_*/L_\lambda$ is itself wavelength-dependent}, such that the FP at longer wavelengths is less dominated by variations in $M_*/L_\lambda$. Interestingly, there is significant contribution from $M_*/L_\lambda$ even at the longest wavelengths. Comparing with the results by \citet{Graves2010III} in the rest-frame $r$-band, we find that for our sample of quiescent galaxies the contribution from stellar populations is $\sim42\%$. This is significantly higher than their measurement of $22\%$ (for their preferred method of estimating $M_*/L_r$), but may be attributed to significant differences in the definition of quiescence: using D$\rm _n$4000 as a proxy for age, if we select the 100 oldest (UVJ) quiescent galaxies in our sample, we find that variation in $M_*/L_r$ contributes $23\%$ to the thickness of the $r$-band FP. 

Importantly, these measurements show that, under the assumption that the effects of dynamical non-homology are small \citep[e.g.,][]{Bolton2008,Schechter2014}, a significant fraction of the intrinsic scatter in the FP must arise variations in $M_{\rm dyn}/M_*$, which may be due to variations in the IMF or the dark matter fraction. Our data currently lack a consistent measurement of the metallicity across the entire redshift range, as well as a measurement of the $\alpha$-element abundance and other IMF-sensitive features \citep[summarized in, e.g.,][]{vanDokkum2012}, and we therefore cannot place constraints on the effect of IMF variations within the FP. On the other hand, we may expect the effect of IMF variations to be approximately as large as the uncertainties in the SED modeling \citep[e.g.,][]{vdSande2015}, which would imply that the intrinsic scatter is dominated by fluctuations in the dark matter content.

For the quiescent LEGA-C galaxies, the significant correlations between the residuals from the luminosity FP and D$\rm _n$4000 or H$\rm\delta_A$, combined with the very weak correlations through the mass FP (Figs.~\ref{fig:lfp_age}~\&~\ref{fig:fp_mass_age}), suggest that galaxies with younger luminosity-weighted ages, due to a later formation time or more extended star-formation history, have marginally higher values of $\Delta \log\Sigma_*$. If the effects of non-homology and IMF variations are small, this result implies that younger quiescent galaxies are slightly more baryon-dominated within $1\,R_{\rm e}$. Although the correlation between age and structure is very weak, in contrast with the strong correlation found by \citet{Graves2010III}, this would be broadly consistent with the proposed scenario in which the truncation time of star formation determines the location of a galaxy within the parameter space of the FP. 

However, the effect of galaxy merging, and how these trends apply to the star-forming population is still unclear. Recently, \citet{Ferrero2020} used cosmological hydrodynamical simulations to show that the tilt of the FP, of both star-forming and quiescent galaxies, can be explained entirely by variations in the dark matter fraction. A quantitative comparison with such simulations is challenging, as there are systematic mismatches between the observed and simulated sizes and velocity dispersions \citep{vdSande2019}. However, hydrodynamical simulations of large volumes do qualitatively reproduce observed galaxy scaling relations, and therefore may also be able to shed light on the physical processes driving the intrinsic scatter in the FP, an analysis that we defer to a future work.

\subsection{Structural non-homology}

We have found that massive star-forming and quiescent galaxies lie on the same mass FP, with a comparable intrinsic scatter about the midplane (Fig.~\ref{fig:fp_mass}). 
Although the star-forming galaxies are typically slightly larger in size at fixed mass, their integrated velocity dispersion or stellar mass surface density tends to be lower, such that they fall on the same FP as the quiescent systems. 
The thickness of the mass FP is, unlike the $g$-band FP, largely uncorrelated with stellar population properties and can be interpreted as variation in $M_{\rm dyn}/M_*$. Under the assumption of a weakly varying IMF, the intrinsic scatter about the FP reflects a variation in the dark matter fraction within the effective radius.

Of particular interest then is the morphology, which we have modeled as a S\'ersic profile. If the value of the S\'ersic index reflects different underlying mass density profiles, we may expect it to correlate with the residuals in the mass FP. However, we find only a very weak correlation within the LEGA-C sample (Fig.~\ref{fig:legac_nsersic}). Interestingly, \citet{Bezanson2015} do find a weak dependence on S\'ersic index within the mass FP at low redshift, for a sample of SDSS galaxies similar to the low-redshift sample considered here. In a different context, \citet{Cappellari2006} and \citet{Taylor2010} also demonstrate the importance of non-homology on the estimation of the dynamical mass of galaxies at $z\sim0$. The lack of a correlation with S\'ersic index in the mass FP in our sample is therefore surprising, as it seems to suggest that the dynamical masses of galaxies at $z\sim1$ are independent of the observed S\'ersic index. Any fluctuations in the dark matter fraction then simply reflect differences in the effective radii of galaxies, rather than the mass distribution itself.

This raises the question of how the difference in the structural dependence at $z\sim0$ and $z\sim1$ can be reconciled. One possibility is that the light profile evolves with redshift, while the underlying mass distribution does not change significantly, such that the mass FP is correlated with S\'ersic index at $z\sim0$, but not at $z\sim1$. This scenario can be tested by measuring the color gradients of galaxies to derive the S\'ersic index and size of the stellar mass profile, instead of the rest-frame $5000\,\angstrom$ sizes used here. \citet{Suess2019a} demonstrate that color gradients are significantly steep especially at high stellar mass and are also dependent on redshift, and may therefore be important to take into account. {\citet{Bernardi2019} show that, for a sample of very massive elliptical galaxies at $z\sim0$, the accounting for stellar population gradients in galaxies can lead to a significant change in the inferred values of and variation in $\MM$. These gradients may then potentially act to wash out any significant dependence on S\'ersic index through the thickness of the FP, although it is unclear how stellar population gradients affect the measurement of $\MM$ for the population of late-type galaxies at $z\sim0$, as well as galaxies at higher redshifts.}

Secondly, if not a difference in the observed morphology, there may be differences in the derivation of the velocity dispersions between the various studies. For example, as opposed to the integrated velocity dispersion within $1\,R_{\rm e}$ used in this work, \citet{Taylor2010} use the central stellar velocity dispersion ($R_{\rm e}/8$); this difference in the aperture may lead to small systematic effects on the measured dispersions (see also Appendix~\ref{sec:apdx_apcor}). Van Houdt et al. (in prep.) demonstrate using axisymmetric Jeans modeling that, at fixed mass, the dynamical masses of the LEGA-C galaxies do depend on S\'ersic index, and do so in the same way as at $z\sim0$. However, they also show that this dependence becomes apparent only when using the major axis size (rather than the circularized size) and after taking into account the effects of the slit aperture and the galaxy inclination (through the observed axis ratio) on the integrated velocity dispersion.

On the other hand, the lack of a residual correlation through the mass FP with S\'ersic index does not imply that non-homology plays no role at all. The FP is tilted with respect to the virial plane, which may (in part) be due to a violation of the assumption of homology. \citet{Bezanson2013} compared the power-law relation between $M_{\rm dyn}/M_*$ and $M_{\rm dyn}$ for two different estimates of $M_{\rm dyn}$, the first having a virial constant $K=5$ (as in Section~\ref{sec:mdynl}) and the second a S\'ersic-dependent virial constant $K(n)$ \citep[derived by][]{Cappellari2006}. At $z\sim0$ the measured relation between $M_{\rm dyn}/M_*$ and $M_{\rm dyn}$ is slightly shallower for the S\'ersic-dependent estimate of $M_{\rm dyn}$, which indicates that non-homology contributes to the tilt of the FP, albeit a small effect \citep[in agreement with findings by][]{Cappellari2006}.{ Moreover, \citet{Bernardi2020} show that by accounting for structural non-homology in their fits of the mass FP, as well as the galaxy inclination, they obtain a plane that is closer to the virial prediction. }

By considering variations in the tilt of the mass FP, we observe a similar, weak effect. Taking the values of the tilt from Section~\ref{sec:mfp_tilt}, we find that the strongest tilt  ($a=1.432$ and $b=-0.736$) produces the weakest correlation with S\'ersic index: $\Delta\log\Sigma_* \propto (-0.016\pm 0.004)\,n$. Conversely, for the virial plane ($a=2$ and $b=-1$) we find $\Delta\log\Sigma_* \propto (-0.026\pm 0.005)\,n$. An evolution in the tilt, such that the mass FP becomes closer to the virial plane at higher redshift, may thus also bring the measurements at $z\sim 0$ and $z\sim 0.8$ into agreement.

\section{Summary and conclusions}\label{sec:conclusion}

We have explored the connection between the structural and stellar kinematic properties of 1419 galaxies in the LEGA-C survey, which form a representative sample of the massive ($\log( M_*/M_\odot) >10.5$) galaxy population at $0.6<z<1$. In addition to the spectral and morphological properties obtained from the LEGA-C spectra and \textit{HST} imaging respectively, we have performed SED modeling of multi-wavelength ($0.2-24\,\micron$) photometry to estimate stellar masses, as well as stellar population properties and the effect of dust attenuation. Separating our sample into star-forming and quiescent galaxies by the rest-frame UVJ colors, we have studied the effect of different structural, environmental and SED properties within the luminosity and mass FP. Our findings can be summarized as follows:

\begin{itemize}
    \item There is significant scatter in the rest-frame $g$-band FP of quiescent galaxies, which exceeds the scatter due to measurement uncertainties. Star-forming galaxies also lie on the $g$-band FP, but with a different zero point and higher intrinsic scatter (Fig.~\ref{fig:fp_gband}). The residuals from the $g$-band FP correlate strongly with spectral age indicators (D$\rm _n$4000 and H$\rm\delta_A$), as well as rest-frame colors ($U-V$, $V-J$). Using SED models, we interpret these correlations as being due to variation in the luminosity-weighted stellar age, and additionally for the star-forming sample, variation in the sSFR and dust attenuation.
    \item Both star-forming and quiescent galaxies lie on the same mass FP, with an approximately equal zero point and a comparable level of intrinsic scatter. In contrast with the $g$-band FP, we find no significant correlations in the residuals from the mass FP with different spectral and SED properties. Moreover, there is only a very weak correlation with S\'ersic index and the observed axis ratio, corresponding to a minimal dependence on morphology for variations in $\MM$ through the thickness of the FP.
    \item We evaluate the effect of environment on the FP, finding a very weak correlation between the residuals from the $g$-band FP and the projected galaxy overdensity, such that galaxies in high density environments have a marginally higher value of $M_{\rm dyn}/L$, in line with previous studies that find galaxies at high overdensity to be slightly older. We find an even weaker correlation within the mass FP, suggesting that there is no significant structural difference between galaxies in low- and high-density environments at fixed size and velocity dispersion. 

\end{itemize}

Overall, we find that variations in the $M_*/L_g$ can account for $\sim 54\%$ of the thickness of the $g$-band FP. The other main contribution comes from variations in the dark matter content within $1\,R_{\rm e}$, or, variations in the IMF. Interestingly, the residuals in $\log \Sigma_*$ in the mass FP do not correlate strongly with morphology (S\'ersic index), suggesting that the effect of structural non-homology is weak. Instead, variations in the galaxy size (at fixed mass) may play a more important role, as this leads to fluctuations in the dark matter fraction. 

Future studies of IMF-sensitive spectral features or abundance measurements are required to quantify the role of IMF variations within the FP. On the other hand, the role of dark matter may well be explored with current cosmological hydrodynamical simulations, which are able to offer insight into the physical processes governing the properties of galaxies throughout the FP and the evolutionary processes that keep galaxies on the mass FP.

\acknowledgments

Based on observations made with ESO Telescopes at the La Silla Paranal Observatory under program ID 194-A.2005 (The LEGA-C Public Spectroscopy Survey). This project has received funding from the European Research Council (ERC) under the European Union’s Horizon 2020 research and innovation program (grant agreement No. 683184). AdG thanks Pieter van Dokkum, Fraser Evans and Mantas Zilinskas for useful discussions. 
JvdS acknowledges support of an Australian Research Council Discovery Early Career Research Award (project number DE200100461) funded by the Australian Government. PFW acknowledges the support of the fellowship from the East Asian Core Observatories Association. We gratefully acknowledge the NWO Spinoza grant.

Funding for the Sloan Digital Sky Survey IV has been provided by the Alfred P. Sloan Foundation, the U.S. Department of Energy Office of Science, and the Participating Institutions. SDSS-IV acknowledges
support and resources from the Center for High-Performance Computing at
the University of Utah. The SDSS web site is www.sdss.org.

SDSS-IV is managed by the Astrophysical Research Consortium for the 
Participating Institutions of the SDSS Collaboration including the 
Brazilian Participation Group, the Carnegie Institution for Science, 
Carnegie Mellon University, the Chilean Participation Group, the French Participation Group, Harvard-Smithsonian Center for Astrophysics, 
Instituto de Astrof\'isica de Canarias, The Johns Hopkins University, Kavli Institute for the Physics and Mathematics of the Universe (IPMU) / 
University of Tokyo, the Korean Participation Group, Lawrence Berkeley National Laboratory, 
Leibniz Institut f\"ur Astrophysik Potsdam (AIP),  
Max-Planck-Institut f\"ur Astronomie (MPIA Heidelberg), 
Max-Planck-Institut f\"ur Astrophysik (MPA Garching), 
Max-Planck-Institut f\"ur Extraterrestrische Physik (MPE), 
National Astronomical Observatories of China, New Mexico State University, 
New York University, University of Notre Dame, 
Observat\'ario Nacional / MCTI, The Ohio State University, 
Pennsylvania State University, Shanghai Astronomical Observatory, 
United Kingdom Participation Group,
Universidad Nacional Aut\'onoma de M\'exico, University of Arizona, 
University of Colorado Boulder, University of Oxford, University of Portsmouth, 
University of Utah, University of Virginia, University of Washington, University of Wisconsin, 
Vanderbilt University, and Yale University.

\vspace{5mm}


\software{Astropy \citep{astropy}, EAZY \citep{Brammer2008}, FAST \citep{Kriek2009}, Galfit \citep{Peng2010}, MAGPHYS \citep{daCunha2008}, Matplotlib \citep{matplotlib}, NumPy \citep{numpy}, pPXF \citep{Cappellari2004,Cappellari2017}, SciPy \citep{scipy}}

\appendix
\section{Comparison of stellar mass estimates}\label{sec:apdx_mstar}

In Section~\ref{sec:photometry}, we ran the MAGPHYS code \citep{daCunha2008} for broad-band photometry from the multi-wavelength catalog by \citet{Muzzin2013a} to model the physical properties of the LEGA-C galaxies. We provide our catalog of derived SED properties in Table~\ref{tab:sed}. Our choice for MAGPHYS is motivated by our aim to minimize the systematic uncertainty in the measurement of the redshift evolution of the mass FP across $0<z<1$, and the public availability of the MAGPHYS modeling results for the SDSS by \citet{Chang2015}. Our SED modeling differs from the results presented previously in \citet{vdWel2016}, who used the FAST code \citep{Kriek2009} with different model assumptions and a different set of photometry.

\begin{deluxetable}{cccccc}
\tablecaption{Results of the MAGPHYS SED modeling \label{tab:sed}}
\tablehead{
\colhead{ID} & \colhead{$\log(M_*/M_\odot$)} & \colhead{$\log({\rm sSFR/ yr^{-1}})$} & \colhead{$\log({\rm age / yr})$} & \colhead{$\rm A_V$ [mag]} & \colhead{D$\rm _n$4000} 
}
\startdata
4792 & $10.52^{+0.13}_{-0.00}$ & $-10.32^{+0.30}_{-0.00}$ & $9.26^{+0.21}_{-0.00}$ & 0.03 & 1.44 \\
5786 &  $11.12^{+0.09}_{-0.05}$  &  $-10.72^{+0.10}_{-0.10}$ & $9.41^{+0.16}_{-0.06}$  & 0.93 & 1.56  \\
6859 & $11.31^{+0.00}_{-0.00}$ & $-11.02^{+0.00}_{-0.00}$ & $9.20^{+0.00}_{-0.00}$ & 0.22 & 1.57 \\
6890 & $11.25^{+0.10}_{-0.09}$ &  $-11.02^{+0.35}_{-0.05}$ &  $9.24^{+0.04}_{-0.08}$ & 1.28 & 1.56 \\
7002 & $10.76^{+0.00}_{-0.09}$ &  $-10.87^{+0.00}_{-0.20}$  & $9.36^{+0.00}_{-0.06}$  & 0.11 & 1.57 
\enddata
\tablecomments{Values and formal error bars for the stellar mass, specific star formation rate, and luminosity-weighted age represent the 16\textsuperscript{th}, 50\textsuperscript{th} and 84\textsuperscript{th} percentiles. The dust attenuation and Lick index D$\rm _n$4000 are measured from the best-fit SED. This table is available in its entirety in machine-readable format.}
\end{deluxetable}

\begin{figure}
    \centering
    \includegraphics[width=\linewidth]{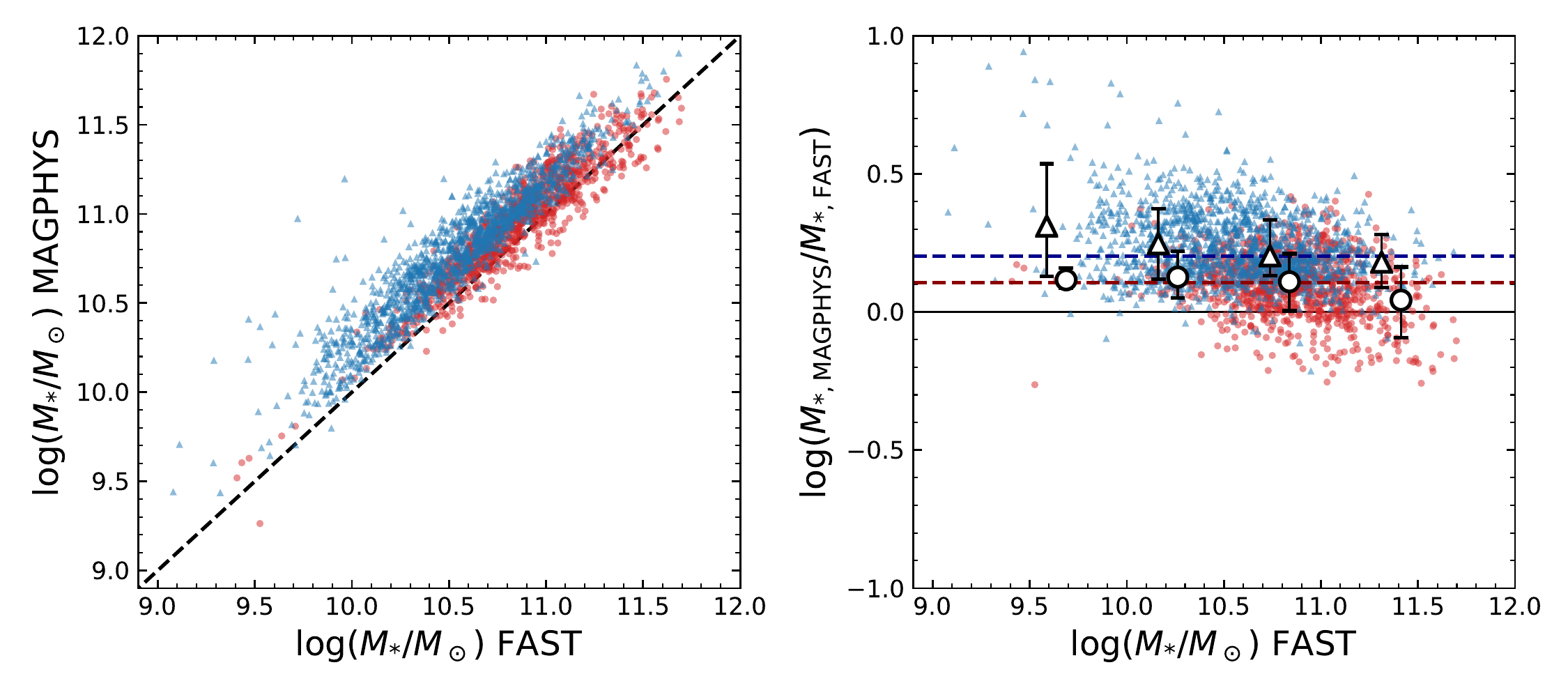}
    \caption{Comparison between the best-fit stellar mass from FAST \citep{vdWel2016} and the median of the stellar mass likelihood distribution from MAGPHYS for the primary sample of LEGA-C galaxies. Blue triangles and red circles indicate star-forming and quiescent galaxies respectively. In the right-hand panel, white markers show the median and 16$^{\rm th}$ and 84$^{\rm th}$ percentiles, with dashed lines indicating the median offset between the two mass estimates for the two populations. Stellar masses estimated with MAGPHYS are systematically larger than those from FAST, due to significant differences in the assumed star formation histories and dust attenuation model. The offset in stellar mass decreases slightly toward higher (FAST-derived) stellar mass and is larger for star-forming galaxies than quiescent galaxies, consistent with the findings by \citet{Leja2019a}.}
    \label{fig:fast_magphys}
\end{figure}

In Fig.~\ref{fig:fast_magphys} we show a comparison between the best-fit stellar mass from  \citet{vdWel2016}, and the median of the likelihood distribution of the stellar mass from MAGPHYS (as provided in Table~\ref{tab:sed}). Red circles and blue triangles show quiescent and star-forming galaxies respectively, with the median shown as black open symbols. There is a clear offset between the two stellar mass estimates (blue and red dashed lines in the right-hand panel), with the masses inferred with MAGPHYS being systematically larger ($\sim 0.1-0.2\,$dex). The offset is particularly significant for star-forming galaxies, and decreases slightly with increasing stellar mass. 

One of the main differences between the modeling with FAST and MAGPHYS is the assumed form of the star formation history (SFH). The SFHs used for the FAST modeling are simply exponentially declining SFRs ($\tau$ models), whereas those for MAGPHYS additionally include random bursts of star formation. This can lead to significant changes in the inferred stellar ages and hence stellar masses, as the fits using the $\tau$ model SFHs can significantly underestimate the stellar mass \citep[see, e.g.,][]{Pforr2012}.

Moreover, the energy balance approach, combined with a different assumed dust model, may also change the inferred stellar mass. Whereas FAST applies a single dust screen, which in this case is the attenuation curve by \citet{Calzetti2000}, MAGPHYS applies a two-component dust model \citep{Charlot2000} with different attenuation for stellar birth clouds and the diffuse interstellar medium, in better accord with observations of local galaxies \citep[e.g.,][]{Calzetti2000}. 

Lastly, there are subtle differences in the photometry used. Although both works use the photometric catalog by \citet{Muzzin2013a}, \citet{vdWel2016} use all available broad-band and medium-band filters, and exclude the \textit{Spitzer}/MIPS $24\,\micron$ data. Since the medium-band filters may suffer from large uncertainties in the zero points, and precise redshifts have already been measured from the LEGA-C spectra, we exclude these filters in our SED fitting. On the other hand, we do include the MIPS photometry, and make use of infrared libraries \citep{daCunha2008} and the energy balance recipe implemented in MAGPHYS to fit the mid-infrared data.

Our findings are broadly consistent with those by \citet{Leja2019a}, who used a Bayesian approach to model the SEDs of galaxies at redshifts $0.5<z<2.5$ with a large number (14) of free parameters. They show that, in comparison with the results from FAST, the more complex model infers older stellar ages and therefore systematically higher stellar masses, by $0.1-0.3\,$dex. Moreover, similar to our result, the discrepancy between the two stellar mass estimates decreases slightly toward higher stellar mass. By using the stellar masses inferred with MAGPHYS, we therefore not only minimize systematic effects in our comparison of the mass FP at $z\sim 0$ and $z\sim0.8$, but also adopt a stellar mass estimate that is likely to agree better with results from more sophisticated modeling.

\section{{Comparison of structural parameter estimates for the SDSS}}\label{sec:apdx_sersic}

{ In Sections~\ref{sec:sdss}, \ref{sec:lfp_tilt} and \ref{sec:mfp_tilt}, we used the structural parameters measured in the $r$-band by \citet{Simard2011}, which relies on imaging from the SDSS DR7, to measure the tilt of the FP of our low-redshift sample. However, by fitting S\'ersic models on improved photometry from the SDSS DR9 for galaxies in the MaNGA survey \citep{Bundy2015,sdss:dr15}, \citet{Fischer2019} show that the size estimates by \citet{Simard2011} may be biased. The resulting FP may therefore also change depending on the photometry and method of S\'ersic modeling used. }

{ Currently, there is no publicly available structural parameter catalog that is based on the SDSS DR9 photometry for the larger spectroscopic sample of the SDSS. A direct assessment of the effect of this improved photometry on the FP is therefore not possible. Nevertheless, \citet{Fischer2017} demonstrate that the structural parameters measured by \citet{Meert2015} are largely unaffected by changes in the photometry, due to a different treatment of the sky background as compared with \citet{Simard2011}.}

\begin{figure}
    \centering
    \includegraphics[width=\linewidth]{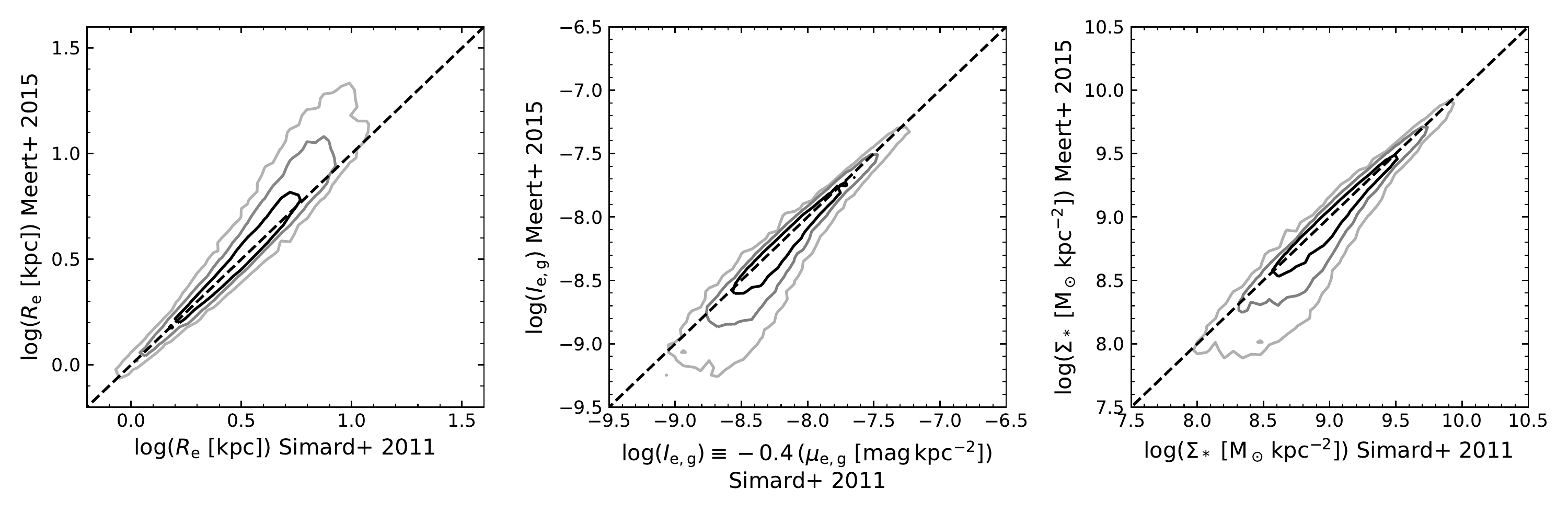}
    \caption{{Comparison between the structural parameter catalogs by \citet{Simard2011} and \citet{Meert2015} for our SDSS sample at $z\approx 0.06$ (Section~\ref{sec:sdss}). Contours enclose $50\%$, $80\%$ and $90\%$ of the total sample, respectively. The two estimates of the effective radius (left) agree well for small galaxies, but become increasingly divergent at large radii. Correspondingly, the surface brightness (middle) and stellar mass surface density (right) are in strongest disagreement at low surface brightness and surface density. Despite these discrepancies, the tilt of the $g$-band and mass FP are unchanged when using the catalog by \citet{Meert2015} rather than the \citet{Simard2011} catalog, which can be attributed to the covariance between the galaxy size and surface brightness or stellar mass surface density (see Fig.~\ref{fig:tilt_cov}).}}
    \label{fig:sdss_sersic}
\end{figure}

{By comparing the structural parameter catalogs by \citet{Simard2011} and \citet{Meert2015}, we can therefore determine the extent to which the measured tilt of the FP depends on the catalog used. Fig.~\ref{fig:sdss_sersic} shows that the effective radii differ significantly between these two different catalogs, with the measurements by \citet{Simard2011} being systematically smaller toward larger radii \citep[in agreement with findings by][]{Fischer2019}. This systematic discrepancy also affects the surface brightness and stellar mass surface density, which deviate most strongly toward low surface brightness or surface density. }

\begin{figure}
    \centering
    \includegraphics[width=0.8\linewidth]{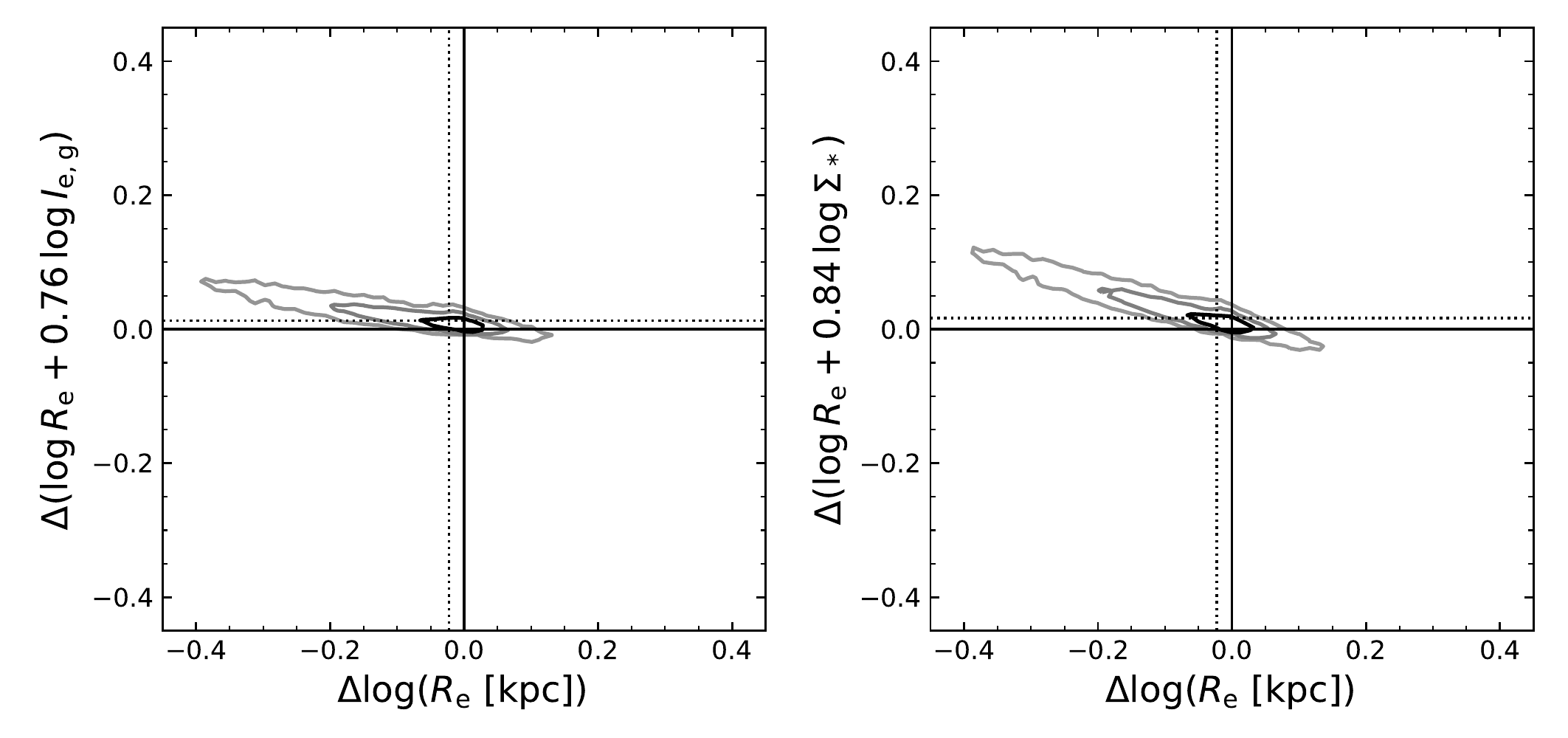}
    \caption{{The change in the $g$-band FP (left) and mass FP (right) due to differences in the size estimates between the \citet{Simard2011} and \citet{Meert2015} catalogs, assuming a fixed tilt from \citet{HydeBenardi2009}. Contours enclose $50\%$, $80\%$ and $90\%$ of the total sample, respectively, and dotted lines indicate the median values. Even a large change in the effective radius results in only a minor difference in the FP, which demonstrates that the uncertainties in $\log R_{\rm e}$ and $\log I_{\rm e,g}$ or $\log\Sigma_*$ are largely correlated along the FP. As a result, the tilt of the $g$-band and mass FP depend only very weakly on the choice of the structural parameter catalog used. }}
    \label{fig:tilt_cov}
\end{figure}

{Next, we evaluate the effect of these differences on the FP. We refit the FP using the catalog by \citet{Meert2015} and following the methodology described in Sections~\ref{sec:lfp_tilt} and \ref{sec:mfp_tilt}. We note that we do not rederive the power-law coefficients of the corrections on the velocity dispersion (Appendix~\ref{sec:apdx_apcor}), as these corrections are very small and therefore are unlikely to have a significant effect on the measurement of the tilt. For the $g$-band FP, we find $a = 1.309 \pm 0.014$ and $b = -0.726 \pm 0.003$, which is in excellent agreement with the results found in Section~\ref{sec:lfp_tilt}, where we used the catalog by \citet{Simard2011}. Similarly, we find good agreement for the mass FP, with $\alpha = 1.437 \pm 0.012$ and $\beta = -0.730 \pm 0.003$. Given the large discrepancies found in Fig.~\ref{fig:sdss_sersic}, this may be surprising. However, in Fig.~\ref{fig:tilt_cov} we show that the change in the FP due to changes in the effective radii are relatively small, which can be explained by the fact that the uncertainties in $\log R_{\rm e}$ and $\log I_{\rm e,g}$ or $\log\Sigma_*$ correlate in a direction that is near-parallel to the FP itself. We therefore conclude that, although there are significant changes in the structural parameters between different catalogs, the FP itself is insensitive to these differences. }

\section{Velocity dispersion aperture corrections}\label{sec:apdx_apcor}

As discussed in Section~\ref{sec:legac}, the integrated velocity dispersion depends on the intrinsic velocity dispersion as well as the rotational velocity of a galaxy. The profiles of these quantities will vary with radius, and the integrated velocity dispersion will therefore depend on the aperture of the spectrum. The spectra of the SDSS galaxies (Section~\ref{sec:sdss}) were obtained with fibers that are $3\arcsec$ in diameter, whereas a typical galaxy in our sample at $z\approx 0.06$ has an effective radius of $r_{\rm e}\approx 5\arcsec$. The variation in galaxy sizes within the sample, and radial gradients in the integrated velocity dispersion may therefore lead to systematic uncertainties in the measured scaling relations. 
To derive a correction for the SDSS fiber velocity dispersions ($\sigma_{\rm fiber}$) to the dispersion within a common physical aperture of $1\,r_{\rm e}$ ($\sigma_{\rm e}$), we investigate the dependence of the integrated velocity dispersion on the aperture size and structural properties using integral field unit (IFU) spectroscopy.

We match the IFU data from the MaNGA survey of the SDSS DR15 \citep{Bundy2015,sdss:dr15} with our catalog from the SDSS DR7, as well as the S\'ersic profile fits by \citet{Simard2011}. We select galaxies in the same way as in Section~\ref{sec:sdss}, but allow for a slightly wider redshift range of $0.04<z<0.08$ (median $z=0.054$), and require that the flags \texttt{DAPQUAL=0} and \texttt{DRP3QUAL=0}, resulting in a selection of 702 galaxies. For each galaxy, we use the publicly available maps of the observed stellar velocity dispersion (corrected for the effect of instrumental resolution) and velocity field \citep{Westfall2019} to calculate the flux-weighted second moment of the line-of-sight velocity:

\begin{equation}
    \sigma^2_{\rm aper} = \frac{\sum_i F_i\left( v_i^2 + \sigma_i^2 \right)}{\sum_i F_i},
    \label{eq:sigma_aper}
\end{equation}
where $F_i$ is the $g$-band flux, $v_i$ the velocity with respect to the galaxy center and $\sigma_i$ the observed
velocity dispersion measured in the $i^{\rm th}$ Voronoi bin. Bins are included only if at least 80\% of their area lies within the specified aperture. We calculate $\sigma_{\rm aper}$ for two different apertures: circular apertures of $3\arcsec$ in diameter ($\sigma_{\rm 3as}$), and elliptical apertures defined by the effective radius ($\sigma_{\rm e}$).

We use the results from the circular apertures to determine whether there are significant systematic effects between the velocity dispersions from the SDSS fiber spectra and MaNGA data, which can be due to differences in the observations themselves or in the reduction and analysis of the spectra. We find good agreement between the two measurements: there is only a small systematic offset, with a median of $(\sigma_{\rm fiber} - \sigma_{\rm 3as})=-3.0\,\rm km\,s^{-1}$, and scatter of 0.065 in the fractional difference ($\Delta \sigma = [\sigma_{\rm fiber} - \sigma_{\rm 3as}]/\sigma_{\rm 3as}$).

\begin{figure}
    \centering
    \includegraphics[width=\linewidth]{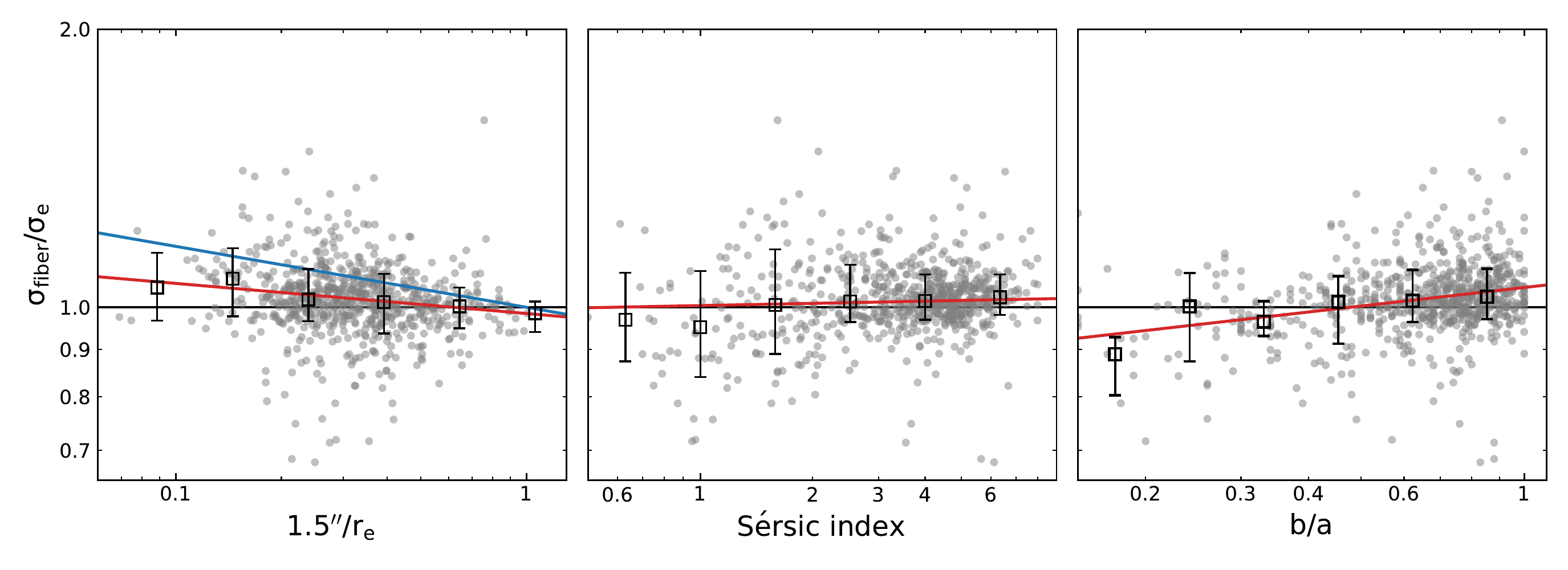}
    \caption{Ratio of the integrated velocity dispersion from the $3\arcsec$ SDSS fiber spectra and the MaNGA spectra within an aperture of one effective radius, calculated as the flux-weighted second velocity moment. Different panels show the dependence of this ratio on the effective radius, S\'ersic index, and axis ratio in gray. Black squares and error bars represent the median and 16$^{\rm th}$ and 84$^{\rm th}$ percentiles. Red lines are the best-fit power laws for each parameter. For comparison, the left-hand panel also shows the result by \citet{Cappellari2006} in blue. }
    \label{fig:manga}
\end{figure}

Next, we use the ratio of $\sigma_{\rm fiber}/\sigma_{\rm e} $ to examine the effect of aperture size. In Fig.~\ref{fig:manga}, we show $\sigma_{\rm fiber}/\sigma_{\rm e} $ for all galaxies (gray symbols) as a function of their structural parameters (the circularized effective radius, S\'ersic index, and axis ratio). Medians and percentiles (16$^{\rm th}$, 84$^{\rm th}$) are shown in black. There is a weak correlation between $\sigma_{\rm fiber}/\sigma_{\rm e} $ and the ratio of the aperture size, indicating a declining profile in the integrated velocity dispersion. On the other hand, for the few galaxies with low S\'ersic index, $\sigma_{\rm fiber}$ appears to be systematically lower than $\sigma_{\rm e} $, which may reflect a missing contribution from the rotational velocity. The third panel demonstrates this effect more clearly: for flattened systems, $\sigma_{\rm fiber}/\sigma_{\rm e}$ is significantly lower than for rounder objects.

Aperture corrections derived in previous studies usually take into account only the dependence on the ratio of the aperture and the effective radius ($r_{\rm aper}/r_{\rm e}$). E.g., \citet{Jorgensen1996} and \citet{Cappellari2006} derive a correction of the form:
\begin{equation}
  \left( \frac{\sigma_{\rm aper}}{\sigma_{\rm e}} \right) =\left( \frac{r_{\rm aper}}{r_{\rm e}}\right)^\alpha\,.
\end{equation}
Here, we use $\sigma_{\rm aper}=\sigma_{\rm fiber}$ and $r_{\rm aper}=1.5\arcsec$, and also fit a power law relation to the S\'ersic index and axis ratio:
\begin{equation}
  \left( \frac{\sigma_{\rm fiber}}{\sigma_{\rm e}} \right) =\left( \frac{4}{n}\right)^\beta\,,
\end{equation}
and
\begin{equation}
  \left( \frac{\sigma_{\rm fiber}}{\sigma_{\rm e}} \right) =\left( \frac{0.6}{b/a}\right)^\gamma\,.
\end{equation}
We fit each parameter separately, and take into account the small systematic offset between $\sigma_{\rm fiber}$ and $\sigma_{\rm 3as}$. The best-fit power law is shown in red in each panel in Fig.~\ref{fig:manga}, which have exponents $\alpha=-0.033\pm0.003 $, $\beta = -0.008\pm0.010$ and $\gamma=-0.067\pm0.012$. We also show the result by \citet{Cappellari2006} in blue, who used IFU spectroscopy for a sample of elliptical and lenticular galaxies and found a steeper relation of $\alpha=-0.066\pm0.035$. Importantly, however, our selection differs significantly from their sample, as we have not selected galaxies by morphology. Finally, we multiply the three correction factors and correct for the systematic offset between the SDSS fiber and MaNGA data, to calculate $\sigma_{\rm e}$ for each SDSS galaxy in our selection in Section~\ref{sec:sdss}. The correction to $\sigma_{\rm e}$ is typically small, with an average of $3\%$.

\section{Tilt of the Fundamental Plane}\label{sec:apdx_tilt}

Throughout this work we have assumed minimal evolution in the tilt of the FP and used a measurement of the tilt at low redshift, as accurate fitting of the FP is highly complex, and our results do not depend strongly on the assumed tilt. However, in Sections~\ref{sec:lfp_tilt}~\&~\ref{sec:mfp_tilt} we showed there is weak evidence for an evolution in the tilt of the FP, particularly so for the $g$-band FP. These measurements relied on a relatively simple planar fit to a subset of the data that is most complete in mass. 
Here, we further examine the redshift evolution of the tilt of the FP for the full sample of LEGA-C galaxies, and additionally take into account the measurement uncertainties and the effect of sample completeness.

We begin by writing the luminosity FP and mass FP as the power-law relations
\begin{equation}
    R_{\rm e} \propto \sigma^a\,I_{\rm e}^b\quad {\rm and} \quad R_{\rm e} \propto \sigma^\alpha\,\Sigma_*^\beta\,,
\end{equation}
where $I_{\rm e }\propto L/R_{\rm e}^2$ and $\Sigma_*\propto M_*/R_{\rm e}^2$. Under the assumption of homology, i.e. $M_{\rm dyn}\propto R_{\rm e}\,\sigma^2$, the FP can be rewritten as a power-law relation between $M_{\rm dyn}$, $R_{\rm e}$, and $M_{\rm dyn}/L$ or $\MM$ \citep[for a full derivation, see][]{Cappellari2006}:
\begin{equation}
    M_{\rm dyn}/L \propto M_{\rm dyn}^d\, R_{\rm e}^f\,,
\end{equation}
or
\begin{equation}
    M_{\rm dyn}/M_* \propto M_{\rm dyn}^\delta \, R_{\rm e}^\eta\,,
    \label{eq:MM_plaw}
\end{equation}
where the exponents $d$ and $f$ (or $\delta$, $\eta$) depend on the tilt of the FP. If $f\ll d$ ($\eta \ll\delta$), then the tilt of the FP reflects the relation between $M_{\rm dyn}/L$ ($\MM$) and mass, as first proposed by \citet{Faber1987}. 

A measurement of the tilt depends strongly on the methodology used \citep[e.g., a direct planar fit versus an orthogonal fit, see][]{HydeBenardi2009}, the sample completeness, and uncertainties on different parameters as well as their covariances \citep[see also][]{Magoulas2012}. However, we can reduce some of these uncertainties by calibrating the relation between $M_{\rm dyn}/L$ and $M_{\rm dyn}$ directly (under the assumption that $f\ll d$, $\eta \ll\delta$), using an estimate of $M_{\rm dyn}$:
\begin{equation}
    M_{\rm dyn} = K \frac{R_{\rm e}\,\sigma^2}{G}\,,
    \label{eq:mdyn}
\end{equation}
where $G$ is the gravitational constant and $K$ is the virial coefficient, which in general depends on the structural properties of the galaxy. We set $K=5$, which was shown by \citet{Cappellari2006} to provide a good approximation for early-type galaxies. This may not provide a good estimate of $K$ for late-type galaxies, however, the effect of the assumed virial constant, e.g. a S\'ersic-dependent virial constant, is small (see also Fig~\ref{fig:legac_nsersic} and Section~\ref{sec:discussion}).

\begin{figure*}
    \centering
    \includegraphics[width=0.8\linewidth]{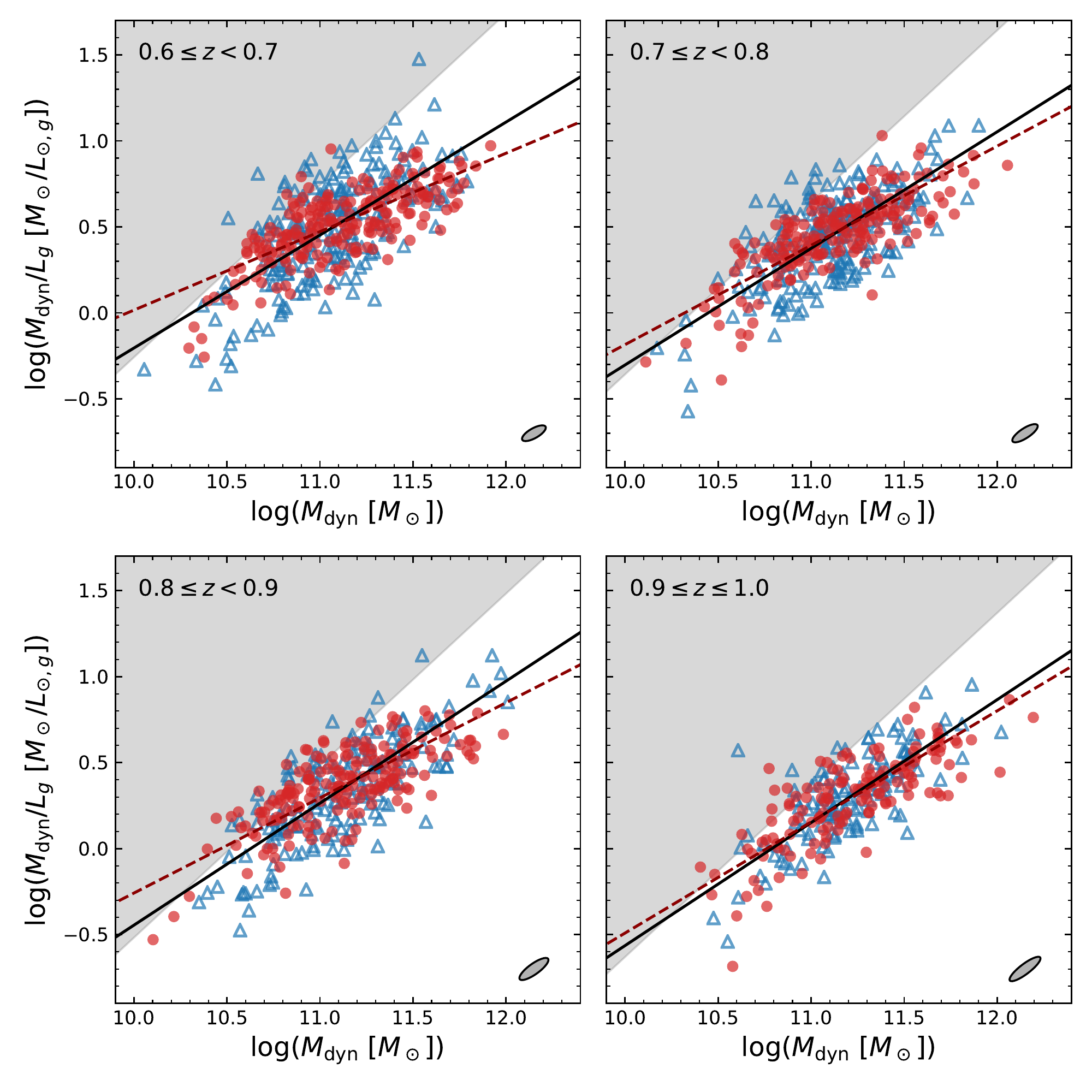}
    \caption{Relation between the dynamical mass-light ratio ($M_{\rm dyn}/L_g$) and dynamical mass as a function of redshift. Dashed lines show linear fits to the quiescent galaxies (red circles) in each redshift bin; black lines show the result for the combined sample of quiescent and star-forming  (blue triangles) galaxies (see Table~\ref{tab:lfp_Mdyn}). Ellipses show the typical measurement uncertainties. Shaded regions mark galaxies of $m_{\rm g,rest} > 22.5$ and illustrate the effect of sample selection in the $I$-band, common in previous studies of the FP, or a S/N limit for the velocity dispersion. 
    The slope of the relation between $M_{\rm dyn}/L_g$ and $M_{\rm dyn}$ varies weakly with redshift, which can partially be attributed to incompleteness at high $M_{\rm dyn}/L_g$ toward higher redshift and lower $M_{\rm dyn}$. Since the slope is analogous to the tilt of the $g$-band FP, there is likely also a weak evolution in the tilt of the FP. Moreover, there is strong evolution in the intercept, as is expected from evolution in the stellar populations \citep[see also][]{deGraaff2020}.}
    \label{fig:MdynL}
\end{figure*}

\subsection{Direct measurement of $M_{\rm dyn}/L_g$ vs. $M_{\rm dyn}$}\label{sec:mdynl}

We show the relation between $M_{\rm dyn}/L_g$ and $M_{\rm dyn}$ in Fig.~\ref{fig:MdynL}, in bins of $\Delta z = 0.10$. Since previous measurements of the FP focused solely on early-type galaxies, we consider both the quiescent population alone (red circles), as well as the combined sample of quiescent and star-forming (blue) galaxies. 

There is a strong correlation between $M_{\rm dyn}/L_g$ and $M_{\rm dyn}$, in part due to the covariance between the two quantities. Moreover, the effect of sample incompleteness becomes apparent from this figure: our S/N selection on the velocity dispersion approximately translates to a selection on the rest-frame $g$-band magnitude, illustrated in Fig.~\ref{fig:MdynL} by shaded regions that cover $m_g > 22.5$. Toward lower $M_{\rm dyn}$ as well as higher redshift, this contributes to an apparent steepening of the observed power-law relation.

To estimate the exponent $d$, we therefore exclude the lowest-mass galaxies, requiring $\log(M_{\rm dyn}/M_\odot) > 10.6$. Fitting in logarithmic space, we use the orthogonal distance regression described by \citet[][Eq.~35]{Hogg2010}, which takes into account the uncertainties in both axes and treats the (Gaussian) intrinsic variance orthogonal to the linear fit as a free parameter. We use the measurement uncertainties to estimate the covariance matrix for each galaxy with 1000 Monte Carlo simulations.

To account for the fact that galaxies of high $M_{\rm dyn}/L_K$ are less likely to be observed, we use the completeness correction Tcor (Section~\ref{sec:lfp_tilt}) to weight the covariance matrices, however, we now firstly renormalize these corrections in bins of $M_{\rm dyn}$. The effect of this normalization is (i) that at fixed $M_{\rm dyn}$ galaxies of high $M_{\rm dyn}/L$ receive a greater weight, and (ii) that galaxies of low $M_{\rm dyn}$, where the completeness in $M_{\rm dyn}/L$ is lowest, do not introduce an extreme bias on the measured exponent.

For the quiescent sample, we find a weak evolution in the exponent between $0.6\leq z<0.9$ (Table~\ref{tab:lfp_Mdyn}), with measurements deviating by $\sim1-2\sigma$ (where uncertainties on the fits are obtained from bootstrapping the data). The largest discrepancy is between the lowest ($0.6\leq z <0.7$) and highest redshift bins ($0.9\leq z \leq 1.0$), which deviate by $2.2\sigma$. However, in the highest redshift bin there are relatively few galaxies at low mass (see also Fig.~\ref{fig:mstar_z}), and the fit is therefore most affected by sample incompleteness. 

Our measurements for the quiescent sample agree well with the results by \citet{Joergensen2013}, who measured $d = 0.44 \pm 0.09$ at $z=0.54$ and $d=0.55 \pm 0.08$ for quiescent cluster galaxies at $z\approx 0.85$ in the rest-frame $B$-band, as well as the work by \citet{Saracco2020}, who found $d=0.6\pm0.1$ at $z\approx 1.3$. Importantly, in both works the galaxy samples were selected in the $I$-band, which introduces a selection effect similar to the shaded regions in Fig.~\ref{fig:MdynL}, and thus can steepen the inferred power-law. 

Lastly, we compare our measured relations with the low-redshift SDSS sample, using Eq.~\ref{eq:mdyn} (with $K=5$) to estimate $M_{\rm dyn}$ and applying the same fitting procedure to estimate $d$. The measurement relation is shallower than the LEGA-C measurements by $2-3\sigma$ ($0.6<z<0.9$), suggesting a weak evolution with redshift or a selection bias against galaxies that are faint in the rest-frame $g$-band, or, most likely, a combination of both.

We find a steepening of the slope when we include both quiescent and star-forming galaxies in the fit. Interestingly, there is no significant evolution within the LEGA-C sample ($0.6<z<1.0$) in this case, suggesting that the effect of sample selection is partially driving the observed evolution in $d$ for the quiescent LEGA-C galaxies. On the other hand, there is weak evolution (at a level of $\approx 3\sigma$) in comparison with the SDSS fit when considering the full sample of star-forming and quiescent galaxies, pointing toward physical differences between the low and high-redshift samples.

In summary, we find evidence for an evolution in the power-law relation between $M_{\rm dyn}/L_g$ and $M_{\rm dyn}$ with redshift, and, by extension, the tilt of the FP. {This evolution can be explained by an increasingly declining $\ML$ for less massive galaxies toward higher redshift:} this evolution is expected, as more galaxies are star-forming at higher redshifts, and this effect is strongest at lower masses \citep[``downsizing'',][]{Cowie1996, Brinchmann2004}. However, we find that the effects of sample selection and completeness also contribute significantly the observed redshift evolution. To determine the extent to which the evolution is of a physical origin, will require a more careful analysis of the selection function of both the SDSS and LEGA-C samples.

\begin{deluxetable}{cccccc}
\tablecaption{Best-fit $M_{\rm dyn}/L_g$ vs. $M_{\rm dyn}$ exponents}
\tablewidth{0pt}
\tablehead{
\colhead{} &\multicolumn{2}{c}{Quiescent} & \multicolumn{3}{c}{Quiescent + star-forming}\\
\colhead{Redshift} & \colhead{$N_{\rm gal}$} & \colhead{$d$}  & \colhead{$N_{\rm gal}$} & \colhead{$d$}  
}
\startdata
$0.05 < z < 0.07$ & 13,468 & $0.386\pm 0.004$ & 20,508 & $0.514\pm 0.005$\\ 
 $0.6\leq z < 0.7$ & 202 & $0.46\pm 0.04$ & 411 & $0.64\pm 0.06$ \\
$0.7\leq z < 0.8$ & 183 & $0.58\pm 0.06$ & 393 & $0.67\pm 0.05$ \\
$0.8\leq z < 0.9$ & 180 & $0.55\pm 0.05$ & 349 & $0.66\pm 0.06$\\
$0.9\leq z \leq 1.0$ & 138& $0.64\pm 0.07$ & 266 &  $0.70\pm 0.06$
\enddata
\tablecomments{Only galaxies of $\log(M_{\rm dyn}/M_\odot) > 10.6$ are included in the fits.}
\label{tab:lfp_Mdyn}
\end{deluxetable}

\subsection{Direct measurement of $M_{\rm dyn}/M_*$ vs. $M_{\rm dyn}$}\label{sec:mdynm}

\begin{figure*}
    \centering
    \includegraphics[width=0.8\linewidth]{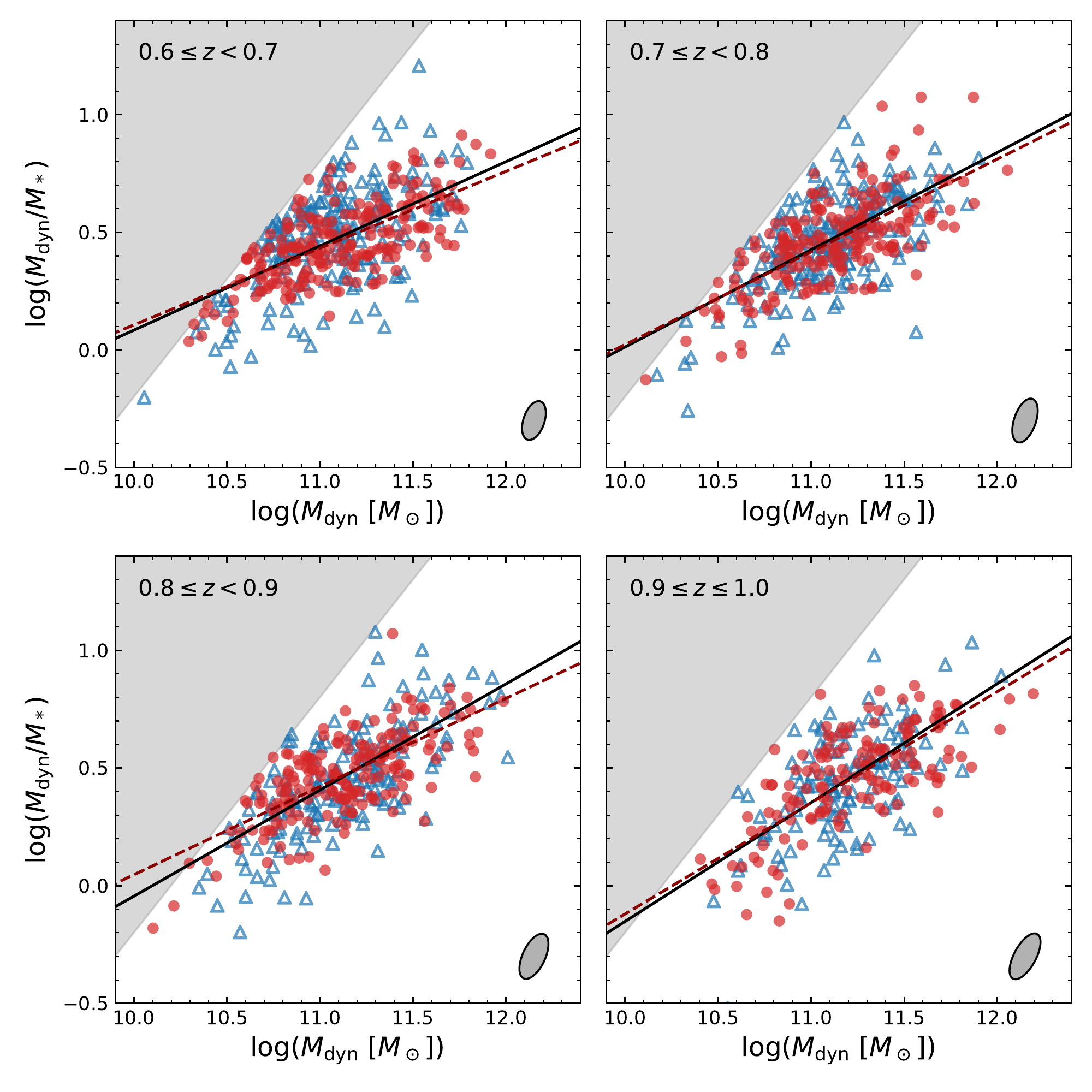}
    \caption{Relation between the dynamical-to-stellar mass ratio ($M_{\rm dyn}/M_*$) and dynamical mass as a function of redshift. Symbols indicate the same as in Fig.~\ref{fig:MdynL}. 
    The effect of our selection on stellar mass ($\log(M_*/M_\odot)\geq10.5$) is indicated by the shaded regions.
    We find no significant variation with redshift in the slope of the relation between $\log(M_{\rm dyn}/M_*)$ and $\log(M_{\rm dyn})$ for the quiescent galaxies, and a weak evolution when considering the full sample (see Table~\ref{tab:mfp_Mdyn}). }
    \label{fig:MdynM}
\end{figure*}

Fig.~\ref{fig:MdynM} shows the relation between $M_{\rm dyn}/M_*$ and $M_{\rm dyn}$ at different redshifts, again with red and blue markers showing the quiescent and star-forming LEGA-C galaxies, respectively. The gray regions now illustrate the effect of our selection on stellar mass, which we used in Section~\ref{sec:sample_selection} to homogenize our sample. 
The best-fit power laws are shown as dashed lines and solid lines (representing fits to the quiescent and full samples, respectively), the exponents ($\delta$) of which are presented in Table~\ref{tab:mfp_Mdyn}. 

For the quiescent galaxies, we find no evolution in $\delta$ between $0.6<z<0.9$ and a reasonable agreement (a deviation $<1.5\sigma$) with the measurement at low redshift. The highest redshift bin does diverge more strongly, but, as is apparent from Fig.~\ref{fig:MdynM}, this measurement is likely strongly affected by an incompleteness in $\MM$ at $\log(M_{\rm dyn}/M_\odot)\lesssim 10.9$. These results are consistent with our findings for the tilt of the FP in Section~\ref{sec:mfp_tilt}, as well as previous work by \citet{Bundy2007}, who found no evolution in the relation between $M_{\rm dyn}$ and $M_*$ between $z\sim0$ and $z\sim1$ for a sample of spheroidal galaxies.

When considering the combined sample of quiescent and star-forming galaxies, we do observe a weak evolution in $\delta$, as we measure a slight steepening with redshift both within the LEGA-C sample itself and in comparison with the SDSS data. Moreover, these exponents are steeper than the fits to the quiescent galaxies for all redshift ranges. {Interestingly, this is opposite to the result of an orthogonal fit to the FP (Section~\ref{sec:mfp_tilt}), where we find that the tilt of the FP is slightly closer to that of the virial plane at $z\sim0.8$ than at $z\sim0$. Additionally, the values of $\alpha$ and $\beta$ are (marginally) closer to the virial prediction for the full sample than for the quiescent sample alone:} $\alpha=1.64\pm0.09$ and $\beta=-0.71\pm0.02$ (LEGA-C, $0.65<z<0.75$), whereas the fit to the quiescent sample gives $\alpha=1.56\pm0.12$ and $\beta=-0.68\pm0.03$ ($\alpha=1.467\pm0.014$ and $\beta=-0.730\pm0.004$ versus $\alpha=1.432\pm0.012$ and $\beta=-0.736\pm0.003$, respectively, for the SDSS). 

This apparent contradiction may reflect an increasing incompleteness in $\MM$ toward higher redshift, with the difference in $\delta$ between the two samples at low redshift being due to the selection on $M_*$ and the maximum allowed uncertainty on the velocity dispersion. We indeed find that the measurements for the SDSS data are in good agreement when we relax our stellar mass limit, with $\delta=0.285\pm0.004$ and $\delta=0.287\pm0.003$ for the quiescent and full SDSS samples, respectively. {Alternatively, whereas the variations in $\ML$ at fixed $M_{\rm dyn}$ are largely due to variations in $M_*/L$, the variations in $\MM$ depend on variations in the IMF \citep[and radial gradients therein; see][]{Bernardi2019} as well as the dark matter content, which in turn depends on the galaxy size (as discussed in Section~\ref{sec:discussion}). Therefore, the discrepancy between the measurement of the tilt and the measurement of the relation between $\MM$ and $M_{\rm dyn}$ may indicate that (i) the effects of stellar population gradients on the measurement of $\MM$ cannot be neglected, or (ii) our assumption of minimal $R_{\rm e}$ dependence ($\eta \ll\delta$; Eq.~\ref{eq:MM_plaw}) no longer holds, and that the measurement of $\delta$ alone therefore is insufficient to draw conclusions on the evolution of the tilt of mass FP.}

\begin{deluxetable}{cccccc}
\tablecaption{Best-fit $M_{\rm dyn}/M_*$ vs. $M_{\rm dyn}$ exponents}
\tablewidth{0pt}
\tablehead{
\colhead{} &\multicolumn{2}{c}{Quiescent} & \multicolumn{3}{c}{Quiescent + star-forming}\\
\colhead{Redshift} & \colhead{$N_{\rm gal}$} & \colhead{$\delta$}  & \colhead{$N_{\rm gal}$} & \colhead{$\delta$}  
}
\startdata
$0.05 < z < 0.07$ & 13,468 & $0.313\pm 0.004$ & 20,508 & $0.330\pm 0.003$\\
 $0.6\leq z < 0.7$ & 202 & $0.33\pm 0.04$ & 411 & $0.36\pm 0.03$ \\
$0.7\leq z < 0.8$ & 183 & $0.39\pm 0.05$ & 393 & $0.41\pm 0.04$ \\
$0.8\leq z < 0.9$ & 180 & $0.37\pm 0.04$ & 349 & $0.45\pm 0.04$\\
$0.9\leq z \leq 1.0$ & 138& $0.50\pm 0.07$ & 266 &  $0.53\pm 0.05$
\enddata
\tablecomments{Only galaxies of $\log(M_{\rm dyn}/M_\odot) > 10.6$ are included in the fits.}
\label{tab:mfp_Mdyn}
\end{deluxetable}

\bibliography{legac}{}

\begin{thebibliography}{}
\expandafter\ifx\csname natexlab\endcsname\relax\def\natexlab#1{#1}\fi
\providecommand{\url}[1]{\href{#1}{#1}}
\providecommand{\dodoi}[1]{doi:~\href{http://doi.org/#1}{\nolinkurl{#1}}}
\providecommand{\doeprint}[1]{\href{http://ascl.net/#1}{\nolinkurl{http://ascl.net/#1}}}
\providecommand{\doarXiv}[1]{\href{https://arxiv.org/abs/#1}{\nolinkurl{https://arxiv.org/abs/#1}}}

\bibitem[{{Abazajian} {et~al.}(2009){Abazajian}, {Adelman-McCarthy},
  {Ag{\"u}eros}, {Allam}, {Allende Prieto}, {An}, {Anderson}, {Anderson},
  {Annis}, {Bahcall}, {Bailer-Jones}, {Barentine}, {Bassett}, {Becker},
  {Beers}, {Bell}, {Belokurov}, {Berlind}, {Berman}, {Bernardi}, {Bickerton},
  {Bizyaev}, {Blakeslee}, {Blanton}, {Bochanski}, {Boroski}, {Brewington},
  {Brinchmann}, {Brinkmann}, {Brunner}, {Budav{\'a}ri}, {Carey}, {Carliles},
  {Carr}, {Castander}, {Cinabro}, {Connolly}, {Csabai}, {Cunha}, {Czarapata},
  {Davenport}, {de Haas}, {Dilday}, {Doi}, {Eisenstein}, {Evans}, {Evans},
  {Fan}, {Friedman}, {Frieman}, {Fukugita}, {G{\"a}nsicke}, {Gates},
  {Gillespie}, {Gilmore}, {Gonzalez}, {Gonzalez}, {Grebel}, {Gunn},
  {Gy{\"o}ry}, {Hall}, {Harding}, {Harris}, {Harvanek}, {Hawley}, {Hayes},
  {Heckman}, {Hendry}, {Hennessy}, {Hindsley}, {Hoblitt}, {Hogan}, {Hogg},
  {Holtzman}, {Hyde}, {Ichikawa}, {Ichikawa}, {Im}, {Ivezi{\'c}}, {Jester},
  {Jiang}, {Johnson}, {Jorgensen}, {Juri{\'c}}, {Kent}, {Kessler}, {Kleinman},
  {Knapp}, {Konishi}, {Kron}, {Krzesinski}, {Kuropatkin}, {Lampeitl},
  {Lebedeva}, {Lee}, {Lee}, {French Leger}, {L{\'e}pine}, {Li}, {Lima}, {Lin},
  {Long}, {Loomis}, {Loveday}, {Lupton}, {Magnier}, {Malanushenko},
  {Malanushenko}, {Mand elbaum}, {Margon}, {Marriner}, {Mart{\'\i}nez-Delgado},
  {Matsubara}, {McGehee}, {McKay}, {Meiksin}, {Morrison}, {Mullally}, {Munn},
  {Murphy}, {Nash}, {Nebot}, {Neilsen}, {Newberg}, {Newman}, {Nichol},
  {Nicinski}, {Nieto-Santisteban}, {Nitta}, {Okamura}, {Oravetz}, {Ostriker},
  {Owen}, {Padmanabhan}, {Pan}, {Park}, {Pauls}, {Peoples}, {Percival}, {Pier},
  {Pope}, {Pourbaix}, {Price}, {Purger}, {Quinn}, {Raddick}, {Re Fiorentin},
  {Richards}, {Richmond}, {Riess}, {Rix}, {Rockosi}, {Sako}, {Schlegel},
  {Schneider}, {Scholz}, {Schreiber}, {Schwope}, {Seljak}, {Sesar}, {Sheldon},
  {Shimasaku}, {Sibley}, {Simmons}, {Sivarani}, {Allyn Smith}, {Smith},
  {Smol{\v{c}}i{\'c}}, {Snedden}, {Stebbins}, {Steinmetz}, {Stoughton},
  {Strauss}, {SubbaRao}, {Suto}, {Szalay}, {Szapudi}, {Szkody}, {Tanaka},
  {Tegmark}, {Teodoro}, {Thakar}, {Tremonti}, {Tucker}, {Uomoto}, {Vanden
  Berk}, {Vandenberg}, {Vidrih}, {Vogeley}, {Voges}, {Vogt}, {Wadadekar},
  {Watters}, {Weinberg}, {West}, {White}, {Wilhite}, {Wonders}, {Yanny},
  {Yocum}, {York}, {Zehavi}, {Zibetti}, \& {Zucker}}]{sdss:dr7}
{Abazajian}, K.~N., {Adelman-McCarthy}, J.~K., {Ag{\"u}eros}, M.~A., {et~al.}
  2009, \apjs, 182, 543, \dodoi{10.1088/0067-0049/182/2/543}

\bibitem[{{Aquino-Ort{\'\i}z} {et~al.}(2020){Aquino-Ort{\'\i}z}, {S{\'a}nchez},
  {Valenzuela}, {Hern{\'a}ndez-Toledo}, {Jin}, {Zhu}, {van de Ven},
  {Barrera-Ballesteros}, {Avila-Reese}, {Rodr{\'\i}guez-Puebla}, \&
  {Tissera}}]{Aquino2020}
{Aquino-Ort{\'\i}z}, E., {S{\'a}nchez}, S.~F., {Valenzuela}, O., {et~al.} 2020,
  \apj, 900, 109, \dodoi{10.3847/1538-4357/aba94e}

\bibitem[{{Astropy Collaboration} {et~al.}(2013){Astropy Collaboration},
  {Robitaille}, {Tollerud}, {Greenfield}, {Droettboom}, {Bray}, {Aldcroft},
  {Davis}, {Ginsburg}, {Price-Whelan}, {Kerzendorf}, {Conley}, {Crighton},
  {Barbary}, {Muna}, {Ferguson}, {Grollier}, {Parikh}, {Nair}, {Unther},
  {Deil}, {Woillez}, {Conseil}, {Kramer}, {Turner}, {Singer}, {Fox}, {Weaver},
  {Zabalza}, {Edwards}, {Azalee Bostroem}, {Burke}, {Casey}, {Crawford},
  {Dencheva}, {Ely}, {Jenness}, {Labrie}, {Lim}, {Pierfederici}, {Pontzen},
  {Ptak}, {Refsdal}, {Servillat}, \& {Streicher}}]{astropy}
{Astropy Collaboration}, {Robitaille}, T.~P., {Tollerud}, E.~J., {et~al.} 2013,
  \aap, 558, A33, \dodoi{10.1051/0004-6361/201322068}

\bibitem[{{Beifiori} {et~al.}(2017){Beifiori}, {Mendel}, {Chan}, {Saglia},
  {Bender}, {Cappellari}, {Davies}, {Galametz}, {Houghton}, {Prichard},
  {Smith}, {Stott}, {Wilman}, {Lewis}, {Sharples}, \& {Wegner}}]{Beifiori2017}
{Beifiori}, A., {Mendel}, J.~T., {Chan}, J. C.~C., {et~al.} 2017, \apj, 846,
  120, \dodoi{10.3847/1538-4357/aa8368}

\bibitem[{{Belli} {et~al.}(2017){Belli}, {Newman}, \& {Ellis}}]{Belli2017}
{Belli}, S., {Newman}, A.~B., \& {Ellis}, R.~S. 2017, \apj, 834, 18,
  \dodoi{10.3847/1538-4357/834/1/18}

\bibitem[{{Bender} {et~al.}(1992){Bender}, {Burstein}, \& {Faber}}]{Bender1992}
{Bender}, R., {Burstein}, D., \& {Faber}, S.~M. 1992, \apj, 399, 462,
  \dodoi{10.1086/171940}

\bibitem[{{Bernardi} {et~al.}(2019){Bernardi}, {Dom{\'\i}nguez S{\'a}nchez},
  {Brownstein}, {Drory}, \& {Sheth}}]{Bernardi2019}
{Bernardi}, M., {Dom{\'\i}nguez S{\'a}nchez}, H., {Brownstein}, J.~R., {Drory},
  N., \& {Sheth}, R.~K. 2019, \mnras, 489, 5633, \dodoi{10.1093/mnras/stz2413}

\bibitem[{{Bernardi} {et~al.}(2020){Bernardi}, {Dom{\'\i}nguez S{\'a}nchez},
  {Margalef-Bentabol}, {Nikakhtar}, \& {Sheth}}]{Bernardi2020}
{Bernardi}, M., {Dom{\'\i}nguez S{\'a}nchez}, H., {Margalef-Bentabol}, B.,
  {Nikakhtar}, F., \& {Sheth}, R.~K. 2020, \mnras, 494, 5148,
  \dodoi{10.1093/mnras/staa1064}

\bibitem[{{Bezanson} {et~al.}(2015){Bezanson}, {Franx}, \& {van
  Dokkum}}]{Bezanson2015}
{Bezanson}, R., {Franx}, M., \& {van Dokkum}, P.~G. 2015, \apj, 799, 148,
  \dodoi{10.1088/0004-637X/799/2/148}

\bibitem[{{Bezanson} {et~al.}(2009){Bezanson}, {van Dokkum}, {Tal},
  {Marchesini}, {Kriek}, {Franx}, \& {Coppi}}]{Bezanson2009}
{Bezanson}, R., {van Dokkum}, P.~G., {Tal}, T., {et~al.} 2009, \apj, 697, 1290,
  \dodoi{10.1088/0004-637X/697/2/1290}

\bibitem[{{Bezanson} {et~al.}(2013){Bezanson}, {van Dokkum}, {van de Sande},
  {Franx}, {Leja}, \& {Kriek}}]{Bezanson2013}
{Bezanson}, R., {van Dokkum}, P.~G., {van de Sande}, J., {et~al.} 2013, \apjl,
  779, L21, \dodoi{10.1088/2041-8205/779/2/L21}

\bibitem[{{Bezanson} {et~al.}(2018{\natexlab{a}}){Bezanson}, {van der Wel},
  {Pacifici}, {Noeske}, {Bari{\v{s}}i{\'c}}, {Bell}, {Brammer}, {Calhau},
  {Chauke}, {van Dokkum}, {Franx}, {Gallazzi}, {van Houdt}, {Labb{\'e}},
  {Maseda}, {Mu{\~n}os-Mateos}, {Muzzin}, {van de Sand e}, {Sobral},
  {Straatman}, \& {Wu}}]{Bezanson2018a}
{Bezanson}, R., {van der Wel}, A., {Pacifici}, C., {et~al.} 2018{\natexlab{a}},
  \apj, 858, 60, \dodoi{10.3847/1538-4357/aabc55}

\bibitem[{{Bezanson} {et~al.}(2018{\natexlab{b}}){Bezanson}, {van der Wel},
  {Straatman}, {Pacifici}, {Wu}, {Bari{\v{s}}i{\'c}}, {Bell}, {Conroy},
  {D'Eugenio}, {Franx}, {Gallazzi}, {van Houdt}, {Maseda}, {Muzzin}, {van de
  Sande}, {Sobral}, \& {Spilker}}]{Bezanson2018b}
{Bezanson}, R., {van der Wel}, A., {Straatman}, C., {et~al.}
  2018{\natexlab{b}}, \apjl, 868, L36, \dodoi{10.3847/2041-8213/aaf16b}

\bibitem[{{Blanton} \& {Roweis}(2007)}]{Blanton2007}
{Blanton}, M.~R., \& {Roweis}, S. 2007, \aj, 133, 734, \dodoi{10.1086/510127}

\bibitem[{{Blanton} {et~al.}(2003){Blanton}, {Hogg}, {Bahcall}, {Baldry},
  {Brinkmann}, {Csabai}, {Eisenstein}, {Fukugita}, {Gunn}, {Ivezi{\'c}},
  {Lamb}, {Lupton}, {Loveday}, {Munn}, {Nichol}, {Okamura}, {Schlegel},
  {Shimasaku}, {Strauss}, {Vogeley}, \& {Weinberg}}]{Blanton2003}
{Blanton}, M.~R., {Hogg}, D.~W., {Bahcall}, N.~A., {et~al.} 2003, \apj, 594,
  186, \dodoi{10.1086/375528}

\bibitem[{{Blanton} {et~al.}(2017){Blanton}, {Bershady}, {Abolfathi},
  {Albareti}, {Allende Prieto}, {Almeida}, {Alonso-Garc{\'{\i}}a}, {Anders},
  {Anderson}, {Andrews}, \& et~al.}]{sdss:dr15}
{Blanton}, M.~R., {Bershady}, M.~A., {Abolfathi}, B., {et~al.} 2017, \aj, 154,
  28, \dodoi{10.3847/1538-3881/aa7567}

\bibitem[{{Bolton} {et~al.}(2008){Bolton}, {Treu}, {Koopmans}, {Gavazzi},
  {Moustakas}, {Burles}, {Schlegel}, \& {Wayth}}]{Bolton2008}
{Bolton}, A.~S., {Treu}, T., {Koopmans}, L. V.~E., {et~al.} 2008, \apj, 684,
  248, \dodoi{10.1086/589989}

\bibitem[{{Brammer} {et~al.}(2008){Brammer}, {van Dokkum}, \&
  {Coppi}}]{Brammer2008}
{Brammer}, G.~B., {van Dokkum}, P.~G., \& {Coppi}, P. 2008, \apj, 686, 1503,
  \dodoi{10.1086/591786}

\bibitem[{{Brammer} {et~al.}(2009){Brammer}, {Whitaker}, {van Dokkum},
  {Marchesini}, {Labb{\'e}}, {Franx}, {Kriek}, {Quadri}, {Illingworth}, {Lee},
  {Muzzin}, \& {Rudnick}}]{Brammer2009}
{Brammer}, G.~B., {Whitaker}, K.~E., {van Dokkum}, P.~G., {et~al.} 2009, \apjl,
  706, L173, \dodoi{10.1088/0004-637X/706/1/L173}

\bibitem[{{Brinchmann} {et~al.}(2004){Brinchmann}, {Charlot}, {White},
  {Tremonti}, {Kauffmann}, {Heckman}, \& {Brinkmann}}]{Brinchmann2004}
{Brinchmann}, J., {Charlot}, S., {White}, S.~D.~M., {et~al.} 2004, \mnras, 351,
  1151, \dodoi{10.1111/j.1365-2966.2004.07881.x}

\bibitem[{{Bruzual} \& {Charlot}(2003)}]{Bruzual2003}
{Bruzual}, G., \& {Charlot}, S. 2003, \mnras, 344, 1000,
  \dodoi{10.1046/j.1365-8711.2003.06897.x}

\bibitem[{{Bundy} {et~al.}(2007){Bundy}, {Treu}, \& {Ellis}}]{Bundy2007}
{Bundy}, K., {Treu}, T., \& {Ellis}, R.~S. 2007, \apjl, 665, L5,
  \dodoi{10.1086/519526}

\bibitem[{{Bundy} {et~al.}(2015){Bundy}, {Bershady}, {Law}, {Yan}, {Drory},
  {MacDonald}, {Wake}, {Cherinka}, {S{\'a}nchez-Gallego}, {Weijmans}, {Thomas},
  {Tremonti}, {Masters}, {Coccato}, {Diamond-Stanic}, {Arag{\'o}n-Salamanca},
  {Avila-Reese}, {Badenes}, {Falc{\'o}n-Barroso}, {Belfiore}, {Bizyaev},
  {Blanc}, {Bland-Hawthorn}, {Blanton}, {Brownstein}, {Byler}, {Cappellari},
  {Conroy}, {Dutton}, {Emsellem}, {Etherington}, {Frinchaboy}, {Fu}, {Gunn},
  {Harding}, {Johnston}, {Kauffmann}, {Kinemuchi}, {Klaene}, {Knapen},
  {Leauthaud}, {Li}, {Lin}, {Maiolino}, {Malanushenko}, {Malanushenko}, {Mao},
  {Maraston}, {McDermid}, {Merrifield}, {Nichol}, {Oravetz}, {Pan}, {Parejko},
  {Sanchez}, {Schlegel}, {Simmons}, {Steele}, {Steinmetz}, {Thanjavur},
  {Thompson}, {Tinker}, {van den Bosch}, {Westfall}, {Wilkinson}, {Wright},
  {Xiao}, \& {Zhang}}]{Bundy2015}
{Bundy}, K., {Bershady}, M.~A., {Law}, D.~R., {et~al.} 2015, \apj, 798, 7,
  \dodoi{10.1088/0004-637X/798/1/7}

\bibitem[{{Burstein} {et~al.}(1990){Burstein}, {Faber}, \&
  {Dressler}}]{Burstein1990}
{Burstein}, D., {Faber}, S.~M., \& {Dressler}, A. 1990, \apj, 354, 18,
  \dodoi{10.1086/168664}

\bibitem[{{Calzetti} {et~al.}(2000){Calzetti}, {Armus}, {Bohlin}, {Kinney},
  {Koornneef}, \& {Storchi-Bergmann}}]{Calzetti2000}
{Calzetti}, D., {Armus}, L., {Bohlin}, R.~C., {et~al.} 2000, \apj, 533, 682,
  \dodoi{10.1086/308692}

\bibitem[{{Cappellari}(2016)}]{Cappellari2016}
{Cappellari}, M. 2016, \araa, 54, 597,
  \dodoi{10.1146/annurev-astro-082214-122432}

\bibitem[{{Cappellari}(2017)}]{Cappellari2017}
---. 2017, \mnras, 466, 798, \dodoi{10.1093/mnras/stw3020}

\bibitem[{{Cappellari} \& {Emsellem}(2004)}]{Cappellari2004}
{Cappellari}, M., \& {Emsellem}, E. 2004, \pasp, 116, 138,
  \dodoi{10.1086/381875}

\bibitem[{{Cappellari} {et~al.}(2006){Cappellari}, {Bacon}, {Bureau}, {Damen},
  {Davies}, {de Zeeuw}, {Emsellem}, {Falc{\'o}n-Barroso}, {Krajnovi{\'c}},
  {Kuntschner}, {McDermid}, {Peletier}, {Sarzi}, {van den Bosch}, \& {van de
  Ven}}]{Cappellari2006}
{Cappellari}, M., {Bacon}, R., {Bureau}, M., {et~al.} 2006, \mnras, 366, 1126,
  \dodoi{10.1111/j.1365-2966.2005.09981.x}

\bibitem[{{Cappellari} {et~al.}(2013){Cappellari}, {Scott}, {Alatalo}, {Blitz},
  {Bois}, {Bournaud}, {Bureau}, {Crocker}, {Davies}, {Davis}, {de Zeeuw},
  {Duc}, {Emsellem}, {Khochfar}, {Krajnovi{\'c}}, {Kuntschner}, {McDermid},
  {Morganti}, {Naab}, {Oosterloo}, {Sarzi}, {Serra}, {Weijmans}, \&
  {Young}}]{Cappellari2013}
{Cappellari}, M., {Scott}, N., {Alatalo}, K., {et~al.} 2013, \mnras, 432, 1709,
  \dodoi{10.1093/mnras/stt562}

\bibitem[{{Chabrier}(2003)}]{Chabrier2003}
{Chabrier}, G. 2003, \pasp, 115, 763, \dodoi{10.1086/376392}

\bibitem[{{Chang} {et~al.}(2015){Chang}, {van der Wel}, {da Cunha}, \&
  {Rix}}]{Chang2015}
{Chang}, Y.-Y., {van der Wel}, A., {da Cunha}, E., \& {Rix}, H.-W. 2015, \apjs,
  219, 8, \dodoi{10.1088/0067-0049/219/1/8}

\bibitem[{{Chang} {et~al.}(2013){Chang}, {van der Wel}, {Rix}, {Holden},
  {Bell}, {McGrath}, {Wuyts}, {H{\"a}ussler}, {Barden}, {Faber}, {Mozena},
  {Ferguson}, {Guo}, {Galametz}, {Grogin}, {Kocevski}, {Koekemoer}, {Dekel},
  {Huang}, {Hathi}, \& {Donley}}]{Chang2013}
{Chang}, Y.-Y., {van der Wel}, A., {Rix}, H.-W., {et~al.} 2013, \apj, 773, 149,
  \dodoi{10.1088/0004-637X/773/2/149}

\bibitem[{{Charlot} \& {Fall}(2000)}]{Charlot2000}
{Charlot}, S., \& {Fall}, S.~M. 2000, \apj, 539, 718, \dodoi{10.1086/309250}

\bibitem[{{Chevance} {et~al.}(2012){Chevance}, {Weijmans}, {Damjanov},
  {Abraham}, {Simard}, {van den Bergh}, {Caris}, \&
  {Glazebrook}}]{Chevance2012}
{Chevance}, M., {Weijmans}, A.-M., {Damjanov}, I., {et~al.} 2012, \apjl, 754,
  L24, \dodoi{10.1088/2041-8205/754/2/L24}

\bibitem[{{Courteau} \& {Rix}(1999)}]{Courteau1999}
{Courteau}, S., \& {Rix}, H.-W. 1999, \apj, 513, 561, \dodoi{10.1086/306872}

\bibitem[{{Cowie} {et~al.}(1996){Cowie}, {Songaila}, {Hu}, \&
  {Cohen}}]{Cowie1996}
{Cowie}, L.~L., {Songaila}, A., {Hu}, E.~M., \& {Cohen}, J.~G. 1996, \aj, 112,
  839, \dodoi{10.1086/118058}

\bibitem[{{da Cunha} {et~al.}(2008){da Cunha}, {Charlot}, \&
  {Elbaz}}]{daCunha2008}
{da Cunha}, E., {Charlot}, S., \& {Elbaz}, D. 2008, \mnras, 388, 1595,
  \dodoi{10.1111/j.1365-2966.2008.13535.x}

\bibitem[{{Darvish} {et~al.}(2017){Darvish}, {Mobasher}, {Martin}, {Sobral},
  {Scoville}, {Stroe}, {Hemmati}, \& {Kartaltepe}}]{Darvish2017}
{Darvish}, B., {Mobasher}, B., {Martin}, D.~C., {et~al.} 2017, \apj, 837, 16,
  \dodoi{10.3847/1538-4357/837/1/16}

\bibitem[{{Darvish} {et~al.}(2015){Darvish}, {Mobasher}, {Sobral}, {Scoville},
  \& {Aragon-Calvo}}]{Darvish2015}
{Darvish}, B., {Mobasher}, B., {Sobral}, D., {Scoville}, N., \& {Aragon-Calvo},
  M. 2015, \apj, 805, 121, \dodoi{10.1088/0004-637X/805/2/121}

\bibitem[{{de Graaff} {et~al.}(2020){de Graaff}, {Bezanson}, {Franx}, {van der
  Wel}, {Bell}, {D'Eugenio}, {Holden}, {Maseda}, {Muzzin}, {Pacifici}, {van de
  Sande}, {Sobral}, {Straatman}, \& {Wu}}]{deGraaff2020}
{de Graaff}, A., {Bezanson}, R., {Franx}, M., {et~al.} 2020, \apjl, 903, L30,
  \dodoi{10.3847/2041-8213/abc428}

\bibitem[{{di Serego Alighieri} {et~al.}(2005){di Serego Alighieri}, {Vernet},
  {Cimatti}, {Lanzoni}, {Cassata}, {Ciotti}, {Daddi}, {Mignoli}, {Pignatelli},
  {Pozzetti}, {Renzini}, {Rettura}, \& {Zamorani}}]{diSerego2005}
{di Serego Alighieri}, S., {Vernet}, J., {Cimatti}, A., {et~al.} 2005, \aap,
  442, 125, \dodoi{10.1051/0004-6361:20053168}

\bibitem[{{Djorgovski} \& {Davis}(1987)}]{Djorgovski1987}
{Djorgovski}, S., \& {Davis}, M. 1987, \apj, 313, 59, \dodoi{10.1086/164948}

\bibitem[{{Dressler} {et~al.}(1987){Dressler}, {Lynden-Bell}, {Burstein},
  {Davies}, {Faber}, {Terlevich}, \& {Wegner}}]{Dressler1987}
{Dressler}, A., {Lynden-Bell}, D., {Burstein}, D., {et~al.} 1987, \apj, 313,
  42, \dodoi{10.1086/164947}

\bibitem[{{Faber} {et~al.}(1987){Faber}, {Dressler}, {Davies}, {Burstein},
  {Lynden Bell}, {Terlevich}, \& {Wegner}}]{Faber1987}
{Faber}, S.~M., {Dressler}, A., {Davies}, R.~L., {et~al.} 1987, in Nearly
  Normal Galaxies. From the Planck Time to the Present, ed. S.~M. {Faber}, 175

\bibitem[{{Ferrero} {et~al.}(2020){Ferrero}, {Navarro}, {Abadi}, {Benavides},
  \& {Mast}}]{Ferrero2020}
{Ferrero}, I., {Navarro}, J.~F., {Abadi}, M.~G., {Benavides}, J.~A., \& {Mast},
  D. 2020, arXiv e-prints, arXiv:2009.03916.
\newblock \doarXiv{2009.03916}

\bibitem[{{Fischer} {et~al.}(2017){Fischer}, {Bernardi}, \&
  {Meert}}]{Fischer2017}
{Fischer}, J.~L., {Bernardi}, M., \& {Meert}, A. 2017, \mnras, 467, 490,
  \dodoi{10.1093/mnras/stx136}

\bibitem[{{Fischer} {et~al.}(2019){Fischer}, {Dom{\'\i}nguez S{\'a}nchez}, \&
  {Bernardi}}]{Fischer2019}
{Fischer}, J.~L., {Dom{\'\i}nguez S{\'a}nchez}, H., \& {Bernardi}, M. 2019,
  \mnras, 483, 2057, \dodoi{10.1093/mnras/sty3135}

\bibitem[{{Forbes} {et~al.}(1998){Forbes}, {Ponman}, \& {Brown}}]{Forbes1998}
{Forbes}, D.~A., {Ponman}, T.~J., \& {Brown}, R. J.~N. 1998, \apjl, 508, L43,
  \dodoi{10.1086/311715}

\bibitem[{{Franx} {et~al.}(2008){Franx}, {van Dokkum}, {F{\"o}rster Schreiber},
  {Wuyts}, {Labb{\'e}}, \& {Toft}}]{Franx2008}
{Franx}, M., {van Dokkum}, P.~G., {F{\"o}rster Schreiber}, N.~M., {et~al.}
  2008, \apj, 688, 770, \dodoi{10.1086/592431}

\bibitem[{{Gargiulo} {et~al.}(2009){Gargiulo}, {Haines}, {Merluzzi}, {Smith},
  {La Barbera}, {Busarello}, {Lucey}, {Mercurio}, \&
  {Capaccioli}}]{Gargiulo2009}
{Gargiulo}, A., {Haines}, C.~P., {Merluzzi}, P., {et~al.} 2009, \mnras, 397,
  75, \dodoi{10.1111/j.1365-2966.2009.14801.x}

\bibitem[{{Graves} \& {Faber}(2010)}]{Graves2010III}
{Graves}, G.~J., \& {Faber}, S.~M. 2010, \apj, 717, 803,
  \dodoi{10.1088/0004-637X/717/2/803}

\bibitem[{{Graves} {et~al.}(2009){Graves}, {Faber}, \&
  {Schiavon}}]{Graves2009II}
{Graves}, G.~J., {Faber}, S.~M., \& {Schiavon}, R.~P. 2009, \apj, 698, 1590,
  \dodoi{10.1088/0004-637X/698/2/1590}

\bibitem[{Harris {et~al.}(2020)Harris, Millman, van~der Walt, Gommers,
  Virtanen, Cournapeau, Wieser, Taylor, Berg, Smith, Kern, Picus, Hoyer, van
  Kerkwijk, Brett, Haldane, del R{'{\i}}o, Wiebe, Peterson,
  G{'{e}}rard-Marchant, Sheppard, Reddy, Weckesser, Abbasi, Gohlke, \&
  Oliphant}]{numpy}
Harris, C.~R., Millman, K.~J., van~der Walt, S.~J., {et~al.} 2020, Nature, 585,
  357, \dodoi{10.1038/s41586-020-2649-2}

\bibitem[{{Hill} {et~al.}(2019){Hill}, {van der Wel}, {Franx}, {Muzzin},
  {Skelton}, {Momcheva}, {van Dokkum}, \& {Whitaker}}]{Hill2019}
{Hill}, A.~R., {van der Wel}, A., {Franx}, M., {et~al.} 2019, \apj, 871, 76,
  \dodoi{10.3847/1538-4357/aaf50a}

\bibitem[{{Hogg} {et~al.}(2010){Hogg}, {Bovy}, \& {Lang}}]{Hogg2010}
{Hogg}, D.~W., {Bovy}, J., \& {Lang}, D. 2010, arXiv e-prints, arXiv:1008.4686.
\newblock \doarXiv{1008.4686}

\bibitem[{{Holden} {et~al.}(2010){Holden}, {van der Wel}, {Kelson}, {Franx}, \&
  {Illingworth}}]{Holden2010}
{Holden}, B.~P., {van der Wel}, A., {Kelson}, D.~D., {Franx}, M., \&
  {Illingworth}, G.~D. 2010, \apj, 724, 714,
  \dodoi{10.1088/0004-637X/724/1/714}

\bibitem[{{Holden} {et~al.}(2012){Holden}, {van der Wel}, {Rix}, \&
  {Franx}}]{Holden2012}
{Holden}, B.~P., {van der Wel}, A., {Rix}, H.-W., \& {Franx}, M. 2012, \apj,
  749, 96, \dodoi{10.1088/0004-637X/749/2/96}

\bibitem[{{Hopkins} {et~al.}(2009){Hopkins}, {Bundy}, {Murray}, {Quataert},
  {Lauer}, \& {Ma}}]{Hopkins2009}
{Hopkins}, P.~F., {Bundy}, K., {Murray}, N., {et~al.} 2009, \mnras, 398, 898,
  \dodoi{10.1111/j.1365-2966.2009.15062.x}

\bibitem[{{Hunter}(2007)}]{matplotlib}
{Hunter}, J.~D. 2007, Computing in Science Engineering, 9, 90,
  \dodoi{10.1109/MCSE.2007.55}

\bibitem[{{Hyde} \& {Bernardi}(2009)}]{HydeBenardi2009}
{Hyde}, J.~B., \& {Bernardi}, M. 2009, \mnras, 396, 1171,
  \dodoi{10.1111/j.1365-2966.2009.14783.x}

\bibitem[{{Ilbert} {et~al.}(2013){Ilbert}, {McCracken}, {Le F{\`e}vre},
  {Capak}, {Dunlop}, {Karim}, {Renzini}, {Caputi}, {Boissier}, {Arnouts},
  {Aussel}, {Comparat}, {Guo}, {Hudelot}, {Kartaltepe}, {Kneib}, {Krogager},
  {Le Floc'h}, {Lilly}, {Mellier}, {Milvang-Jensen}, {Moutard}, {Onodera},
  {Richard}, {Salvato}, {Sanders}, {Scoville}, {Silverman}, {Taniguchi},
  {Tasca}, {Thomas}, {Toft}, {Tresse}, {Vergani}, {Wolk}, \&
  {Zirm}}]{Ilbert2013}
{Ilbert}, O., {McCracken}, H.~J., {Le F{\`e}vre}, O., {et~al.} 2013, \aap, 556,
  A55, \dodoi{10.1051/0004-6361/201321100}

\bibitem[{{Joachimi} {et~al.}(2015){Joachimi}, {Singh}, \&
  {Mandelbaum}}]{Joachimi2015}
{Joachimi}, B., {Singh}, S., \& {Mandelbaum}, R. 2015, \mnras, 454, 478,
  \dodoi{10.1093/mnras/stv1962}

\bibitem[{{J{\o}rgensen} \& {Chiboucas}(2013)}]{Joergensen2013}
{J{\o}rgensen}, I., \& {Chiboucas}, K. 2013, \aj, 145, 77,
  \dodoi{10.1088/0004-6256/145/3/77}

\bibitem[{{Jorgensen} {et~al.}(1996){Jorgensen}, {Franx}, \&
  {Kjaergaard}}]{Jorgensen1996}
{Jorgensen}, I., {Franx}, M., \& {Kjaergaard}, P. 1996, \mnras, 280, 167,
  \dodoi{10.1093/mnras/280.1.167}

\bibitem[{{J{\o}rgensen} {et~al.}(2019){J{\o}rgensen}, {Hunter}, {O'Neill},
  {Chiboucas}, {Cole}, {Toft}, \& {Schiavon}}]{Jorgensen2019}
{J{\o}rgensen}, I., {Hunter}, L.~C., {O'Neill}, C.~R., {et~al.} 2019, \apj,
  881, 42, \dodoi{10.3847/1538-4357/ab2d9d}

\bibitem[{{Kauffmann} {et~al.}(2003){Kauffmann}, {Heckman}, {White}, {Charlot},
  {Tremonti}, {Peng}, {Seibert}, {Brinkmann}, {Nichol}, {SubbaRao}, \&
  {York}}]{Kauffmann2003}
{Kauffmann}, G., {Heckman}, T.~M., {White}, S. D.~M., {et~al.} 2003, \mnras,
  341, 54, \dodoi{10.1046/j.1365-8711.2003.06292.x}

\bibitem[{{Kriek} {et~al.}(2009){Kriek}, {van Dokkum}, {Labb{\'e}}, {Franx},
  {Illingworth}, {Marchesini}, \& {Quadri}}]{Kriek2009}
{Kriek}, M., {van Dokkum}, P.~G., {Labb{\'e}}, I., {et~al.} 2009, \apj, 700,
  221, \dodoi{10.1088/0004-637X/700/1/221}

\bibitem[{{La Barbera} {et~al.}(2010){La Barbera}, {Lopes}, {de Carvalho}, {de
  La Rosa}, \& {Berlind}}]{LaBarbera2010}
{La Barbera}, F., {Lopes}, P.~A.~A., {de Carvalho}, R.~R., {de La Rosa}, I.~G.,
  \& {Berlind}, A.~A. 2010, \mnras, 408, 1361,
  \dodoi{10.1111/j.1365-2966.2010.17273.x}

\bibitem[{{Laigle} {et~al.}(2016){Laigle}, {McCracken}, {Ilbert}, {Hsieh},
  {Davidzon}, {Capak}, {Hasinger}, {Silverman}, {Pichon}, {Coupon}, {Aussel},
  {Le Borgne}, {Caputi}, {Cassata}, {Chang}, {Civano}, {Dunlop}, {Fynbo},
  {Kartaltepe}, {Koekemoer}, {Le F{\`e}vre}, {Le Floc'h}, {Leauthaud}, {Lilly},
  {Lin}, {Marchesi}, {Milvang-Jensen}, {Salvato}, {Sanders}, {Scoville},
  {Smolcic}, {Stockmann}, {Taniguchi}, {Tasca}, {Toft}, {Vaccari}, \&
  {Zabl}}]{Laigle2016}
{Laigle}, C., {McCracken}, H.~J., {Ilbert}, O., {et~al.} 2016, \apjs, 224, 24,
  \dodoi{10.3847/0067-0049/224/2/24}

\bibitem[{{Leja} {et~al.}(2019{\natexlab{a}}){Leja}, {Tacchella}, \&
  {Conroy}}]{Leja2019b}
{Leja}, J., {Tacchella}, S., \& {Conroy}, C. 2019{\natexlab{a}}, \apjl, 880,
  L9, \dodoi{10.3847/2041-8213/ab2f8c}

\bibitem[{{Leja} {et~al.}(2019{\natexlab{b}}){Leja}, {Johnson}, {Conroy}, {van
  Dokkum}, {Speagle}, {Brammer}, {Momcheva}, {Skelton}, {Whitaker}, {Franx}, \&
  {Nelson}}]{Leja2019a}
{Leja}, J., {Johnson}, B.~D., {Conroy}, C., {et~al.} 2019{\natexlab{b}}, \apj,
  877, 140, \dodoi{10.3847/1538-4357/ab1d5a}

\bibitem[{{Magoulas} {et~al.}(2012){Magoulas}, {Springob}, {Colless}, {Jones},
  {Campbell}, {Lucey}, {Mould}, {Jarrett}, {Merson}, \&
  {Brough}}]{Magoulas2012}
{Magoulas}, C., {Springob}, C.~M., {Colless}, M., {et~al.} 2012, \mnras, 427,
  245, \dodoi{10.1111/j.1365-2966.2012.21421.x}

\bibitem[{{McCracken} {et~al.}(2012){McCracken}, {Milvang-Jensen}, {Dunlop},
  {Franx}, {Fynbo}, {Le F{\`e}vre}, {Holt}, {Caputi}, {Goranova}, {Buitrago},
  {Emerson}, {Freudling}, {Hudelot}, {L{\'o}pez-Sanjuan}, {Magnard}, {Mellier},
  {M{\o}ller}, {Nilsson}, {Sutherland}, {Tasca}, \& {Zabl}}]{McCracken2012}
{McCracken}, H.~J., {Milvang-Jensen}, B., {Dunlop}, J., {et~al.} 2012, \aap,
  544, A156, \dodoi{10.1051/0004-6361/201219507}

\bibitem[{{Meert} {et~al.}(2015){Meert}, {Vikram}, \& {Bernardi}}]{Meert2015}
{Meert}, A., {Vikram}, V., \& {Bernardi}, M. 2015, \mnras, 446, 3943,
  \dodoi{10.1093/mnras/stu2333}

\bibitem[{{Mo} {et~al.}(1998){Mo}, {Mao}, \& {White}}]{Mo1998}
{Mo}, H.~J., {Mao}, S., \& {White}, S. D.~M. 1998, \mnras, 295, 319,
  \dodoi{10.1046/j.1365-8711.1998.01227.x}

\bibitem[{{Mowla} {et~al.}(2019){Mowla}, {van Dokkum}, {Brammer}, {Momcheva},
  {van der Wel}, {Whitaker}, {Nelson}, {Bezanson}, {Muzzin}, {Franx},
  {MacKenty}, {Leja}, {Kriek}, \& {Marchesini}}]{Mowla2019}
{Mowla}, L.~A., {van Dokkum}, P., {Brammer}, G.~B., {et~al.} 2019, \apj, 880,
  57, \dodoi{10.3847/1538-4357/ab290a}

\bibitem[{{Muzzin} {et~al.}(2013{\natexlab{a}}){Muzzin}, {Marchesini},
  {Stefanon}, {Franx}, {McCracken}, {Milvang-Jensen}, {Dunlop}, {Fynbo},
  {Brammer}, {Labb{\'e}}, \& {van Dokkum}}]{Muzzin2013b}
{Muzzin}, A., {Marchesini}, D., {Stefanon}, M., {et~al.} 2013{\natexlab{a}},
  \apj, 777, 18, \dodoi{10.1088/0004-637X/777/1/18}

\bibitem[{{Muzzin} {et~al.}(2013{\natexlab{b}}){Muzzin}, {Marchesini},
  {Stefanon}, {Franx}, {Milvang-Jensen}, {Dunlop}, {Fynbo}, {Brammer},
  {Labb{\'e}}, \& {van Dokkum}}]{Muzzin2013a}
---. 2013{\natexlab{b}}, \apjs, 206, 8, \dodoi{10.1088/0067-0049/206/1/8}

\bibitem[{{Naab} {et~al.}(2009){Naab}, {Johansson}, \& {Ostriker}}]{Naab2009}
{Naab}, T., {Johansson}, P.~H., \& {Ostriker}, J.~P. 2009, \apjl, 699, L178,
  \dodoi{10.1088/0004-637X/699/2/L178}

\bibitem[{{Newman} {et~al.}(2018){Newman}, {Belli}, {Ellis}, \&
  {Patel}}]{Newman2018}
{Newman}, A.~B., {Belli}, S., {Ellis}, R.~S., \& {Patel}, S.~G. 2018, \apj,
  862, 126, \dodoi{10.3847/1538-4357/aacd4f}

\bibitem[{{Peng} {et~al.}(2010){Peng}, {Ho}, {Impey}, \& {Rix}}]{Peng2010}
{Peng}, C.~Y., {Ho}, L.~C., {Impey}, C.~D., \& {Rix}, H.-W. 2010, \aj, 139,
  2097, \dodoi{10.1088/0004-6256/139/6/2097}

\bibitem[{{Pforr} {et~al.}(2012){Pforr}, {Maraston}, \& {Tonini}}]{Pforr2012}
{Pforr}, J., {Maraston}, C., \& {Tonini}, C. 2012, \mnras, 422, 3285,
  \dodoi{10.1111/j.1365-2966.2012.20848.x}

\bibitem[{{Prichard} {et~al.}(2017){Prichard}, {Davies}, {Beifiori}, {Chan},
  {Cappellari}, {Houghton}, {Mendel}, {Bender}, {Galametz}, {Saglia}, {Stott},
  {Wilman}, {Lewis}, {Sharples}, \& {Wegner}}]{Prichard2017}
{Prichard}, L.~J., {Davies}, R.~L., {Beifiori}, A., {et~al.} 2017, \apj, 850,
  203, \dodoi{10.3847/1538-4357/aa96a6}

\bibitem[{{Renzini} \& {Ciotti}(1993)}]{Renzini1993}
{Renzini}, A., \& {Ciotti}, L. 1993, \apjl, 416, L49, \dodoi{10.1086/187068}

\bibitem[{{Roberts} \& {Haynes}(1994)}]{Roberts1994}
{Roberts}, M.~S., \& {Haynes}, M.~P. 1994, \araa, 32, 115,
  \dodoi{10.1146/annurev.aa.32.090194.000555}

\bibitem[{{Romanowsky} \& {Fall}(2012)}]{Romanowsky2012}
{Romanowsky}, A.~J., \& {Fall}, S.~M. 2012, \apjs, 203, 17,
  \dodoi{10.1088/0067-0049/203/2/17}

\bibitem[{{Saglia} {et~al.}(2010){Saglia}, {S{\'a}nchez-Bl{\'a}zquez},
  {Bender}, {Simard}, {Desai}, {Arag{\'o}n-Salamanca}, {Milvang-Jensen},
  {Halliday}, {Jablonka}, {Noll}, {Poggianti}, {Clowe}, {De Lucia},
  {Pell{\'o}}, {Rudnick}, {Valentinuzzi}, {White}, \& {Zaritsky}}]{Saglia2010}
{Saglia}, R.~P., {S{\'a}nchez-Bl{\'a}zquez}, P., {Bender}, R., {et~al.} 2010,
  \aap, 524, A6, \dodoi{10.1051/0004-6361/201014703}

\bibitem[{{Saglia} {et~al.}(2016){Saglia}, {S{\'a}nchez-Bl{\'a}zquez},
  {Bender}, {Simard}, {Desai}, {Arag{\'o}n-Salamanca}, {Milvang-Jensen},
  {Halliday}, {Jablonka}, {Noll}, {Poggianti}, {Clowe}, {De Lucia},
  {Pell{\'o}}, {Rudnick}, {Valentinuzzi}, {White}, \& {Zaritsky}}]{Saglia2016}
---. 2016, \aap, 596, C1, \dodoi{10.1051/0004-6361/201014703e}

\bibitem[{{Saracco} {et~al.}(2020){Saracco}, {Gargiulo}, {La Barbera},
  {Annunziatella}, \& {Marchesini}}]{Saracco2020}
{Saracco}, P., {Gargiulo}, A., {La Barbera}, F., {Annunziatella}, M., \&
  {Marchesini}, D. 2020, \mnras, 491, 1777, \dodoi{10.1093/mnras/stz3109}

\bibitem[{{Schechter} {et~al.}(2014){Schechter}, {Pooley}, {Blackburne}, \&
  {Wambsganss}}]{Schechter2014}
{Schechter}, P.~L., {Pooley}, D., {Blackburne}, J.~A., \& {Wambsganss}, J.
  2014, \apj, 793, 96, \dodoi{10.1088/0004-637X/793/2/96}

\bibitem[{{Scoville} {et~al.}(2007){Scoville}, {Abraham}, {Aussel}, {Barnes},
  {Benson}, {Blain}, {Calzetti}, {Comastri}, {Capak}, {Carilli}, {Carlstrom},
  {Carollo}, {Colbert}, {Daddi}, {Ellis}, {Elvis}, {Ewald}, {Fall},
  {Franceschini}, {Giavalisco}, {Green}, {Griffiths}, {Guzzo}, {Hasinger},
  {Impey}, {Kneib}, {Koda}, {Koekemoer}, {Lefevre}, {Lilly}, {Liu},
  {McCracken}, {Massey}, {Mellier}, {Miyazaki}, {Mobasher}, {Mould}, {Norman},
  {Refregier}, {Renzini}, {Rhodes}, {Rich}, {Sanders}, {Schiminovich},
  {Schinnerer}, {Scodeggio}, {Sheth}, {Shopbell}, {Taniguchi}, {Tyson}, {Urry},
  {Van Waerbeke}, {Vettolani}, {White}, \& {Yan}}]{Scoville2007}
{Scoville}, N., {Abraham}, R.~G., {Aussel}, H., {et~al.} 2007, \apjs, 172, 38,
  \dodoi{10.1086/516580}

\bibitem[{{Simard} {et~al.}(2011){Simard}, {Mendel}, {Patton}, {Ellison}, \&
  {McConnachie}}]{Simard2011}
{Simard}, L., {Mendel}, J.~T., {Patton}, D.~R., {Ellison}, S.~L., \&
  {McConnachie}, A.~W. 2011, \apjs, 196, 11, \dodoi{10.1088/0067-0049/196/1/11}

\bibitem[{{Somerville} {et~al.}(2008){Somerville}, {Barden}, {Rix}, {Bell},
  {Beckwith}, {Borch}, {Caldwell}, {H{\"a}u{\ss}ler}, {Heymans}, {Jahnke},
  {Jogee}, {McIntosh}, {Meisenheimer}, {Peng}, {S{\'a}nchez}, {Wisotzki}, \&
  {Wolf}}]{Somerville2008}
{Somerville}, R.~S., {Barden}, M., {Rix}, H.-W., {et~al.} 2008, \apj, 672, 776,
  \dodoi{10.1086/523661}

\bibitem[{{Straatman} {et~al.}(2018){Straatman}, {van der Wel}, {Bezanson},
  {Pacifici}, {Gallazzi}, {Wu}, {Noeske}, {Bari{\v s}i{\'c}}, {Bell},
  {Brammer}, {Calhau}, {Chauke}, {Franx}, {van Houdt}, {Labb{\'e}}, {Maseda},
  {Mu{\~n}oz-Mateos}, {Muzzin}, {van de Sande}, {Sobral}, \&
  {Spilker}}]{Straatman2018}
{Straatman}, C.~M.~S., {van der Wel}, A., {Bezanson}, R., {et~al.} 2018, \apjs,
  239, 27, \dodoi{10.3847/1538-4365/aae37a}

\bibitem[{{Suess} {et~al.}(2019){Suess}, {Kriek}, {Price}, \&
  {Barro}}]{Suess2019a}
{Suess}, K.~A., {Kriek}, M., {Price}, S.~H., \& {Barro}, G. 2019, \apj, 877,
  103, \dodoi{10.3847/1538-4357/ab1bda}

\bibitem[{{Taylor} {et~al.}(2010){Taylor}, {Franx}, {Brinchmann}, {van der
  Wel}, \& {van Dokkum}}]{Taylor2010}
{Taylor}, E.~N., {Franx}, M., {Brinchmann}, J., {van der Wel}, A., \& {van
  Dokkum}, P.~G. 2010, \apj, 722, 1, \dodoi{10.1088/0004-637X/722/1/1}

\bibitem[{{Taylor} {et~al.}(2015){Taylor}, {Hopkins}, {Baldry},
  {Bland-Hawthorn}, {Brown}, {Colless}, {Driver}, {Norberg}, {Robotham},
  {Alpaslan}, {Brough}, {Cluver}, {Gunawardhana}, {Kelvin}, {Liske},
  {Conselice}, {Croom}, {Foster}, {Jarrett}, {Lara-Lopez}, \&
  {Loveday}}]{Taylor2015}
{Taylor}, E.~N., {Hopkins}, A.~M., {Baldry}, I.~K., {et~al.} 2015, \mnras, 446,
  2144, \dodoi{10.1093/mnras/stu1900}

\bibitem[{{Toft} {et~al.}(2017){Toft}, {Zabl}, {Richard}, {Gallazzi},
  {Zibetti}, {Prescott}, {Grillo}, {Man}, {Lee}, {G{\'o}mez-Guijarro},
  {Stockmann}, {Magdis}, \& {Steinhardt}}]{Toft2017}
{Toft}, S., {Zabl}, J., {Richard}, J., {et~al.} 2017, \nat, 546, 510,
  \dodoi{10.1038/nature22388}

\bibitem[{{Trujillo} {et~al.}(2004){Trujillo}, {Burkert}, \&
  {Bell}}]{Trujillo2004}
{Trujillo}, I., {Burkert}, A., \& {Bell}, E.~F. 2004, \apjl, 600, L39,
  \dodoi{10.1086/381528}

\bibitem[{{Tully} \& {Fisher}(1977)}]{TullyFisher1977}
{Tully}, R.~B., \& {Fisher}, J.~R. 1977, \aap, 500, 105

\bibitem[{{van de Sande} {et~al.}(2014){van de Sande}, {Kriek}, {Franx},
  {Bezanson}, \& {van Dokkum}}]{vdSande2014}
{van de Sande}, J., {Kriek}, M., {Franx}, M., {Bezanson}, R., \& {van Dokkum},
  P.~G. 2014, \apjl, 793, L31, \dodoi{10.1088/2041-8205/793/2/L31}

\bibitem[{{van de Sande} {et~al.}(2015){van de Sande}, {Kriek}, {Franx},
  {Bezanson}, \& {van Dokkum}}]{vdSande2015}
---. 2015, \apj, 799, 125, \dodoi{10.1088/0004-637X/799/2/125}

\bibitem[{{van de Sande} {et~al.}(2013){van de Sande}, {Kriek}, {Franx}, {van
  Dokkum}, {Bezanson}, {Bouwens}, {Quadri}, {Rix}, \& {Skelton}}]{vdSande2013}
{van de Sande}, J., {Kriek}, M., {Franx}, M., {et~al.} 2013, \apj, 771, 85,
  \dodoi{10.1088/0004-637X/771/2/85}

\bibitem[{{van de Sande} {et~al.}(2018){van de Sande}, {Scott},
  {Bland-Hawthorn}, {Brough}, {Bryant}, {Colless}, {Cortese}, {Croom},
  {d'Eugenio}, {Foster}, {Goodwin}, {Konstantopoulos}, {Lawrence}, {McDermid},
  {Medling}, {Owers}, {Richards}, \& {Sharp}}]{vdSande2018}
{van de Sande}, J., {Scott}, N., {Bland-Hawthorn}, J., {et~al.} 2018, Nature
  Astronomy, 2, 483, \dodoi{10.1038/s41550-018-0436-x}

\bibitem[{{van de Sande} {et~al.}(2019){van de Sande}, {Lagos}, {Welker},
  {Bland-Hawthorn}, {Schulze}, {Remus}, {Bah{\'e}}, {Brough}, {Bryant},
  {Cortese}, {Croom}, {Devriendt}, {Dubois}, {Goodwin}, {Konstantopoulos},
  {Lawrence}, {Medling}, {Pichon}, {Richards}, {Sanchez}, {Scott}, \&
  {Sweet}}]{vdSande2019}
{van de Sande}, J., {Lagos}, C. D.~P., {Welker}, C., {et~al.} 2019, \mnras,
  484, 869, \dodoi{10.1093/mnras/sty3506}

\bibitem[{{van der Wel} {et~al.}(2004){van der Wel}, {Franx}, {van Dokkum}, \&
  {Rix}}]{vdWel2004}
{van der Wel}, A., {Franx}, M., {van Dokkum}, P.~G., \& {Rix}, H.~W. 2004,
  \apjl, 601, L5, \dodoi{10.1086/381887}

\bibitem[{{van der Wel} {et~al.}(2005){van der Wel}, {Franx}, {van Dokkum},
  {Rix}, {Illingworth}, \& {Rosati}}]{vdWel2005}
{van der Wel}, A., {Franx}, M., {van Dokkum}, P.~G., {et~al.} 2005, \apj, 631,
  145, \dodoi{10.1086/430464}

\bibitem[{{van der Wel} {et~al.}(2012){van der Wel}, {Bell}, {H{\"a}ussler},
  {McGrath}, {Chang}, {Guo}, {McIntosh}, {Rix}, {Barden}, {Cheung}, {Faber},
  {Ferguson}, {Galametz}, {Grogin}, {Hartley}, {Kartaltepe}, {Kocevski},
  {Koekemoer}, {Lotz}, {Mozena}, {Peth}, \& {Peng}}]{vdWel2012}
{van der Wel}, A., {Bell}, E.~F., {H{\"a}ussler}, B., {et~al.} 2012, \apjs,
  203, 24, \dodoi{10.1088/0067-0049/203/2/24}

\bibitem[{{van der Wel} {et~al.}(2014{\natexlab{a}}){van der Wel}, {Franx},
  {van Dokkum}, {Skelton}, {Momcheva}, {Whitaker}, {Brammer}, {Bell}, {Rix},
  {Wuyts}, {Ferguson}, {Holden}, {Barro}, {Koekemoer}, {Chang}, {McGrath},
  {H{\"a}ussler}, {Dekel}, {Behroozi}, {Fumagalli}, {Leja}, {Lundgren},
  {Maseda}, {Nelson}, {Wake}, {Patel}, {Labb{\'e}}, {Faber}, {Grogin}, \&
  {Kocevski}}]{vdWel2014}
{van der Wel}, A., {Franx}, M., {van Dokkum}, P.~G., {et~al.}
  2014{\natexlab{a}}, \apj, 788, 28, \dodoi{10.1088/0004-637X/788/1/28}

\bibitem[{{van der Wel} {et~al.}(2014{\natexlab{b}}){van der Wel}, {Chang},
  {Bell}, {Holden}, {Ferguson}, {Giavalisco}, {Rix}, {Skelton}, {Whitaker},
  {Momcheva}, {Brammer}, {Kassin}, {Martig}, {Dekel}, {Ceverino}, {Koo},
  {Mozena}, {van Dokkum}, {Franx}, {Faber}, \& {Primack}}]{vdWel2014b}
{van der Wel}, A., {Chang}, Y.-Y., {Bell}, E.~F., {et~al.} 2014{\natexlab{b}},
  \apjl, 792, L6, \dodoi{10.1088/2041-8205/792/1/L6}

\bibitem[{{van der Wel} {et~al.}(2016){van der Wel}, {Noeske}, {Bezanson},
  {Pacifici}, {Gallazzi}, {Franx}, {Mu{\~n}oz-Mateos}, {Bell}, {Brammer},
  {Charlot}, {Chauk{\'e}}, {Labb{\'e}}, {Maseda}, {Muzzin}, {Rix}, {Sobral},
  {van de Sande}, {van Dokkum}, {Wild}, \& {Wolf}}]{vdWel2016}
{van der Wel}, A., {Noeske}, K., {Bezanson}, R., {et~al.} 2016, \apjs, 223, 29,
  \dodoi{10.3847/0067-0049/223/2/29}

\bibitem[{{van Dokkum} \& {Conroy}(2012)}]{vanDokkum2012}
{van Dokkum}, P.~G., \& {Conroy}, C. 2012, \apj, 760, 70,
  \dodoi{10.1088/0004-637X/760/1/70}

\bibitem[{{van Dokkum} \& {Franx}(1996)}]{vanDokkum1996}
{van Dokkum}, P.~G., \& {Franx}, M. 1996, \mnras, 281, 985,
  \dodoi{10.1093/mnras/281.3.985}

\bibitem[{{van Dokkum} \& {Franx}(2001)}]{VanDokkum2001_pbias}
---. 2001, \apj, 553, 90, \dodoi{10.1086/320645}

\bibitem[{{van Dokkum} {et~al.}(2001){van Dokkum}, {Franx}, {Kelson}, \&
  {Illingworth}}]{VanDokkum2001}
{van Dokkum}, P.~G., {Franx}, M., {Kelson}, D.~D., \& {Illingworth}, G.~D.
  2001, \apjl, 553, L39, \dodoi{10.1086/320502}

\bibitem[{{van Dokkum} \& {van der Marel}(2007)}]{VanDokkum2007}
{van Dokkum}, P.~G., \& {van der Marel}, R.~P. 2007, \apj, 655, 30,
  \dodoi{10.1086/509633}

\bibitem[{Virtanen {et~al.}(2020)Virtanen, Gommers, Oliphant, Haberland, Reddy,
  Cournapeau, Burovski, Peterson, Weckesser, Bright, {van der Walt}, Brett,
  Wilson, Millman, Mayorov, Nelson, Jones, Kern, Larson, Carey, Polat, Feng,
  Moore, {VanderPlas}, Laxalde, Perktold, Cimrman, Henriksen, Quintero, Harris,
  Archibald, Ribeiro, Pedregosa, {van Mulbregt}, \& {SciPy 1.0
  Contributors}}]{scipy}
Virtanen, P., Gommers, R., Oliphant, T.~E., {et~al.} 2020, Nature Methods, 17,
  261, \dodoi{10.1038/s41592-019-0686-2}

\bibitem[{{Westfall} {et~al.}(2019){Westfall}, {Cappellari}, {Bershady},
  {Bundy}, {Belfiore}, {Ji}, {Law}, {Schaefer}, {Shetty}, {Tremonti}, {Yan},
  {Andrews}, {Brownstein}, {Cherinka}, {Coccato}, {Drory}, {Maraston},
  {Parikh}, {S{\'a}nchez-Gallego}, {Thomas}, {Weijmans}, {Barrera-Ballesteros},
  {Du}, {Goddard}, {Li}, {Masters}, {Ibarra Medel}, {S{\'a}nchez}, {Yang},
  {Zheng}, \& {Zhou}}]{Westfall2019}
{Westfall}, K.~B., {Cappellari}, M., {Bershady}, M.~A., {et~al.} 2019, arXiv
  e-prints.
\newblock \doarXiv{1901.00856}

\bibitem[{{Whitaker} {et~al.}(2011){Whitaker}, {Labb{\'e}}, {van Dokkum},
  {Brammer}, {Kriek}, {Marchesini}, {Quadri}, {Franx}, {Muzzin}, {Williams},
  {Bezanson}, {Illingworth}, {Lee}, {Lundgren}, {Nelson}, {Rudnick}, {Tal}, \&
  {Wake}}]{Whitaker2011}
{Whitaker}, K.~E., {Labb{\'e}}, I., {van Dokkum}, P.~G., {et~al.} 2011, \apj,
  735, 86, \dodoi{10.1088/0004-637X/735/2/86}

\bibitem[{{Wu} {et~al.}(2018){Wu}, {van der Wel}, {Gallazzi}, {Bezanson},
  {Pacifici}, {Straatman}, {Franx}, {Bari{\v{s}}i{\'c}}, {Bell}, {Brammer},
  {Calhau}, {Chauke}, {van Houdt}, {Maseda}, {Muzzin}, {Rix}, {Sobral},
  {Spilker}, {van de Sande}, {van Dokkum}, \& {Wild}}]{Wu2018}
{Wu}, P.-F., {van der Wel}, A., {Gallazzi}, A., {et~al.} 2018, \apj, 855, 85,
  \dodoi{10.3847/1538-4357/aab0a6}

\bibitem[{{Wuyts} {et~al.}(2004){Wuyts}, {van Dokkum}, {Kelson}, {Franx}, \&
  {Illingworth}}]{Wuyts2004}
{Wuyts}, S., {van Dokkum}, P.~G., {Kelson}, D.~D., {Franx}, M., \&
  {Illingworth}, G.~D. 2004, \apj, 605, 677, \dodoi{10.1086/381746}

\bibitem[{{Wyder} {et~al.}(2007){Wyder}, {Martin}, {Schiminovich}, {Seibert},
  {Budav{\'a}ri}, {Treyer}, {Barlow}, {Forster}, {Friedman}, {Morrissey},
  {Neff}, {Small}, {Bianchi}, {Donas}, {Heckman}, {Lee}, {Madore}, {Milliard},
  {Rich}, {Szalay}, {Welsh}, \& {Yi}}]{Wyder2007}
{Wyder}, T.~K., {Martin}, D.~C., {Schiminovich}, D., {et~al.} 2007, \apjs, 173,
  293, \dodoi{10.1086/521402}

\bibitem[{{Zahid} {et~al.}(2016){Zahid}, {Damjanov}, {Geller}, {Hwang}, \&
  {Fabricant}}]{Zahid2016}
{Zahid}, H.~J., {Damjanov}, I., {Geller}, M.~J., {Hwang}, H.~S., \&
  {Fabricant}, D.~G. 2016, \apj, 821, 101, \dodoi{10.3847/0004-637X/821/2/101}

\bibitem[{{Zaritsky} {et~al.}(2008){Zaritsky}, {Zabludoff}, \&
  {Gonzalez}}]{Zaritsky2008}
{Zaritsky}, D., {Zabludoff}, A.~I., \& {Gonzalez}, A.~H. 2008, \apj, 682, 68,
  \dodoi{10.1086/529577}

\bibitem[{{Zwaan} {et~al.}(1995){Zwaan}, {van der Hulst}, {de Blok}, \&
  {McGaugh}}]{Zwaan1995}
{Zwaan}, M.~A., {van der Hulst}, J.~M., {de Blok}, W.~J.~G., \& {McGaugh},
  S.~S. 1995, \mnras, 273, L35, \dodoi{10.1093/mnras/273.1.L35}

\end{thebibliography}
\bibliographystyle{aasjournal}



\end{document}